\documentclass[fleqn,usenatbib]{mnras}

\usepackage{newtxtext,newtxmath}

\usepackage[T1]{fontenc}
\usepackage{ae,aecompl}

\usepackage{graphicx}	

\usepackage{color}
\usepackage{url}
\usepackage{xspace}
\usepackage[normalem]{ulem}

\newcommand{\msun}{{\rm M}$_{\odot}$\xspace}
\newcommand{\ors}{{\sc Outer Rim} simulation\xspace}

\newcommand{\Eq}[1]{Eq.~\ref{#1}}
\newcommand{\Eqs}[1]{Eqs.~\ref{#1}}
\newcommand{\Fig}[1]{Figure~\ref{#1}}

\newcommand{\Sec}[1]{\S~\ref{#1}}
\newcommand{\Tab}[1]{Table~\ref{#1}}

\newcommand{\parts}{PART}
\newcommand{\nfw}{\textsc{NFW}}
\newcommand{\fsat}{$f_{\rm sat}$}
\newcommand{\vinfall}{$v^{\rm infall}$}

\newcommand{\ANLHEP}{HEP Division, Argonne National Laboratory, Lemont, IL 60439, USA}

\newcommand{\ARI}{Astrophysics Research Institute, Liverpool John Moores University, 146 Brownlow Hill, Liverpool L3 5RF, UK}
\newcommand{\AstroUC}{Instituto de Astrof\'isica, Pontificia Universidad Cat\'{o}lica de Chile, Santiago, Chile}

\newcommand{\CCAPP}{Center for Cosmology and AstroParticle Physics, The Ohio State University, Columbus, OH 43212}

\newcommand{\DFT}{Departamento de F\'isica Te\'orica, Facultad de Ciencias, Universidad Aut\'onoma de Madrid, 28049 Cantoblanco, Madrid, Spain}

\newcommand{\ICC}{ICC, University of Barcelona, IEEC-UB, Mart\' i i Franqu\` es, 1, E08028 Barcelona, Spain}

\newcommand{\ICG}{Institute of Cosmology \& Gravitation, University of Portsmouth, Dennis Sciama Building, Portsmouth PO1 3FX, UK.}

\newcommand{\IfA}{Institute for Astronomy, University of Edinburgh, Royal Observatory, EH9 3HJ Edinburgh, United Kingdom}

\newcommand{\IFT}{Instituto de F\'isica Teorica UAM-CSIC, Universidad Aut\'onoma de Madrid, 28049 Cantoblanco, Madrid, Spain}

\newcommand{\IRFU}{IRFU, CEA, Universit\'e Paris-Saclay, F-91191 Gif-sur-Yvette, France}

\newcommand{\PSU}{The Pennsylvania State University, University Park, PA 16802}

\newcommand{\UWaterloo}{Department of Physics and Astronomy, University of Waterloo, 200 University Ave W, Waterloo, ON N2L 3G1, Canada}

\vspace{-1.5cm}  
\title[HOD for eBOSS ELGs]{
\begin{minipage}{7.03 in}
\vskip -0.4 in
\begin{flushright}
{\rm \small IFT-UAM/CSIC-20-109}
\end{flushright}
\end{minipage}\\
\vskip 0.0in
\vspace{-0.65cm}     
The Completed SDSS-IV extended Baryon Oscillation Spectroscopic Survey: exploring the Halo Occupation Distribution model for Emission Line Galaxies}

\author[Avila et al.]{                          
\parbox{\textwidth}{
\Large  
S. Avila $^{1,\ 2,\ 3}$\thanks{E-mail: santiago.avila@uam.es},
V. Gonzalez-Perez$^{4,\ 3}$\thanks{E-mail: violetagp@protonmail.com}, F. G. Mohammad$^{5,\ 6}$, A. de Mattia$^{7}$, C. Zhao$^{8}$,  
A. Raichoor$^{8}$, A. Tamone$^{8}$, S. Alam$^{9}$ , J. Bautista$^3$, D. Bianchi$^{10\ 11}$, E. Burtin$^{7}$, M. J. Chapman $^{5,\ 6}$, Chia-Hsun Chuang,$^{12}$, J. Comparat$^{13}$, K. Dawson$^{14}$, T. Divers$^3$, H. du Mas des Bourboux$^{14}$, H. Gil-Marin$^{10,\ 11}$, E. M. Mueller$^{15}$,  S. Habib$^{16}$, K. Heitmann$^{16}$, V. Ruhlmann-Kleider$^{7}$, N. Padilla$^{17}$, W. J. Percival$^{5,\ 6,\ {18}}$, A. J. Ross$^{19}$, H. J. Seo$^{19}$, D. P. Schneider$^{20}$,  G. Zhao$^{21,\ 3}$
}
\\
\\
\parbox{\textwidth}{
\scriptsize
$^{1}$\DFT\\
$^{2}$\IFT\\
$^{3}$\ICG\\
$^{4}$\ARI\\     
$^{5}$Waterloo Centre for Astrophysics, University of Waterloo, 200 University Ave W, Waterloo, ON N2L 3G1, Canada\\  
$^{6}$\UWaterloo\\  
$^{7}$\IRFU\\  
$^{8}$Institute of Physics, Laboratory of Astrophysics, Ecole Polytechnique Federale de Lausanne (EPFL), 1290 Versoix, Switzerland\\  
$^{9}$\IfA\\
$^{10}$\ICC\\
$^{11}$Institut  d{\textquotesingle}Estudis  Espacials  de  Catalunya  (IEEC),  E08034  Barcelona,  Spain\\
$^{12}$Kavli Institute for Particle Astrophysics and Cosmology, Stanford University, 452 Lomita Mall, Stanford, CA 94305, USA\\
$^{13}$Max-Planck Institut f{\"u}r extraterrestrische Physik, Postfach 1312, D-85741 Garching bei M{\"u}nchen, Germany \\
$^{14}$University of Utah, Department of Physics and Astronomy, 115 S 1400 E, Salt Lake City, UT 84112, USA\\ 
$^{15}$University of Oxford, Oxford OX1~3RH, United Kingdom\\
$^{16}$\ANLHEP\\
{$^{17}$\AstroUC\\}
{$^{18}$Perimeter Institute for Theoretical Physics, 31 Caroline St. North, Waterloo, ON N2L 2Y5, Canada}\\
$^{19}$\CCAPP\\
$^{20}$\PSU\\
$^{21}$National Astronomy Observatories, Chinese Academy of Science, Beijing, 100012, P.R. China 
}
}

\pubyear{2020}
\begin{document}
\label{firstpage}
\pagerange{\pageref{firstpage}--\pageref{lastpage}}
\vspace{-5.5cm}  
\maketitle
\vspace{-5.5cm}

\begin{minipage}{7.03 in}
\vskip 0.4 in
\begin{flushleft}
{\rm \small $\star$  E-mail:  santiago.avila@uam.es  \\ 
$\dag$ E-mail: violetagp@protonmail.com}
\end{flushleft}
\end{minipage}

\begin{abstract}
We study the modelling of the Halo Occupation Distribution (HOD) for the eBOSS DR16 Emission Line Galaxies (ELGs). Motivated by previous theoretical and observational studies, we consider different physical effects that can change how ELGs populate haloes. 
We explore the shape of the average HOD, the fraction of satellite galaxies, their probability distribution function (PDF), and their density and velocity profiles.
Our baseline HOD shape was fitted to a semi-analytical model of galaxy formation and evolution, with a decaying occupation of central ELGs at high halo masses. 
We consider Poisson and sub/super-Poissonian PDFs for satellite assignment. We model both NFW and particle profiles for satellite positions, also allowing for decreased concentrations. We model velocities with the virial theorem and particle velocity distributions. Additionally, we introduce a velocity bias and a net infall velocity. 
We study how these choices impact the clustering statistics while keeping the number density and bias fixed to that from eBOSS ELGs. The projected correlation function, $w_p$, captures most of the effects from the PDF and satellites profile. The quadrupole, $\xi_2$, captures most of the effects coming from the velocity profile. We find that the impact of the mean HOD shape is subdominant relative to the rest of choices. 
We fit the clustering of the eBOSS DR16 ELG data under different combinations of the above assumptions. 
The catalogues presented here have been analysed in companion papers, showing that eBOSS RSD+BAO measurements are insensitive to the details of galaxy physics considered here. These catalogues are made publicly available.
\end{abstract}

\vspace{-4.5cm}  


%
\vspace*{-1.55cm}
\hspace{-9.4cm}
\parbox{\textwidth}{
\scriptsize        

}

\newpage

\clearpage

\section{Introduction}

The large scale structure of the Universe contains a wealth of information on cosmology. The spectroscopic galaxy surveys via studies of galaxy clustering have measured the scale of Baryonic Acoustic Oscillations (BAO), Redshift Space Distortions (RSD) and constrained Primordial Non-Gaussianities among other cosmological probes. Whereas previous wide angle surveys such as BOSS have targeted Luminous Red Galaxies (LRG) at low redshift (z<0.6) achieving unprecedented constrains on Cosmology from the Large Scale Structure \citep{BOSS}, this type of galaxies becomes harder to target at higher redshifts. That is why recently the focus has turned to new targets such as the star-forming Emission Line Galaxies (ELGs) that can be targeted at higher redshifts, such as eBOSS 0.6<z<1.1 \citep{dawson16} and DESI 0.6<z<1.6 \citep{desi16}, or observed via slit-less spectroscopy \citep[Euclid, $0.9<z<1.8$][]{laureijs11}.
                                          
 The eBOSS survey, part of the SDSS-IV program \citep{SDSSIV}, has created the largest spectroscopic sample of star-forming ELGs to date with a final sample of $173,736$ ELGs in the redshift range $0.6<z<1.1$ \citep{elg}. This was achieved after measuring the spectra of ELG targets selected from the DECaLS photometric survey. Additionally, eBOSS has surveyed close to half a  million of LRGs in the range $0.6<z<1.0$ and $\sim 330,000$ QSOs in the range $0.8<z<2.0$ \citep{ross20,qso}.

$N$-body simulations play an important role in the Large-Scale Structure analysis in order to validate theoretical tools used for data analysis. One unknown is the way a certain type of galaxies relates to the underlying dark matter distribution. One way to explore this relation is with Semi-Analytical Models (SAMs) of galaxy formation and evolution. However, these models require dark matter halo merger trees and high mass resolution, rarely available in simulations able to probe beyond the $({\rm Gpc}/h)^3$-scale volume. For the very large scale simulations, typically, dark matter haloes are populated with Halo Occupation Distribution (HOD) models \citep[e.g.][]{seljak00, cooray02, berlind02, zheng05, zehavi05} or, alternatively, with (Sub-) Halo Abundance Matching techniques \citep[e.g.][]{favole16}. In the original HOD models, galaxy properties are determined from the halo mass of the host. More sophisticated HOD models try to encapsulate the assembly bias, i.e. dependence of halo clustering on properties other than halo mass, by introducing secondary parameters~\citep[e.g.][]{hearin2016, zehavi2018}. We defer an exploration of the assembly bias for ELGs for future studies.

In this paper, we applied a series of ELG HOD models motivated from previous theoretical studies to produce galaxy catalogues based on the \ors dark-matter only simulation. We then compare and fit the clustering statistics of these mocks to ELG data from eBOSS.

The produced mocks are used in companion papers \citep{alam2020,tamone2020,demattia2020,elg} to test the robustness of theoretical models of galaxy anisotropic clustering against variations in the HOD model, finding that those models can be trusted at least (with a conservative budget computation) to within 1.8\%, 1.5\% 3.3\% for, respectively, $\{\alpha_\parallel, \alpha_\perp, f\sigma_8\}$ (Alcock-Paczynski and growth rate parameters), well below the statistical errors for eBOSS. 

 Our baseline model takes the shape of the average HOD ($\langle N (M) \rangle$) presented in \citet{vgp2018} 
 for ELGs, which we approximate by a step-wise Gaussian plus a decaying power-law for central galaxies, whereas we model satellites following a power-law above a certain halo mass. We complete the baseline model with the following usual assumptions in the generation of galaxy mock catalogues regarding the assignment of satellites \citep[e.g. ][]{Carretero15, halotools}: the number of satellites $N_{\rm sat}$ is drawn from a Poisson distribution, their spatial distribution follows a NFW profile and the velocity profiles can be inferred from the virial theorem. 
 
 We then create alternative models by varying each of the baseline assumptions of the HOD and studying their effect on clustering via monopole, quadrupole and projected correlation functions ($\xi_0$, $\xi_2$, $w_p$). 
 We explore the choice of a Gaussian or a smooth step-function as an alternative for the shape of the mean HOD for central galaxies, $\langle N_{\rm cen} (M) \rangle$, options explored in other HOD studies \citep{zehavi05,favole16,guo19}. 
 Motivated by the study in \citet{jimenez2019}, we also study non-Poissonian Probability Distribution Functions (PDF, $P(N|\langle N \rangle)$) for populating haloes with satellite galaxies: the nearest-integer and the negative binomial distribution.

 As an alternative to NFW profiles, we also use the particle distribution within haloes and allowed for a rescaling of the halo concentrations. The latter is motivated by some studies predicting ELGs should be in the outskirts of haloes \citep[e.g.][]{orsi2018}.
 With respect to satellite velocities, in the case of NFW profiles, velocities follow a Gaussian distribution with a dispersion predicted by the virial theorem, whereas in the case of using particles, satellites take the particle velocity. We also include a velocity bias parameter that modulates the dispersion of satellite velocities with respect to the halo velocity. Another ingredient for our alternative models consists on adding a net infall velocity as motivated by \citet{orsi2018}.

This study is part of a coordinated release of the final eBOSS measurements of BAO and RSD in the clustering of not only ELGs \citep{elg,tamone2020,demattia2020}, but also luminous red galaxies \citep{LRG_corr,gil-marin20}, and quasars \citep{hou20a,neveux20a}. An essential component of these studies is the construction of data catalogues \citep{ross20,qso}, approximated mock catalogues \citep{lin20a,EZmocks}, and N-body simulations based mock catalogues for assessing systematic errors \citep{alam2020,rossi2020,smith2020}, as the ones presented here. At the highest redshifts ($z>2.1$), the coordinated release of final eBOSS measurements includes measurements of BAO in the $Ly_{\alpha}$ forest \citep{2020duMasdesBourbouxH}. The cosmological interpretation of these results in combination with the final BOSS results and other probes is found in
\citet{eBOSS_Cosmology}. \footnote{A description of eBOSS and a link to its associated publications can be found in \url{https://www.sdss.org/surveys/eboss/}}

The plan of this paper is as follows. In \Sec{sec:sim} we introduce the \ors. In \Sec{sec:data} we describe the input observational data: ELG catalogue, summary statistics and correlation functions. In \Sec{sec:mocks} we describe the different halo occupation models we use to generate mock catalogues varying the mean halo occupation distribution for central and satellite galaxies (\Sec{sec:hod}), the probability distribution function (\Sec{sec:pdf}), the radial (\Sec{sec:profile}) and velocity (\Sec{sec:velocities}) distributions of satellite galaxies. In \Sec{sec:results} we present mocks that best fit the data under different assumptions. Finally, we conclude in \Sec{sec:conclusions} and discuss our results and future prospects.

  
\section{The \ors}\label{sec:sim}

\begin{figure}  
\includegraphics[width=0.485\textwidth]{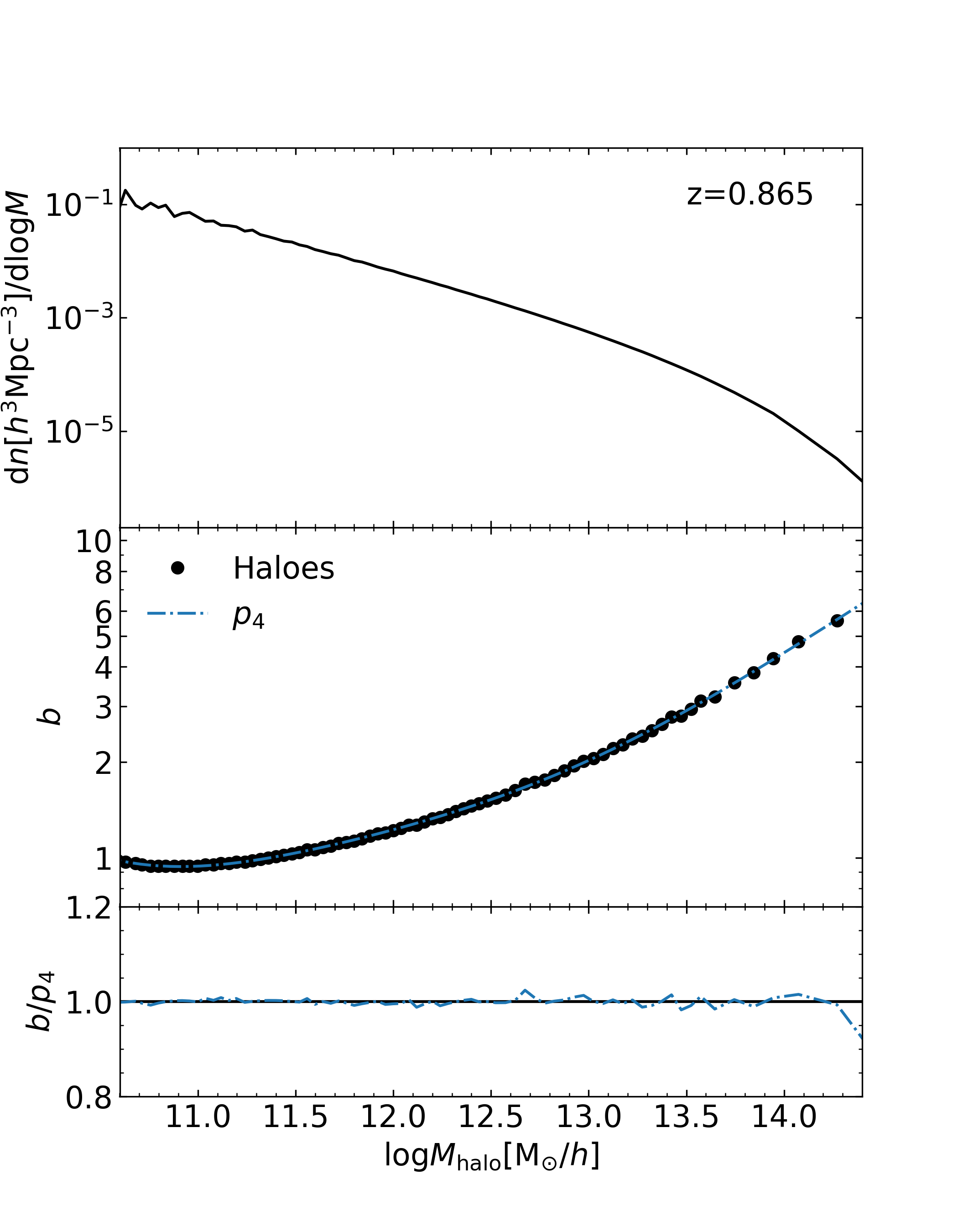}
\caption{
{\it Top:} The halo mass function from the \ors output at $z=0.865$. {\it Middle:} The halo bias function, 
fitted for separations $20 \leq r(h^{-1}{\rm Mpc}) \leq 80$ for the \ors dark matter haloes as a function of the halo mass (see text), black circles. A fourth order polynomial fit, $p_4$, to the bias function, $b(M)$ is also shown. {\it Bottom:} Ratio of the bias, $b(M)$ tothe polynomial fit, $p_4$.
}
\label{fig:biasf} 
\end{figure}

\begin{table} 
\begin{center}
\begin{tabular}{|l|r|}
\hline
   $\Omega_{\rm cdm}$             &    0.220 \\
   $\Omega_{\rm b}$              & 0.0448 \\
   $\Omega_\Lambda$             &    0.7352 \\
   $h \equiv H_0$/(100 km\,s$^{-1}$\,Mpc$^{-1})$ &  0.71 \\
   $\sigma_8$                   &    0.8 \\
   $n_{\rm s}$                  &    0.963 \\
   \hline
   $z_{\rm snap}$  & 0.865  \\
   Volume       &   $(3000 h^{-1}{\rm Mpc})^3$  \\
   $N$             &    10240$^3$ \\
   $m_{\rm p}$    &    $1.85\times10^{9}h^{-1}$\msun \\
\hline
\end{tabular}
\end{center}
  \caption{The \ors cosmological and setup parameters. The cosmological parameters \citep{wmap7}:  $\Omega_{\rm cdm}$, $\Omega_{\rm b}$ and $\Omega_\Lambda$ are the average densities of cold dark matter, baryonic matter and vacuum energy  in units of the critical density today, $H_0$ is the Hubble parameter, $\sigma_8$ is the 
  rms of the matter fluctuations at $8~h^{-1}{\rm Mpc}$ and $n_{\rm s}$ is the spectral index of the primordial power spectrum. Simulation parameters: Redshift of the snapshot used in this paper, volume of the simulation box, number of particles in the simulation and particle mass resolution. 
  }
  \label{tab:sim}
\end{table}

The \ors~\citep{heitmann2019} was run assuming a cosmology consistent with the 7$^{\rm th}$ year release from WMAP \citep{wmap7}, as summarised in \Tab{tab:sim}, which will be our fiducial cosmology throughout the paper. The \ors has outputs at 99 redshifts (34 between $1<z<3.5$). 
Haloes with at least 20 particle members, were identified at $0<z<10$ using the Friends-of-Friends (FoF) algorithm \citep{Davis1985} with a linking length of $b=0.168$. Technical properties of the \ors are summarised in the second part of \Tab{tab:sim}, some of these are similar to those from its predecessor, the Q Continuum simulation \citep{heitmann15}.

For this study, we use a snapshot at fixed $z=0.865$, close to the effective redshift of the distribution of eBOSS/ELGs \citep[$z_{\rm eff}=0.845$, ][]{elg}.
Two key functions for the Halo Occupation Distribution Model (HOD) are the halo mass function and the bias function of the \ors shown in~\Fig{fig:biasf}. We will see in \Sec{sec:mocks} how their integrals are used to put basic constraints  on the HOD parameters. When refering to the bias in this study we will always refer to the linear local bias, shown to describe accurately the clustering of haloes/galaxies at large scales. 

In order to compute the \ors bias function we split the simulation box in 27 cubes of $l=1{\rm Gpc}/h$ side. We further split the halo catalogues by mass in logarithmic mass bins with $\Delta ( {\rm log} M) = 0.1$\footnote{Throughout this paper we use ${ log}$ for the decimal logarithm and take its argument in units of $M_{\odot}/h$} 
 for ${\rm log}M< 12.5$, and making larger intervals at higher masses in order to decrease the shot noise. We compute the correlation function (details in  \Sec{sec:2PCF}) in real space $\xi_{M_i}(r)$ for each mass bin and subbox. Then, we compute the mean $\bar{\xi}_{M_i}$, and corresponding standard deviation, $\sigma( \xi_{M_i})$ for the 27 subboxes. We find the bias $b_i$ that minimises the $\chi^2$ defined as:

\begin{equation}
    \chi^2(b_i) = \sum_r \Bigg( \frac{\xi_{\rm lin}(r)\cdot b_i^2 -\bar{\xi}_{M_i}(r) }{\sigma( \xi_{M_i})(r)} \Bigg)^2 \, .
    \label{eq:chi2_sim}
\end{equation}

The summation above is done in the range $20\leq r\leq80 \ {\rm Mpc}/h$ (with this choice we avoid non-linearities affecting small scales and the BAO feature). We found (here, and also in \Sec{sec:stat}, \Eq{eq:chi2_data}) that the fits 
are more stable against noise and scale cuts, when using logarithmic binning for the data/mock correlation functions and most stable in the selected range of scales. We compute $\xi_{\rm lin} (r)$ by Hankel transforming the linear power spectrum obtained from \textsc{camb} \citep{camb}:

\begin{equation}
   \xi_{\rm lin} (r) = \frac{1}{2\pi^2}\int_0^\infty P_{\rm lin}(k) j_0(k r) k^2 {\rm d}k \ .
\end{equation}

 The bias as a function of halo mass, $b({\rm log}M)$, is then fitted to polynomials of orders from 2 to 5. We discard the points beyond ${\rm log}M=14.4$ as they yield a poor fit. At those masses the binning becomes too coarse if we want to keep low contribution from shot noise and the definition of the bin centre becomes ambiguous as the halo mass function decays exponentially within the bin. We find a good fit with a polynomial of order $k=4$ or larger. For the rest of the work we use the fourth order polynomial fit, $p_4$, to approximate the bias function. This fit is shown in \Fig{fig:biasf}.

\subsection{Correlation Functions from {\sc Outer Rim} mocks}
\label{sec:2PCF}

    When computing 2-point correlation functions (2PCF) in the simulation, we will always assume periodic conditions for the subboxes. This will introduce a small error in the boundaries, but we expect it to be negligible, given that our subbox size is much larger than the maximum scale used here.

A way to correct for the boundary conditions would be to introduce a random catalogue in order to account for the geometry of the subbox, as  we will do for the survey geometry (see Section \ref{sec:PIP}). However, this process would significantly slow down the 2PCF calculations as random catalogues are typically required to have at least 10 times more objects than the data catalogue in order to avoid introducing extra noise. 

In the cases we compute correlations in redshift space, we use 

\begin{equation}
    \vec{s}=\vec{r}+\frac{1+z}{H(z)}    \frac{\vec v \cdot \vec r}{| \vec r |}\frac{\vec r}{|\vec r|}\, ,
\end{equation}
with $\vec s$ representing the halo/galaxy position in redshift space, $\vec r$ in real space, $\vec v$ its comoving velocity and $H(z)$ the Hubble parameter at redshift $z$. We adopt the plane-parallel approximation and assume the $Z$-axis as the line-of-sigh.

Given the assumed boundary conditions, $\xi(r,\mu)$, is calculated using simply the natural estimator: $1+\xi(r,\mu)= DD(r,\mu)/(n \Delta V)$, where DD is the number of galaxy/halo pairs with separation between $r$ and $r + \Delta r$ and an orientation between $\mu$ and $\mu+\Delta \mu$ (where $\mu$ is the cosine of the angle with respect to the line of sight) and the denominator is the average number of galaxies found in the volume $\Delta V$ of the spherical shell of radius $r$ and thickness $dr$. 
This was computed with a modified version of the code \textsc{cute} ~\citep{cute} \footnote{\url{https://github.com/damonge/CUTE}}.

We then compute the multipoles, integrating over the Legendre polynomials $L_\ell$:

\begin{equation}
    \xi_\ell(s) = (2\ell+1)\int_0^1 \xi(s,\mu)L_\ell(\mu)d\mu \, .
    \label{eq:multipoles}
\end{equation}

The projected two-point correlation function, $w_p(r_p)$, is obtained with the publicly available Python package \texttt{Corrfunc}\footnote{\url{https://github.com/manodeep/Corrfunc}~\citep{corrfunc}}. This correlation function removes most of the effect of peculiar velocities by integrating along the line of sight.

\begin{equation}
    w_p (r_p) = 2 \int^{\pi_{\rm max}}_0 \xi(r_p,\pi) d\pi \, ,
    \label{eq:wp}
\end{equation}
where we use $\pi_{\rm max}=80 {\rm Mpc}/h$.
In this case, $\xi$ is also computed using the natural estimator, but counting pairs in bins of comoving distance parallel ($\pi$) and perpendicular ($r_p$) to the line of sight.


\section{The eBOSS ELG data}
\label{sec:data}

 The eBOSS survey \citep{dawson16} has observed a spectroscopic sample of star-forming emission line galaxies~\citep[ELGs,][]{raichoor2016,raichoor2017} in fields both in the North and South Galactic Caps (NGC, SGC). The redshift of these galaxies has been identified using the $\left[\ion{O}{\,\sc II}\right]$ doublet, with rest-frame wavelengths of $\lambda=3727,3729$\AA.

We aim at reproducing the eBOSS ELG number density and linear bias with the mock catalogues we generate in this work. We describe below how these quantities are measured from the data. We also describe the correlation functions of the data used as an input for \Sec{sec:results}.

\subsection{eBOSS LSS ELG catalogues}
\label{sec:catalogue}

Optical and near-infrared cosmological surveys are targeting star-forming ELGs at $0.5<z<2$, as these galaxies can provide a high enough effective volume to measure the BAO with high precision~\citep[e.g.][]{comparat16}. Star-forming ELGs present strong spectral emission lines that allow for a robust determination of their redshifts in a small observing time, maximising the volume covered by the survey~\citep[e.g.][]{okada16}. Strong spectral lines can also be produced by galaxies with nuclear activity, AGNs. These are expected to be hosted by different average dark matter haloes than star-forming galaxies. No broad band lines have been found among the eBOSS ELG sample and only a small fraction of eBOSS ELGs are expected to be AGNs in the redshift range under study, based on previous studies~\citep{comparat2013}.

Here we use the data from the DR16 ELG clustering catalogue, described in \citet{elg}, which only includes ELGs with a good redshift determination. There are 173,736 eBOSS ELGs, within the redshift range $0.6<z<1.1$ and with an effective redshift of $z_{\rm eff}= 0.845$.  The effective areas in the North and South fields are: $A_{\rm NGC,eff}=369{\rm deg}^2$, 
$A_{\rm SGC,eff}=358{\rm deg}^2$. The catalogue also contains the weights to correct individual galaxies for systematic errors:

\begin{equation}
    w_{\rm ELG}=w_{\rm sys}\ w_{\rm no\, z}\ w_{\rm CP} \, ,
\end{equation}\label{eq:weights}
due to the photometric target selection, $w_{\rm sys}$; redshift failures, $w_{\rm no\, z}$; and $w_{\rm CP}$ includes the 'close pairs' fiber collision correction adopted in eBOSS cosmological analysis~\citep{elg,ross20,demattia2020}. We will not use the $w_{CP}$ weights in this study as the fiber collision effect will be accounted for by the PIP + ANG weights, which are more accurate, specially at small scales (see \Sec{sec:PIP}). Additionally, the standard inverse variance weights $w_{\rm FKP}$ are applied, in order to improve signal to noise ratio \citep{FKP}. 

\subsection{Number density of the eBOSS ELGs}
\label{sec:stat}

We aim to reproduce the abundance and clustering of the eBOSS ELGs in the \ors.
The number density and bias derived from the data will need to rely in the assumed cosmology, that in this case is that of the \ors (\Tab{tab:sim}). 
\footnote{Note that the latter is different from the cosmology used for the main data analysis (\citet{demattia2020,tamone2020}, with namely $\Omega_M=0.31$).}

First, we compute the number density of ELGs for the NGC, the SGC and the combination of both: 

\begin{equation}
\begin{split}
    & \bar{n} = \frac{N_{\rm eff}}{V_{\rm eff}} \, \, \ \ \ \  \bar{n}_{\rm eBOSS}=  2.187 \cdot 10^{-4}  ({\rm Mpc}/h)^{-3}    \\
    & \bar{n}_{\rm SGC} = 2.267 \cdot 10^{-4} ({\rm Mpc}/h)^{-3}, \, \bar{n}_{\rm NGC}=  2.110 \cdot 10^{-4} ({\rm Mpc}/h)^{-3}, 
\label{eq:ndata}
\end{split}
\end{equation}

 with
\begin{equation}
 V= \frac{1}{3} \Bigg( \chi( z_{\rm max} )^3- \chi( z_{\rm min} )^3  \Bigg) \cdot 
    A_{\rm eff}\ \Bigg( \frac{\pi}{180 {\rm deg} }\Bigg)^2
 \end{equation} 

giving a volume of $V_{\rm SGC,eff}=0.410 ({\rm Gpc}/h)^3$, $V_{\rm NGC,eff}=0.424 ({\rm Gpc}/h)^3$, $V_{\rm tot,eff}=0.834 ({\rm Gpc}/h)^3$. Note that this $V_{\rm eff}$ refers purely to the observational volume taken into account, given the effective area and redshift range. In other  studies $V_{\rm eff}$ may refer to the equivalent volume of a cosmic variance limited sample with the same power spectrum variance (i.e. the shot noise being interpreted as a reduction of the effective volume).  

The eBOSS number density $\bar{n}_{\rm eBOSS}$ will be used as a reference throughout the paper. In some occasions we will use a factor 7 or 10 higher density in order to measure more accurately the clustering of the mocks. 

\subsection{Linear bias of eBOSS ELGs}
The second quantity that we want to measure from the data is the large scale bias $b$. A simple way is to compute the bias from the monopole of the data, and fit it using the Kaiser factor \citep{Kaiser} together with linear perturbation theory:

\begin{equation}
    \xi_{0, \rm lin} (s) =\Big(b + \frac{2}{3}b f + \frac{1}{5}f^2 \Big) \xi_{\rm lin} (s)\, ,
    \label{eq:Kaiser}
\end{equation}
with $f$ the log-derivative of the growth factor. Within the standard assumption of General Relativity, $d{\rm log}D/d{\rm log}a\approx \Omega_m(a)^{0.545}$ \citep{Peebles80,gamma}. We fix $f$ using this approximation for the {\sc Outer Rim} cosmology. 

In earlier versions of the (data and mock) catalogues we used the approach described above. For the mocks presented in this paper, we decided to fit the bias of the data by rescaling the shape of $\xi_0(s)$ of a mock with similar amplitude as that of the data (derived in earlier versions of the mocks):

\begin{equation}
    \xi_{0} (b,s) = \frac{\big(b + \frac{2}{3}b f + \frac{1}{5}f^2\big)}{(b_{\rm mock} + \frac{2}{3}b_{\rm mock} f + \frac{1}{5}f^2)} \xi_{0, \rm mock} (s)
    \label{eq:fit_bias_mock}
\end{equation}

This method encapsulates better the non-linearities, as they are present in the simulation and are expected in the data. 

To measure the large scale linear bias, we follow a similar approach as in \Sec{sec:sim} and split the simulation in 27 subboxes of size $l=1 Gpc/$h and comparable volume $V_{\rm mock}=l^3$ to the total eBOSS ELG effective volume. The $b_{\rm mock}$ is computed as explained in \Sec{sec:hod} (\Eq{eq:bias}) and validated using the same approach as in \Sec{sec:sim} (\Eq{eq:chi2_sim}) for the haloes. We compute the monopole in each subbox and the global mean $\xi_{0, \rm mock}$, to be input to \Eq{eq:fit_bias_mock}, and standard deviation $\sigma_{\rm mock}$, so that we can compute: 

\begin{equation}
    \chi^2(b) = \sum_s \Big( \frac{\xi_0(b,s) -\xi_{0,{\rm data}}(s) }{\sigma_{\rm mock}(s)  \sqrt{V_{\rm mock}/V_{\rm eBOSS}}}\Big)^2\,,
     \label{eq:chi2_data}
\end{equation}

and minimise $\chi^2$ to get the best fit bias $b$ with a 1-$\sigma$ confidence interval corresponding to $\Delta \chi^2=1$. 
The fit to the data bias is more stable when using logarithmic binning for both the analytical approximation and the mocks. We use the range of scales $20<s<55 {\rm Mpc}/h$, where the $p$-values are good for all cases (combining linear versus logarithmic scale, using the linear theory or mocks for the fits and using either or both galactic caps). Following this procedure we obtain: 

\begin{equation}
   \begin{split}
    b_{\rm eBOSS} &=1.320 \pm 0.014 \, , \\
    b_{\rm SGC} &= 1.310  \pm 0.020 \, , \,
    b_{\rm NGC} = 1.330  \pm 0.020 \, .
    \end{split}
\end{equation}

As we find that the NGC and SGC have compatible biases, in the remainder of this paper we use the combined data, unless otherwise specified. Note again that this differs from bias values found in complementary studies, where the assumed cosmology was different.  

\subsection{Weighted correlation functions}
\label{sec:PIP}

eBOSS measured spectra using optical fibres positioned in pre-drilled plates at the 2.5m Sloan Telescope~\citep{gunn2006, smee2013}. The plates have a field of view of $\sim 7{\deg}^2$ and can hold up to 1000 fibres. Typically 100 fibres are used for calibration and 900 for science targets. Each fibre plus its ferrule has a diameter of 62''. The fibre collision scale imposes a minimum separation between objects that can be observed simultaneously. The eBOSS survey repeats observations of the same regions of the sky, allowing to observe some of the target objects, missed at previous passes, due to fibre collision. However, not all the targets will be (spectroscopically) observed once the survey is finished.

In this work, we find the best HOD models by fitting different clustering statistics measured from the model catalogues to the observed ones (details of this procedure can be found in~\S\ref{sec:results}). The data derived from observations has been corrected for the effect of missing spectra for photometric targets using the Pairwise-inverse-probability (PIP) weighting and the angular up-weighting (ANG) techniques~\citep{pip}. When pairs of galaxies are counted for calculating the correlation functions, these techniques modify the standard fibre collision correction, $w_{\rm CP}$ (see \Eq{eq:weights}).

The PIP weight of a given pair of galaxies is defined as the inverse of the probability of this pair being assigned a fibre within the ensemble set from which the survey undertaken is considered to be randomly drawn~\citep{pip, pipeBOSS}. This weighting scheme does not take into account the fraction of colliding pairs of galaxies that fall in single pass regions. The small-scale clustering, affected by fibre collisions, can be recovered using the angular up-weighting scheme proposed by~\citet{percival2017}. This scheme assumes that the set of un-observed pairs is statistically equivalent to the observed one. The angular up-weighting scheme (ANG) is applied to the counts of pairs both of observed galaxies and observed-random ones. This method gives a statistically unbiased estimator for the clustering even at scales below the fibre collision.

Additionally, for large scales $s>25$Mpc/$h$, there are some photometric angular systematics in the quadrupole that are not accounted for by any of the schemes described above. This is why in \citet{tamone2020} they have decided to remove the $0.6<z<0.7$ data and use a \textit{modified} correlation function that removes angular power. This is also the reason for our work to use the quadrupole data only for $s<25$, see \Sec{sec:results}. 
Alternatively, we could  used power spectrum multipoles, where these angular systematics are nulled with a pixelisation scheme. We leave this for a future study, as comparing data and mocks in Fourier Space requires a careful window function treatment of the mocks. Moreover, the information is spread differently in Fourier space, with 1-halo and 2-halo terms more entangled (see Appendix \ref{sec:fourier}).

For the data we compute the correlation function from data-data (DD), data-random (DR) and random-random (RR) pairs using the Landy-Szalay estimator \citep{LS}.

\begin{equation}
\xi(x,y) = \frac{DD(x,y) - 2DR(x,y) +RR(x,y)}{RR(x,y)}
\end{equation}

This gives us $\xi(s,\mu)$  and $\xi(r_p,\pi)$ in order to obtain, respectively,  the multipoles (\Eq{eq:multipoles}) and $w_p$ (\Eq{eq:wp}). We also use this approach when computing correlations of the {\sc EZmocks} in \Sec{sec:results}, as they also include the survey geometry.




\section{The mock catalogues}\label{sec:mocks}

Within the Halo Occupation Distribution (HOD) Model framework, we assume that a galaxy mock catalogue can be constructed directly from a halo catalogue containing just the halo positions, velocities and masses. Only in some specified examples below (\Sec{sec:profile}, \Sec{sec:velocities}), we will use additional information from particles within haloes. 

The HOD models used here have contributions from two galaxy populations: centrals and satellites, with $\langle N _{\rm cen} (M) \rangle$  and  $\langle N_{\rm sat} (M) \rangle$ their expected number of galaxies per halo of mass $M$. The number density of the total galaxy sample in the model catalogues is calculated as follows:

\begin{equation}
\bar{n}_{\rm gal} = \int \frac{{\rm d}n(M)}{{\rm d} M} \big[  \langle N_{\rm cen} (M) \rangle    +    \langle   N_{\rm sat} (M) \rangle \big] {\rm d}M \, ,
\label{eq:n}
\end{equation}
with $\frac{{\rm d}n(M)}{{\rm d} M}$ the differential halo mass function. 

The clustering of the resulting model galaxy sample has two contributions: one coming from galaxies on the same halo, the 1-halo term, and one coming from correlations of galaxies hosted by different haloes, the 2-halo term.  The clustering at large scales, which is dominated by the 2-halo term, can be described almost completely by the linear bias, which depends only on $\langle N _{\rm tot}(M)\rangle = \langle N _{\rm cen} (M)\rangle +  \langle N _{\rm sat} (M) \rangle)$~\citep[e.g.][]{berlind02}: 

\begin{equation}
b_{\rm gal} = \frac{1}{\bar{n}_{\rm gal} }  \int \frac{{\rm d}n(M)}{{\rm d} M} \cdot b(M) \big[ \langle N_{\rm cen} (M)\rangle +  \langle N_{\rm sat} (M)\rangle ) \big] {\rm d}M \, .
\label{eq:bias}
\end{equation}

 The clustering at small scales is dominated by the 1-halo term, which is affected by a range of properties beyond the linear bias of the sample. Below, we list the modelling of a set of properties that can have a strong impact on the 1-halo term of the clustering:

\begin{itemize}
\item The split between satellite and central galaxies, and the specific HOD shape $\langle N _{\rm cen} (M) \rangle$,  $\langle N_{\rm sat} (M) \rangle$ (\Sec{sec:hod}). \\

\item The probability distribution function $P(N|\langle N \rangle)$ (\Sec{sec:pdf}). \\

\item The radial profile of satellites $\rho_{\rm sat}(r)$ (\Sec{sec:profile}). \\

\item The velocity profile of the satellites $\phi(v_{r})$ (\Sec{sec:velocities}). \\
\end{itemize}

In the following subsections we describe the choices we make about those properties, and how we vary them to explore their influence on the clustering.

\subsection{Mean halo occupation distribution for centrals and satellites }\label{sec:hod}

\begin{figure}  \includegraphics[width=0.5\textwidth]{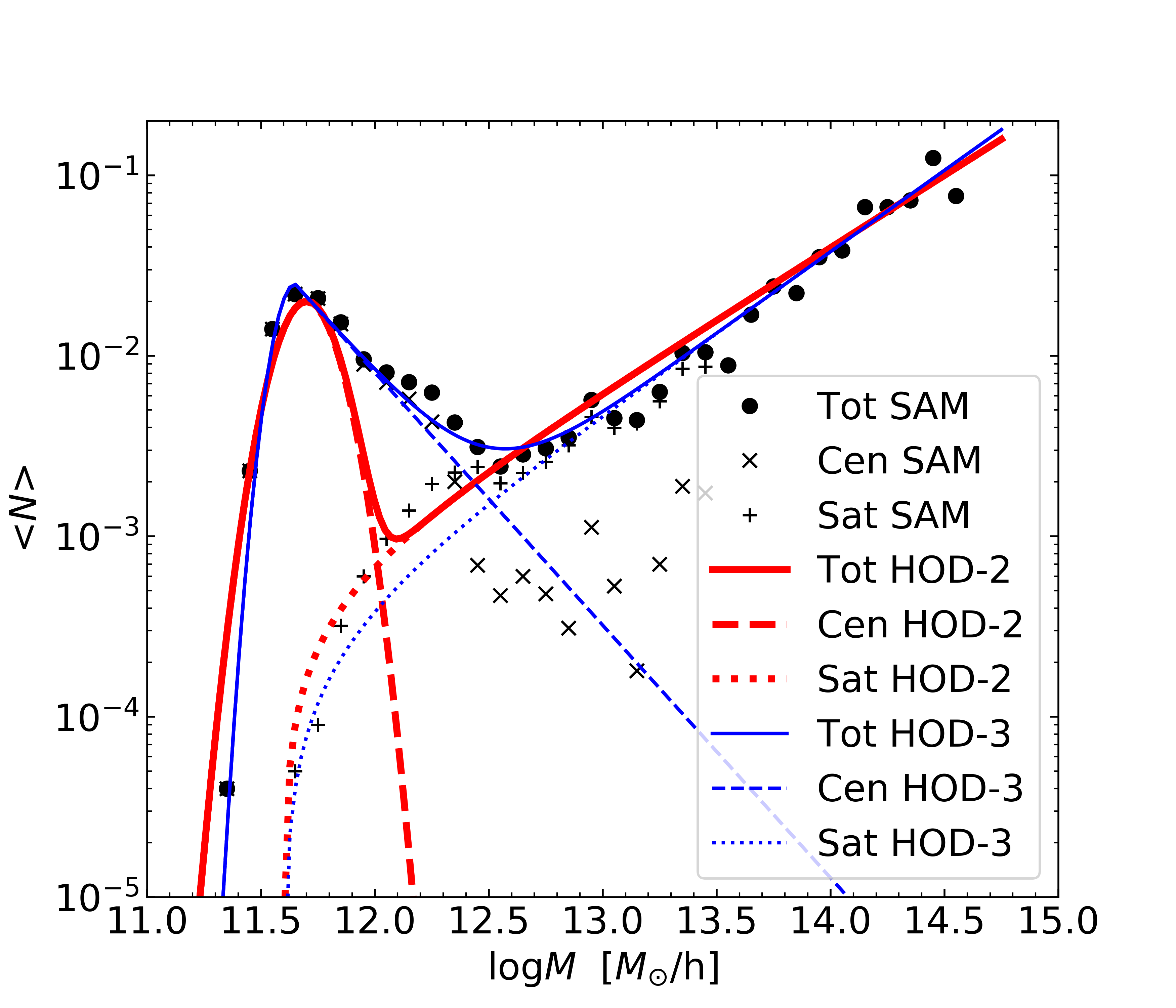}
\caption{
The mean eBOSS ELG HOD from a semi-analytical model (SAM) of galaxy formation~\citep{vgp2018} for central ($\times$), satellites ($+$) and all galaxies (circles). The SAM HOD has been fitted using  HOD-2, (\Eq{eq:hod2}, red thick lines) and HOD-3 (\Eq{eq:hod3} , blue thin lines). The solid lines show the fit to the total mean HOD, the dashed lines the contribution from centrals and the dotted lines that from satellite galaxies. 
}\label{fig:hod} 
\end{figure}

\begin{table*}
    \centering
    \begin{tabular}{l|c|c|c|c|c|c|c|c}
             $\ $   & $A_c$ & $A_s$ & $\mu$ & log$M_0$ & log$M_1$ & $\sigma$ & $\alpha$ &$\gamma$ \\
             \hline
          \hline
          HOD-1  & 0.00723 ($\bar{n}_{\rm gal}$) & 0.01294 ($f_{\rm sat}$)  & 11.110 ($b_{\rm gal}$) & $\mu$ & $\mu+1.3$ & 0.15 & 1.0 & -- \\
          HOD-2 & 0.01185 ($\bar{n}_{\rm gal}$)   &  0.009008 ($f_{\rm sat}$)    & 11.707 ($b_{\rm gal}$)  & $\mu-0.1$ & $\mu+0.3$ & 0.12 & 0.8 & -- \\
          HOD-3  & 0.00537 ($\bar{n}_{\rm gal}$) & 0.005301 ($f_{\rm sat}$) & 11.515 ($b_{\rm gal}$) & $\mu-0.05$ & $\mu+0.35$  & 0.08   & 0.9 & -1.4 \\
             \hline
             \hline
    \end{tabular}
    \caption{List of  parameters used in the three HOD models described in \Eq{eq:hod1}, \Eq{eq:hod2} \& \Eq{eq:hod3} 
    $A_c$, $A_s$, $\mu$ are free parameters that mostly control the quantity indicated in brackets, although one has to simultaneously fit the three free parameters to obtain the target $\{\bar{n}_{\rm gal}, f_{\rm sat}, b_{\rm gal} \}$. The values shown correspond to $\{\bar{n}_{\rm gal} = n_{\rm eBOSS}, f_{\rm sat}=0.30, b_{\rm gal}=b_{\rm eBOSS} \}$. The rest of mass parameters (log$M_0$,log$M_1$) have a fixed offset with respect to $\mu$ and the scaling parameters ($\sigma$, $\alpha$ and $\gamma$) are fixed. Both the offsets and scaling parameters are derived from the fits shown in \Fig{fig:hod} for HOD-2 and HOD-3, whereas for HOD-1 we take the values from \citet{zehavi05}.}
    \label{tab:HOD}
\end{table*}

The mean halo occupation distribution (HOD), $\langle N_i(M) \rangle$, encapsulates the average distribution of a given type of galaxy hosted per halo of a certain mass $M$. 
The analytical description of the mean HODs has been derived from either semi-analytical or hydro-dynamical simulations for galaxy formation and evolution~\citep[e.g.][]{berlind2003,zheng05}. The shape of the model mean HOD depends on the properties of the selected galaxies. When galaxies are selected by their magnitude or stellar mass, the $\langle N_{\rm cen}\rangle$ can be described as a smoothed step function (erf$(x)$) and a power law for satellite galaxies. This is the most commonly used shape for the mean HOD. Here we label this shape as HOD-1: 

{\bf HOD-1} centrals:
\begin{equation}
\langle N_{\rm cen}(M) \rangle =
 \frac{1}{2} A_c \left( 1 + {\rm erf} \left(  \frac{\rm log(M)- \mu}{\sigma} \right) \right)\, .
\label{eq:hod1}
\end{equation}

Satellites (all HODs):

\begin{equation}
    \langle N_{\rm sat}(M) \rangle = A_s \Bigg(\frac{M - M_0}{M_1} \Bigg)^\alpha \, .
\label{eq:sat}
\end{equation}

 This shape has been shown to describe well the abundance of galaxies for a magnitude limited sample~\citep[e.g.][]{zehavi2011}. In a complete sample, we would have $A_c=1$ and the number of central galaxies would transition from 0 to 1 at ${\rm log}M\sim \mu$ with a smoothing scale of $\sigma$. This means that for high enough masses all haloes are expected to have a central galaxy. For samples that are quite incomplete in mass, such as ELGs, QSO or colour-selected samples, one could have $A_c<1.0$~\citep[e.g.][]{geach2012,smith2020}. Note that even if the mean HOD of model LRGs does not follow HOD-1 exactly~\citep{hernandez2020}, these samples have been shown to be well described with such a parametrisation~\citep[e.g.][]{gil-marin20}.
Here, we define the completeness as the ratio between the number of galaxies in a given sample and the total number of galaxies. In a more general case $A_c$ could vary with mass, changing the shape of the mean HOD. In the literature $\mu$ is usually denoted as ${\rm log}M_{\rm min}$, but we choose this nomenclature for consistency with models HOD-2 and HOD-3 (see below). 

In \Eq{eq:sat}, the satellite occupation follows an increasing power law, implying that the more massive the halo, the more satellite galaxies we expect to find, with $\alpha$ controlling the mass-richness relation. For $A_s=1.0$ and $M_0\ll M_1$, $M_1$ represents the mass at which we expect 1 satellite per halo. We note that $A_s$ is completely degenerate with $M_1$, but we keep both parameters to separate their physical meaning and interpret $A_s$ as the completeness of the satellites.

In \Eq{eq:hod1}, central galaxies have a constant probability to be found in haloes above a certain mass. However, this is at odds with the results derived for observed star-forming ELGs~\citep{geach2012,cochrane2017,guo19} and model ones~\citep{cochrane2018eagle,vgp2018,favole2020}.  More generally, the soft step function shape is not representative of samples of model galaxies selected by their age or their star formation rate~\citep{zheng05,contreras2013}. The star formation of galaxies depends in a non-trivial way on their stellar mass and environment. Massive galaxies tend to have lower star formation rates per stellar mass unit~\citep[e.g.][]{davies2019}. ELGs are on average less massive than samples such as LRGs, and are mostly found in filaments~\citep[e.g.][]{darvish2014,vgp2020}. Galaxy formation processes  affecting star-forming galaxies impact the expected shape of their HOD.

\Fig{fig:hod} shows the HOD for eBOSS ELGs from the semi-analytical model (SAM) of galaxy formation and evolution presented by~\citet{vgp2018}. Here we have fit the shape for centrals with two mean HOD models:

{\bf HOD-2} centrals: 
\begin{equation}
\begin{split}
\langle N_{\rm cen} (M) \rangle &= \frac{A_c}{\sqrt{2\pi}\sigma} \cdot e^{-\frac{({\rm log}M-\mu)^2}{2\sigma^2}} 
\end{split}
\label{eq:hod2}
\end{equation}

{\bf HOD-3} centrals (default):
\begin{equation}
\begin{split}
\langle N_{\rm cen} (M) \rangle &=
   \begin{cases} 
      \frac{A_c}{\sqrt{2\pi}\sigma} \cdot e^{-\frac{({\rm log}M-\mu)^2}{2\sigma^2}}  & {\rm log}M\leq \mu \\
      \frac{A_c}{\sqrt{2\pi}\sigma} \cdot \Bigg( \frac{M}{10^\mu}\Bigg)^\gamma & {\rm log}M\geq  \mu \\
   \end{cases}
   \\
\end{split}
\label{eq:hod3}
\end{equation}

The HOD-2 has a simpler expression for centrals, being a Gaussian with amplitude $A_c$, mean $\mu$ and variance $\sigma^2$. However, it fails to describe the asymmetry at ${\rm log}M>\mu$. That is why HOD-3 introduces a decaying power-law for ${\rm log}M>\mu$. 
Note that all three HODs have the same functional shape for the satellites, although their parameters may be different. 
The functional form that best describes the HOD from the SAM is the HOD-3, which we will consider as our default model in this work. The other two HOD shapes will be considered as variations of our model in which we increase (HOD-1) or decrease (HOD-2) the central galaxy occupation on the high halo mass end. 

For every HOD model, we first apply the constraints of $\bar{n}_{\rm gal}=n_{\rm eBOSS}$ (or a multiple of it) and $b_{\rm gal}=b_{\rm eBOSS}$, using \Eqs{eq:n} \& \Eq{eq:bias}. Then, we set the fraction of satellites with:

\begin{equation}
f_{\rm sat} = \frac{1}{\bar{n}_{\rm gal}} \int \frac{{\rm d}n(M)}{{\rm d} M}    \langle   N_{\rm sat} (M) \rangle  {\rm d}M \, .
\label{eq:fsat}
\end{equation}

\begin{figure*}
    \centering
    \includegraphics[width=0.5\linewidth]{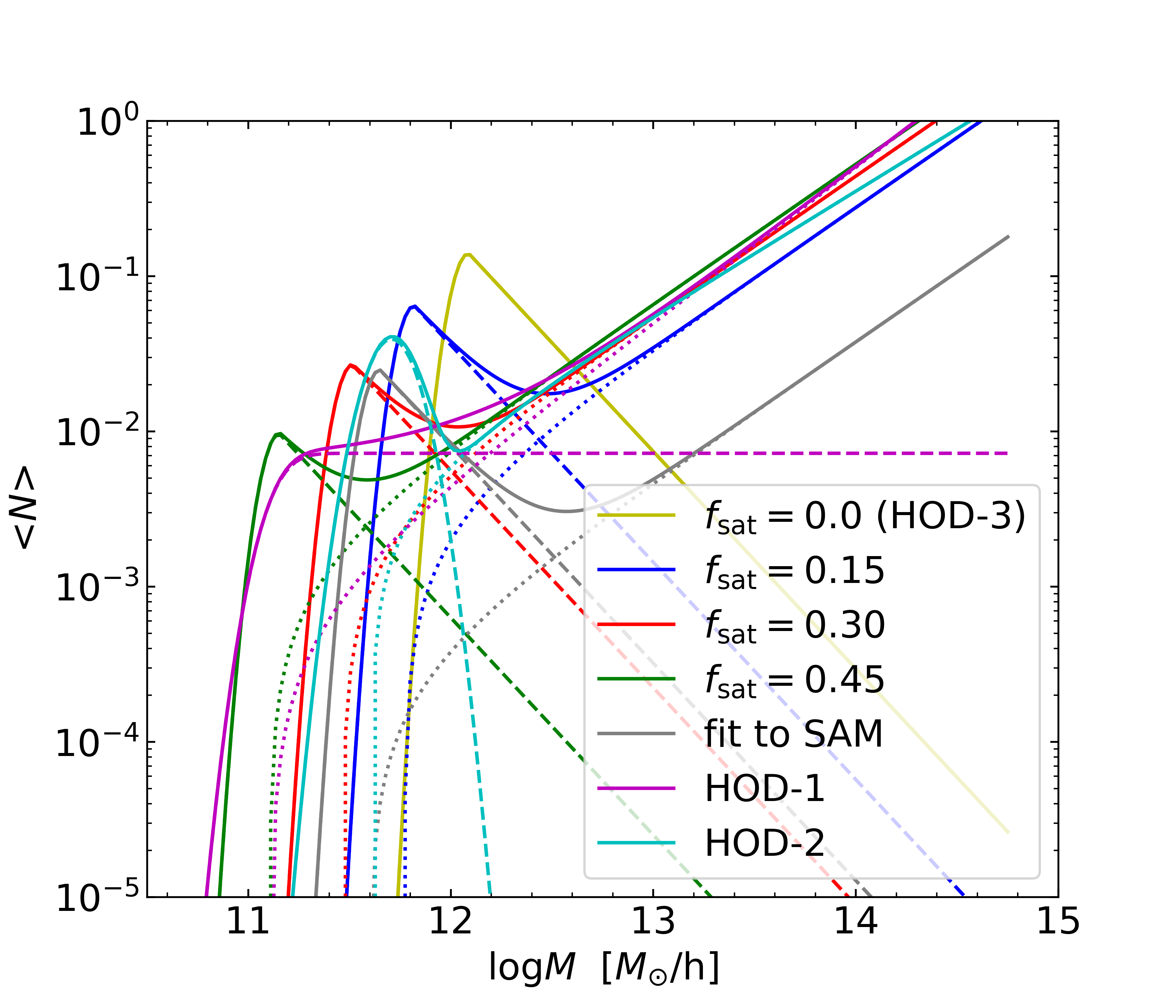}\includegraphics[width=0.5\linewidth]{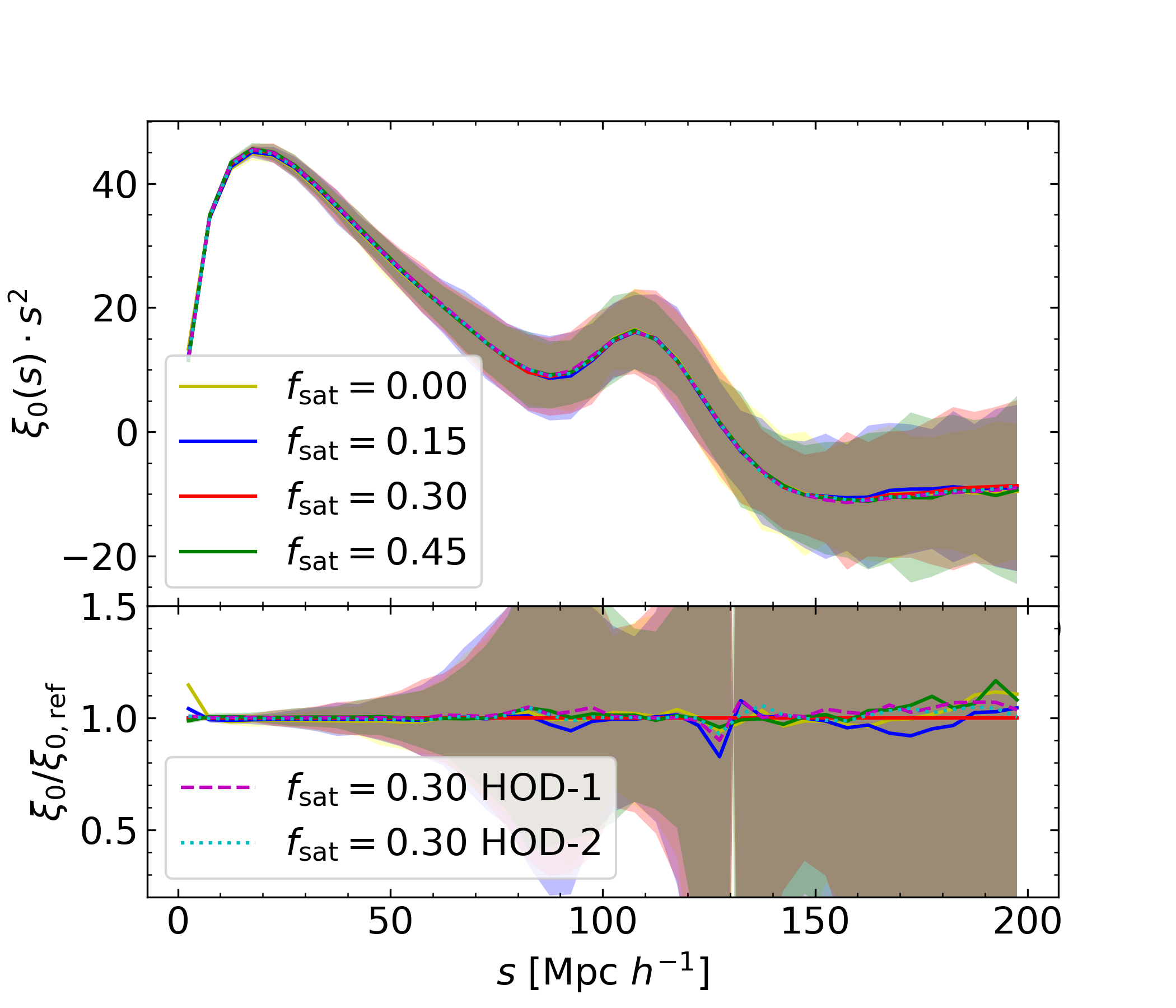}
    \includegraphics[width=0.5\linewidth]{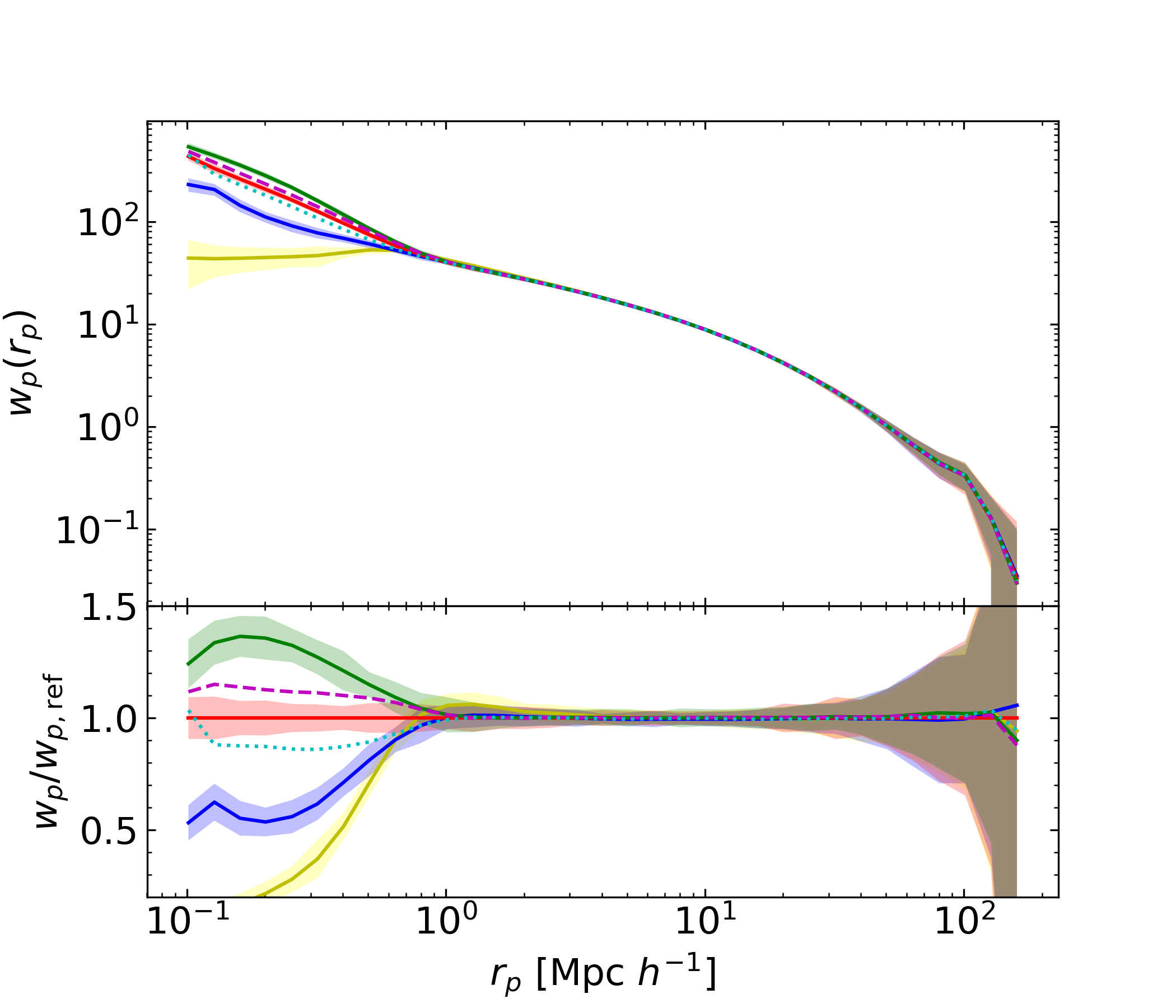}\includegraphics[width=0.5\linewidth]{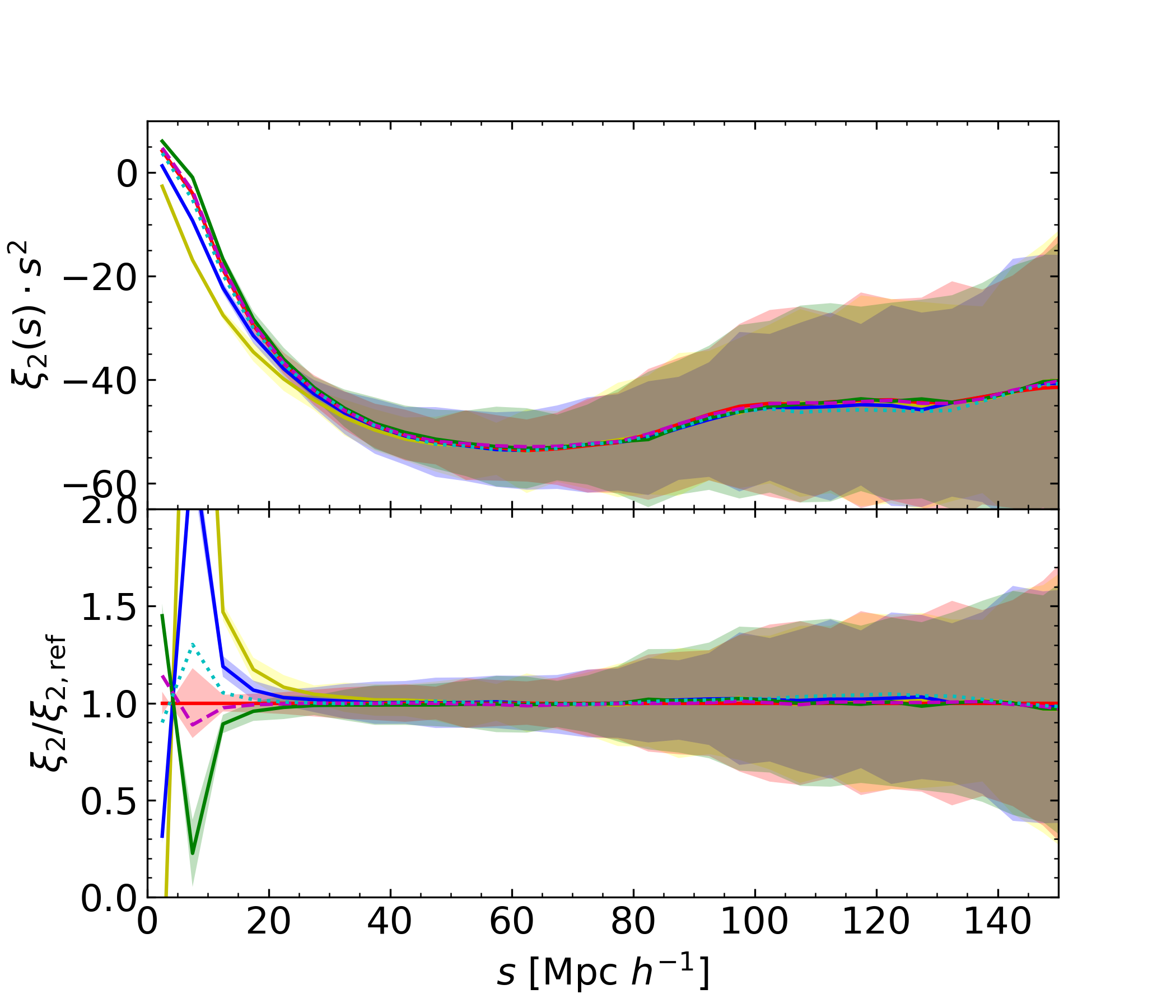}
    \caption{{\it Top-Left}: Mean Halo Occupation Distribution for different models (\Eqs{eq:hod1},\ref{eq:hod2},\ref{eq:hod3}) and different fractions of satellites as indicated in the label. All HODs have been fitted to have the same number density and bias as the data, except for the grey curve (labelled \textit{SAM}) that was fitted in \Fig{fig:hod} to the SAM, and is shown here for comparison. We use the same line styles as in \Fig{fig:hod}: dashed for centrals, dotted for satellites and solid for all. {\it Top-right:} Monopole of the 2PCF for the HODs shown in the Top-Left sub-figure. {\it Bottom-Left:} Projected correlation function of the different HODs. {\it Bottom-Right:} Quadrupole of the 2PCF of the different HODs.
    The mock catalogues shown here were constructed assuming that satellites follow a Poisson distribution ($\beta = 0$, see~\S\ref{sec:pdf}), were distributed in haloes following NFW ($K=1$, see~\S\ref{sec:profile})  and that their velocities follow the virial theorem ($\alpha=1$, see~\S\ref{sec:velocities}). These choices, together with HOD-3 and \fsat$=0.3$ are the default choices, and will be used unless otherwise specified. The reference model (used for the ratios and labelled as \textit{ref}) has all the default choices and is the same across Figs. \ref{fig:2PCF_fsat}, \ref{fig:beta}, \ref{fig:K} \& \ref{fig:vel}. For the clustering sub-figures we have used an enhanced number density whereas the shaded area corresponds to the error expected for eBOSS data (see text for details). We use the same approach for Figs. \ref{fig:beta}, \ref{fig:K} \& \ref{fig:vel}.
    The legend is consistent across the different clustering sub-figures. 
    }
    \label{fig:2PCF_fsat}
\end{figure*}

We fix the offset between $\mu$ and the mass parameters, log$M_0$ and log$M_1$. These fixed offsets and the parameters $\sigma$, $\alpha$ and $\gamma$ are derived from fitting the HOD-2 and HOD-3 equations to the ELG mean HOD derived in ~\citet{vgp2018} and shown in \Fig{fig:hod}. For the case of HOD-1 those values are taken from \citet{zehavi05}. All the choices made for the different HOD parameters are shown in Table \ref{tab:HOD}. For a given HOD (of the 3 described above) and the fixed choices of parameters just described, any given choice of $\{A_c,A_s,\mu\}$ yields a a set of $\{\bar{n}_{\rm gal},f_{\rm sat}, b_{\rm gal} \}$, and vice-versa.

On top-left of \Fig{fig:2PCF_fsat} we present together the HOD shape of HOD-1, HOD-2 and HOD-3 for $\bar{n}_{\rm gal}=n_{\rm eBOSS}$, $b_{\rm gal}=b_{\rm eBOSS}$ and $f_{\rm sat}=0.30$, as well as HOD-3 with  $f_{\rm sat}=0,0.15,0.45$ (and the same $\bar{n}_{\rm gal}$,  $b_{\rm gal}$) and the original fit to the SAM with HOD-3, for comparison. We also show the corresponding monopole $\xi_0(s)$, quadrupole $\xi_2(s)$ and projected correlations $w_p(r_{p})$. 

In order to compute more accurately the correlation functions, we increase the number density by a factor of 10 (7 in the case of $f_{\rm sat}=0$ in order to avoid hitting the $\langle N_{\rm cen} \rangle=1$ limit). The shaded area represents the one $\sigma_{\rm eBOSS}$ region expected for eBOSS computed as $\sigma_{\rm eBOSS} = \sqrt{\frac{1({\rm Gpc}/h)^3}{V_{\rm eBOSS}}} \sigma_{\rm mock}$, with  $\sigma_{\rm mock}$ the standard deviation over the 27 $l=1Gpc/h$ subboxes of a mock realisation with the eBOSS number density. We will follow this approach for all other figures in this section. 

We find that the differences on the shown scales for the monopole are negligible. This is expected as we fixed the bias and we are looking at linear or quasi-linear scales, so we will not show the monopole in the following subsections. We would  also find differences if we explored lower scales on the monopole using logarithmic binning. 
However, we find more illustrative to study the projected correlation function and the quadrupole, which offer complementary information of the effects of positions and velocities of satellites (as we will see along this section), respectively, whereas for the monopole on logarithmic binning those effects appear entangled (as it happens in Fourier space, see Appendix \ref{sec:fourier}).   

The projected correlation shows the expected trend: a higher signal at small scales as we increase $f_{\rm sat}$, increasing the contribution of the 1-halo term. For the quadrupole, we find differences already at $s\sim 25 Mpc/h$ with lower (closer to zero, we will use this terminology in the remainder) signal for larger fractions of satellites, due to an increase of the Finger-of-God effect \citep{FoG,Peebles80}. For the lowest point, some of the mocks invert their quadrupole's sign.  

It is remarkable that the differences introduced by the choice of HOD shape are much less significant than the value of the satellite fraction, or the choices detailed in the subsections below. We also note that the clustering of the HOD-3 is approximately half way between that of HOD-1 and that of HOD-2, confirming our interpretation of HOD-3 being bracketed by models HOD-1 and HOD-2.


\subsection{Probability Distribution Function}\label{sec:pdf}

\begin{figure}
    \centering
    \setlength\lineskip{0pt}
    \includegraphics[width=1.02\linewidth,scale=1]{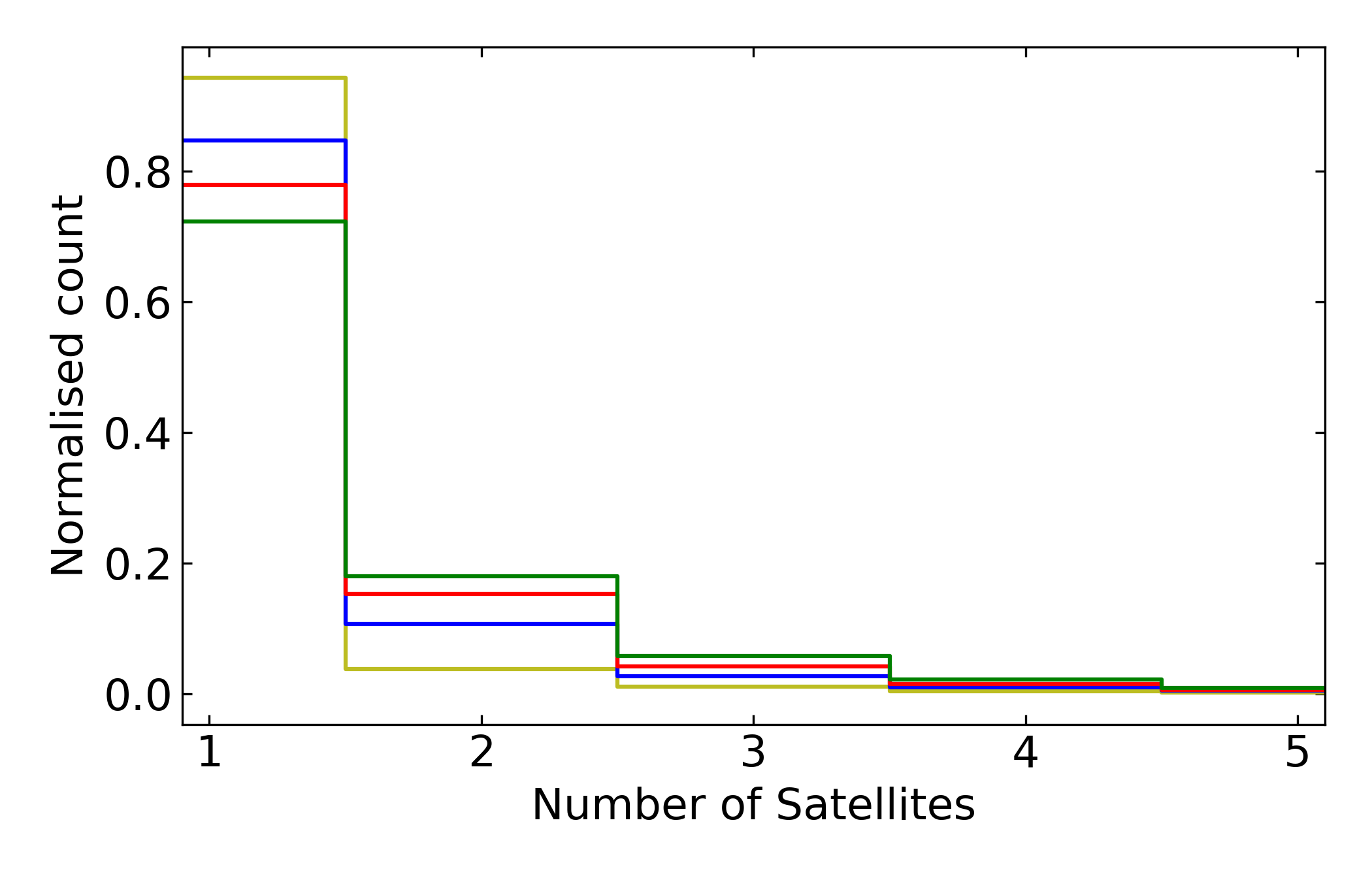} \\
    \includegraphics[trim=0 5 45 40,clip,width=1\linewidth,scale=1]{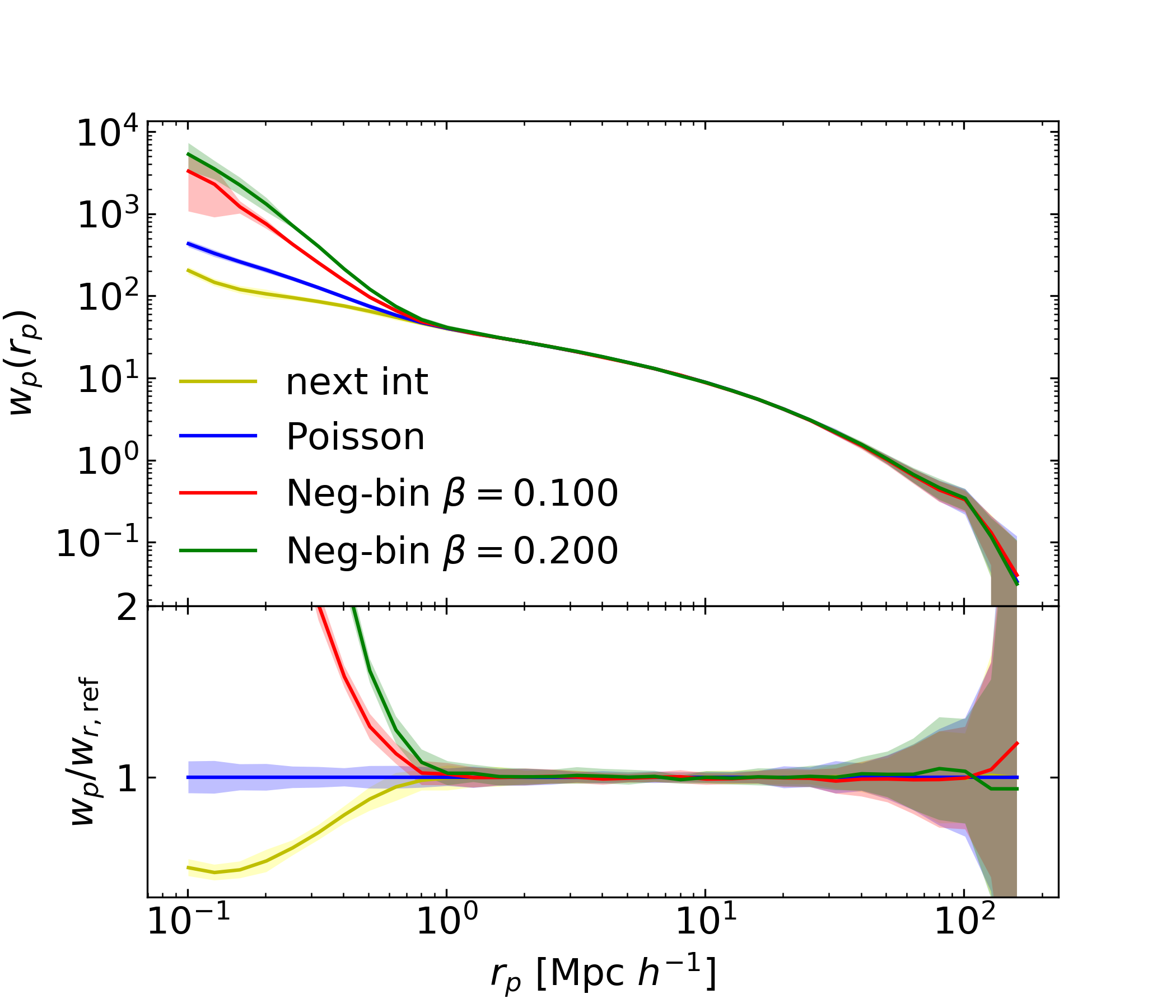} \\
    \includegraphics[trim=0 5 45 40,clip,width=1\linewidth,scale=1]{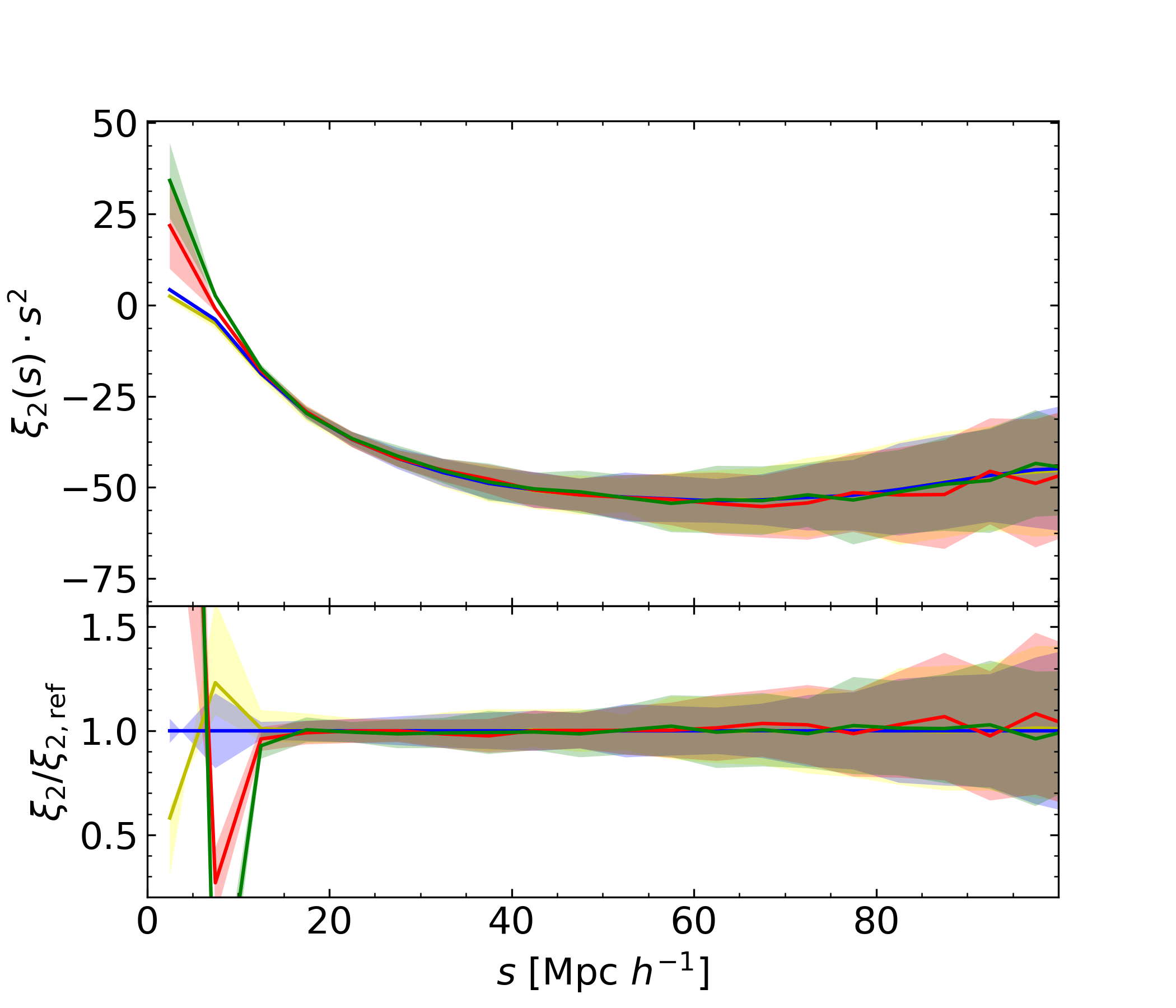}
    \caption{Effect of PDF of satellite assignment, considering: Poisson (\Eq{eq:poisson}), nearest-integer (\Eq{eq:ni}), negative binomial with $\beta=0.1$ and $\beta=0.2$ (\Eq{eq:binomial}). 
    {\it Top:} Counts of haloes occupied by a given number of satellite galaxies $N \geq 1$  in the full ELG mock sample (with contributions from all halo masses and their corresponding mean occupations $\langle N_{\rm sat} (M) \rangle$, \Eq{eq:sat}), divided by the total number of counts.
    {\it Middle:} Projected correlation function $w_p$ for the same mocks, as indicated in the legend, and the ratios with respect to the mocks with the Poisson distribution. {\it Bottom:} Quadrupole and ratios.
    Besides the PDF specified, we take the default choices:  $f_{\rm sat}=0.30$, NFW profile ($K=1.0$, see \Sec{sec:profile}) and the virial theorem for the velocities ($\alpha=1.0$, see \Sec{sec:velocities}).  
    }\label{fig:beta}
\end{figure}

In \Sec{sec:hod} we have studied the mean halo occupation distribution of satellites and centrals. Here we study how we go from the mean value $\langle N \rangle$ to a given realisation $N$, of the (integer) number of galaxies in a halo. This is given by the Probability Distribution Function (PDF), $P(N|\langle N\rangle)$.

By definition, for central galaxies $N$ can only be $0$ or $1$. If satellite galaxy formation were a random uncorrelated process, they would follow a Poisson distribution. However, galaxy formation could affect their PDF, increasing or decreasing the scatter~\citep{jimenez2019}. The three PDFs for satellites that we consider are: 

\begin{itemize}
    \item \textbf{Poisson distribution} (default) 
    
    \begin{equation}
        P(N|\lambda)= \frac{e^{-N}\lambda^N}{N!}\, ,
         \label{eq:poisson}
    \end{equation}
    with $\lambda \equiv \langle N\rangle$  ($=  \langle P(N|\lambda)\rangle$) and $\sigma=\sqrt{\langle N\rangle}$
    
    The Poisson distribution will be used for assigning satellite galaxies to haloes, unless otherwise stated. 
    
    \item \textbf{Nearest integer distribution}. 
    
    This function only allows two possible values for $N$, which are the two closest integers to $\lambda=\langle N \rangle$:
     \begin{equation}
        P(N|\lambda) = 
    \begin{cases}
        1 - (\lambda - \textsc{int}(\lambda)) & N=\textsc{int}(\lambda) \\
        \lambda - \textsc{int}(\lambda)  & N=\textsc{int}(\lambda) +1  \, . \\
        0 & {\rm else}    \\
    \end{cases}
     \label{eq:ni}
    \end{equation}
    
    The function $\textsc{int}(x)$ represents the truncation of $x$ to the nearest lower integer.  
    This distribution is always used for the centrals (for which only the $0$ and $1$ values are allowed) and it will be used for the satellites only when specified. 
    This function has a lower scatter than the Poisson distribution: $\sigma= \sqrt{\Delta(1-\Delta)}$, with $\Delta = \lambda - \textsc{int}(\lambda)$.
    
    \item \textbf{Negative Binomial Distribution}. 
    
     This function allows for a larger scatter than the Poisson distribution. The parameter $\beta$ represents the relative increment of the standard deviation with respect to the Poisson distribution with  $\lambda=\sqrt{N}$ and $\sigma = \lambda (1+\beta)$. Here we follow a similar notation as in \citet{jimenez2019}. 
    \begin{equation}
    \begin{split}
        P(N|r,p) = \frac{\Gamma(N+r)}{\Gamma(r)\Gamma(N+1)} p^r (1-p)^N \, \, {\rm with} \\
        p=\frac{1}{(1+\beta)^2}, \,\, r = \frac{\lambda}{\beta(1+2\beta)} \, .
        \label{eq:binomial}
    \end{split}
    \end{equation}
    The Poisson distribution is recovered in the limit $\beta \to 0$, after solving a few indeterminations. 
    
\end{itemize}

In \Fig{fig:beta} we show how the different PDFs introduced here affect the overall distribution $N_{\rm sat}$ (top sub-figure) and how they affect the galaxy clustering. 
Changing the PDF has a small impact on the quadrupole, except for small scales below $10$Mpc$/h$. However, the effect on the projected correlation function is large at scales below $1$Mpc$/h$, increasing the signal as we increase the scatter. This is expected as when the scatter is increased, the probability of having a pair (or more) of satellites in the same halo increases. 


\subsection{Spatial distribution of satellite galaxies}
\label{sec:profile}

We now study how to spatially distribute galaxies within a halo. The central galaxies are always placed at the position of the host halo. However, here we consider three ways to place satellite galaxies within their host haloes:

\begin{itemize}

    \item \textbf{NFW profile} (default)
    
    We place satellite galaxies following a Navarro-Frenk-White profile \citep{NFW}:
    \begin{equation}
    \begin{split}
        \rho(x) \propto \frac{1}{x\cdot(1+x)^2} \quad
        {\rm with}\quad x = c \frac{r}{r_{\rm vir}} \, ,
        \label{eq:NFW}
    \end{split}
    \end{equation}
    where $c$ is the concentration of the halo, which we take from the values tabulated in \citet{klypin2016}: $c(M)=c_{\rm kly}(M)$. 
    The virial radius from \Eq{eq:NFW}, $r_{\rm vir}$, is computed following a common approach \citep[e.g.][]{Carretero15, Avila18} based on the spherical collapse model \citep{SphericalCollapse_oCDM}:
    
    \begin{equation}
          r_{\rm vir} = \Bigg( \frac{3}{4\rho_{\rm crit} \Delta_{\rm vir} M} \Bigg)^{1/3}\, ,
          \label{eq:rvir}
    \end{equation}
    and 
    
    \begin{equation}
          \Delta_{\rm vir} = 18 \pi^2 + 82 (1 - \Omega_M(z))  -39 (1-\Omega_M(z))^2 \, .
          \label{eq:Dvir}
    \end{equation}
    
    \item \textbf{Modified NFW}. 
    
    Observations find star-forming galaxies preferentially in the outskirts of filaments~\citep[e.g.][]{chen2017,kraljic2018}. Semi-analytical models of galaxy formation and evolution have found that star-forming galaxies tend to be found in the outskirts of haloes~\citep{orsi2018}. This suggests that star-forming ELGs will also be preferably located in the outskirts of haloes. We model this effect by placing satellite ELGs following a  less concentrated profile.
    In this case, we modify the halo concentration from \Eq{eq:NFW} by a factor $K<1$: 
    \begin{equation}
            c(M) = K \cdot c_{\rm kly}(M) \, .
            \label{eq:K}
    \end{equation}

    \item \textbf{Particles}.
    
    We pick a random particle within the halo and assign that position to the satellite galaxy, we will denote this choice as PART. This is computationally expensive. Additionally, since we only transferred 1\% of the halo particles randomly selected, with a minimum of 5 particles per halo, we find a few cases in which we run out of particles. The number of cases is really small, always fewer than 50/217,000 cases for $n_{\rm eBOSS}$ number density in extreme parameter choices. When that happens, we assign an additional satellite to the next halo. Since we apply the HOD to halos ordered by decreasing mass, the mass of the next halo is expected to be similar to that of the previous one. In this way, the original number of galaxies is maintained at the expense of modifying very slightly the PDF and HOD. Since the numbers are very small, we do not expect this choice to impact our results.
    
    \item \textbf{Particles with a modified profile}.
   In this case, we use particles positions, but also model the  ELGs preference to be in the outskirts of haloes. To accomplish this, once the satellite positions are assigned to random particles, these are perturbed following:    

    \begin{equation}
        \begin{split}
            \vec{r}_{\rm sat} =  \vec{r}_{\rm h} + \frac{1}{K} (\vec{r}_{\rm DM} -   \vec{r}_{\rm h}) \, ,
            \label{eq:K_part}
        \end{split}
    \end{equation}
    with $\vec{r}_{\rm sat},\ \vec{r}_{\rm h},\ \vec{r}_{\rm DM}$ the position of the satellite galaxy, the halo and the  dark mater particle respectively. This prescription is equivalent to the rescaling of concentrations for the NFW case, in both cases we are rescaling the profiles by $1/K$.
\end{itemize}

\begin{figure}
    \centering
    \setlength\lineskip{0pt}
    \includegraphics[width=1.02\linewidth]{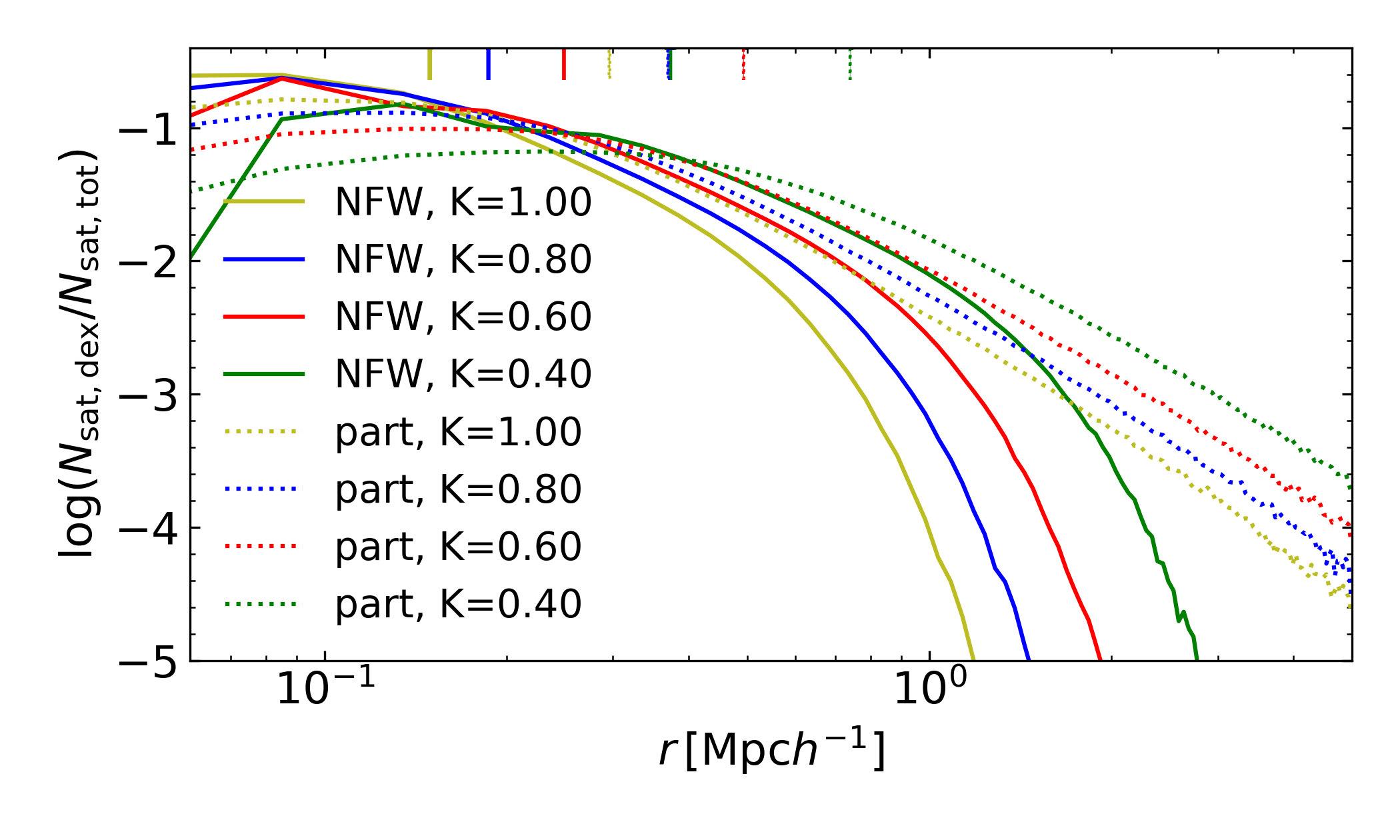}
    \includegraphics[trim=0 5 45 40,clip,width= 1\linewidth]{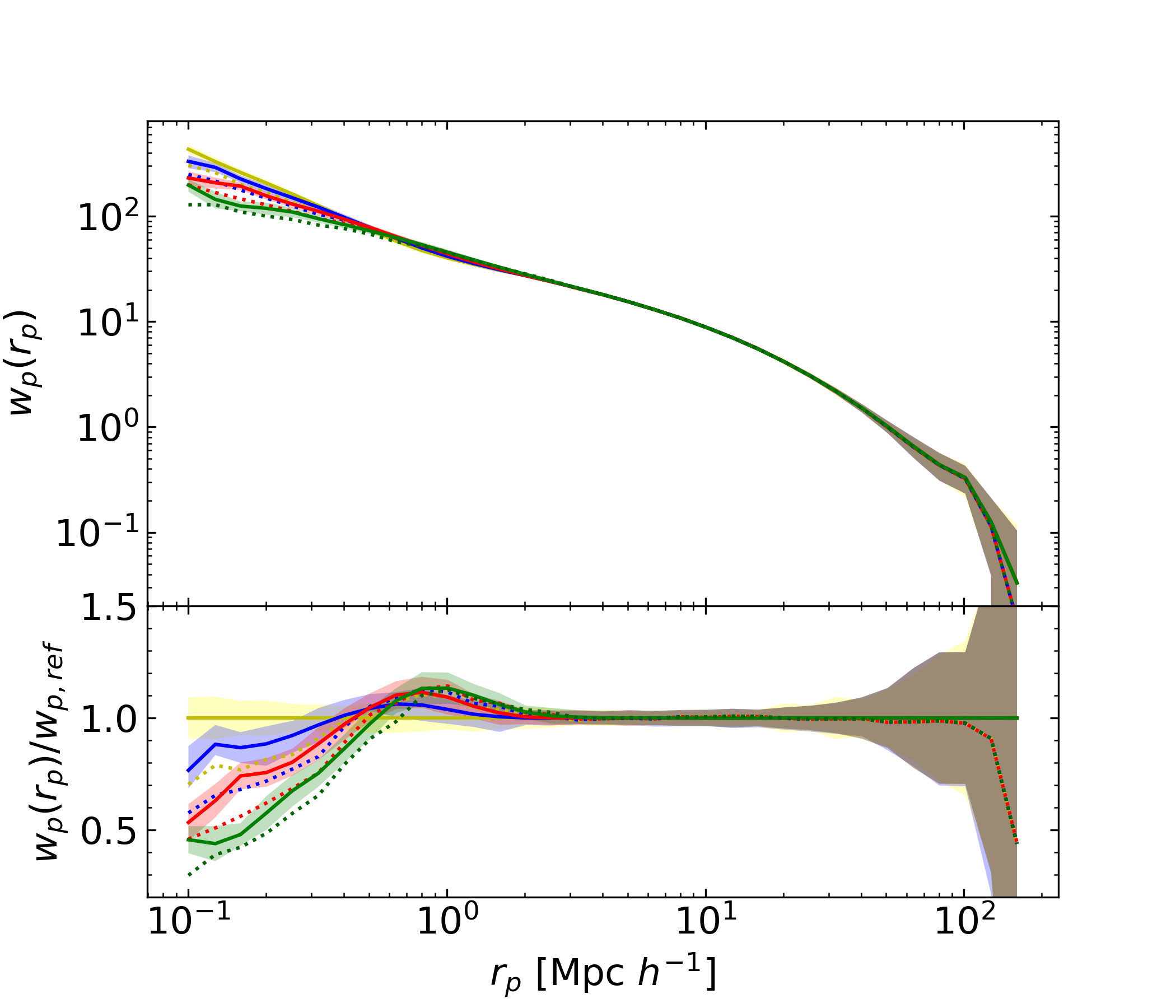}
    \includegraphics[trim=0 5 45 40, clip,width=1\linewidth]{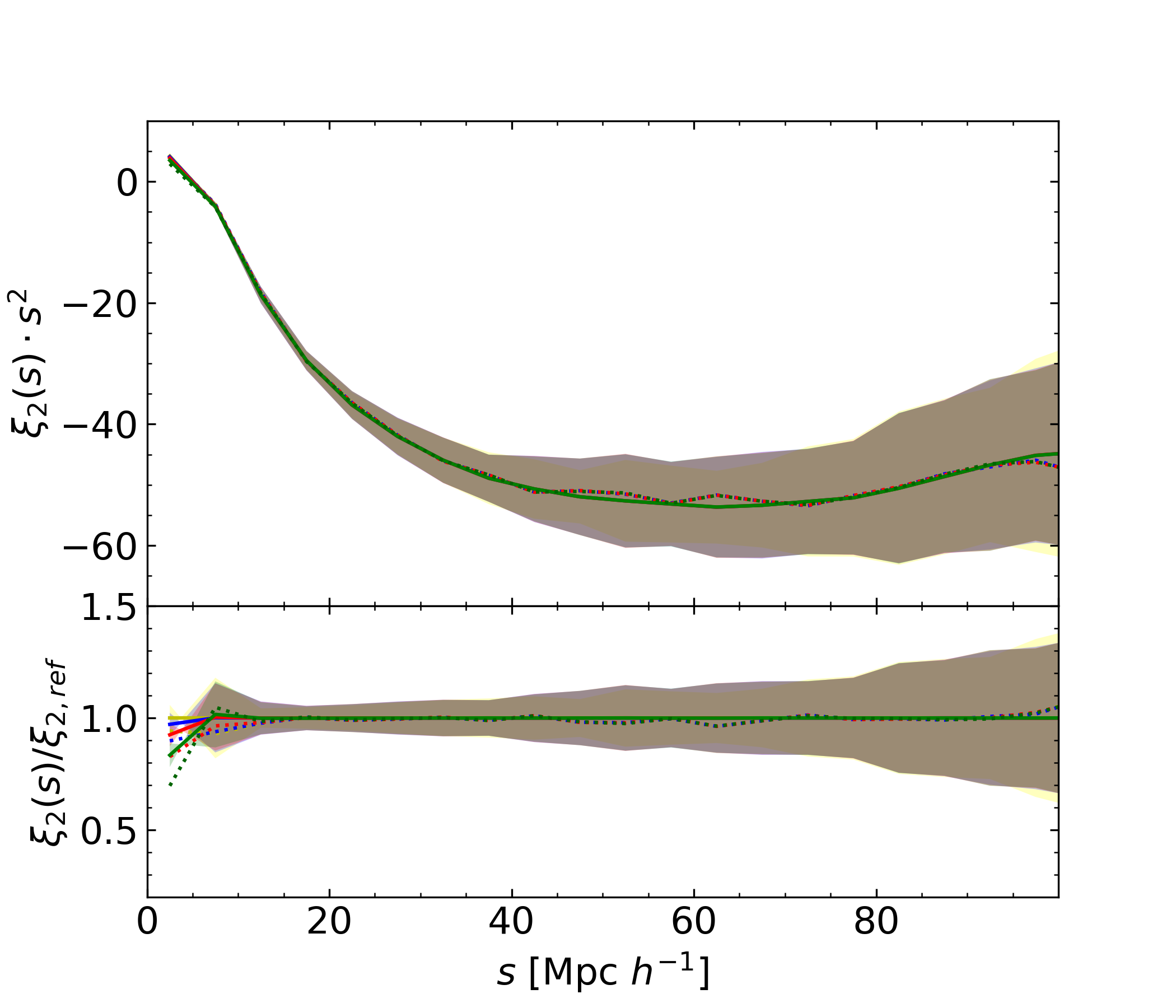}
    \caption{Similar to Fig.~\ref{fig:beta} but for the effect of the density profile of the satellites. We consider a NFW profile (NFW), profiles following the distribution of particles (PART), and both of them with modified concentration profiles ($K\neq1$,\Eq{eq:K}). {\it Top:} Normalised mock ELG satellite count as a function of the distance from the halo centres $r$ for the mocks indicated in the legend. The short vertical lines indicate the mean values of $r$. {\it Middle:} Projected two-point correlation function and ratios to the NFW with $K=1$ case. {\it Bottom:} Quadrupole and ratios. Note the legends are consistent across sub-figures.  
    }
    \label{fig:K}
\end{figure}

In the top panel of \Fig{fig:K} we show the profile of satellite galaxies for mock catalogues with different $K$ values for both NFW profiles and particle position assignments.  There are clear differences between the profiles from mocks constructed assuming either NFW profiles or using the particle information. We can explain these differences first, because only relaxed dark matter have profiles that can be described analytically~\citep[e.g.][]{wang2019}, and at $z\sim 1$, only half of the dark matter haloes in the \ors are relaxed~\citep{child2018}. Second, relaxed dark matter haloes are better described by~\citet{Einasto} profiles, rather than a NFW one~\citep[e.g.][]{gao2008,child2018}. Third, due to reduced access to the information, we derive the halo concentration from their FOF mass using \citet{klypin2016}. In general, in order to compute accurately the concentration of a halo one would fit the distribution of particles with a given profile (for the NFW case, Equation \ref{eq:NFW}).

Given that assuming NFW galaxy profiles 
is a common practice in mock catalogue generation and that the concentrations are defined in a more straightforward way in this case~\citep[e.g.][]{klypin2016}, we continue to use the NFW galaxy assignment and compare their results to using particle profiles. 
In fact, despite the differences seen in the top panel of \Fig{fig:K}, the differences in the projected clustering ($w_p$, middle panel) are much smaller. At very small scales $r \sim 0.1 Mpc/h$, the particle profiles flattens and becomes lower than the NFW profile, giving also a smaller correlation at those scales. However, at $r \sim 1.0 Mpc/h$ the situation is inverted, the NFW profile ($K=1.0$) has nearly reached the tail of $r_{\rm vir}$, with $r_{\rm vir}=0.78Mpc/h$ for log$M=14$, having only $2.6\%$ of satellites beyond that mass. 

Finally, we note that the impact of changing the density profiles is negligible for the quadrupole. The differences found between the PART and NFW profiles, are mostly due to having different velocities, as described below. This confirms our initial claim that these two 2PCF statistics ($w_p$ and $\xi_2$) have very complementary information. 

\begin{figure}
    \centering
    \setlength\lineskip{0pt}
    \includegraphics[trim=20 5 10 10,clip,width=1.06\linewidth]{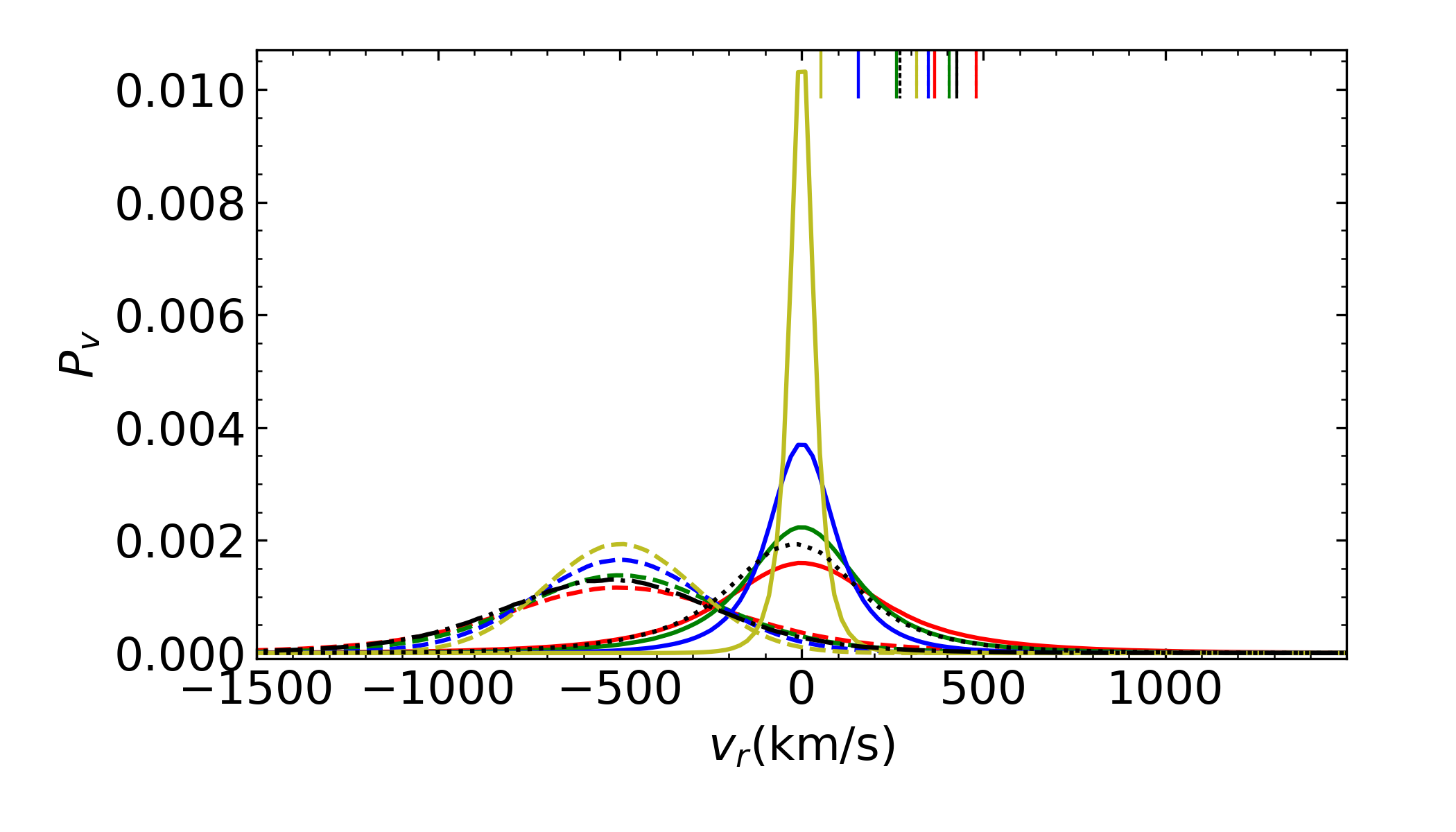}
    \includegraphics[trim=0 5 45 40,clip,width=1\linewidth]{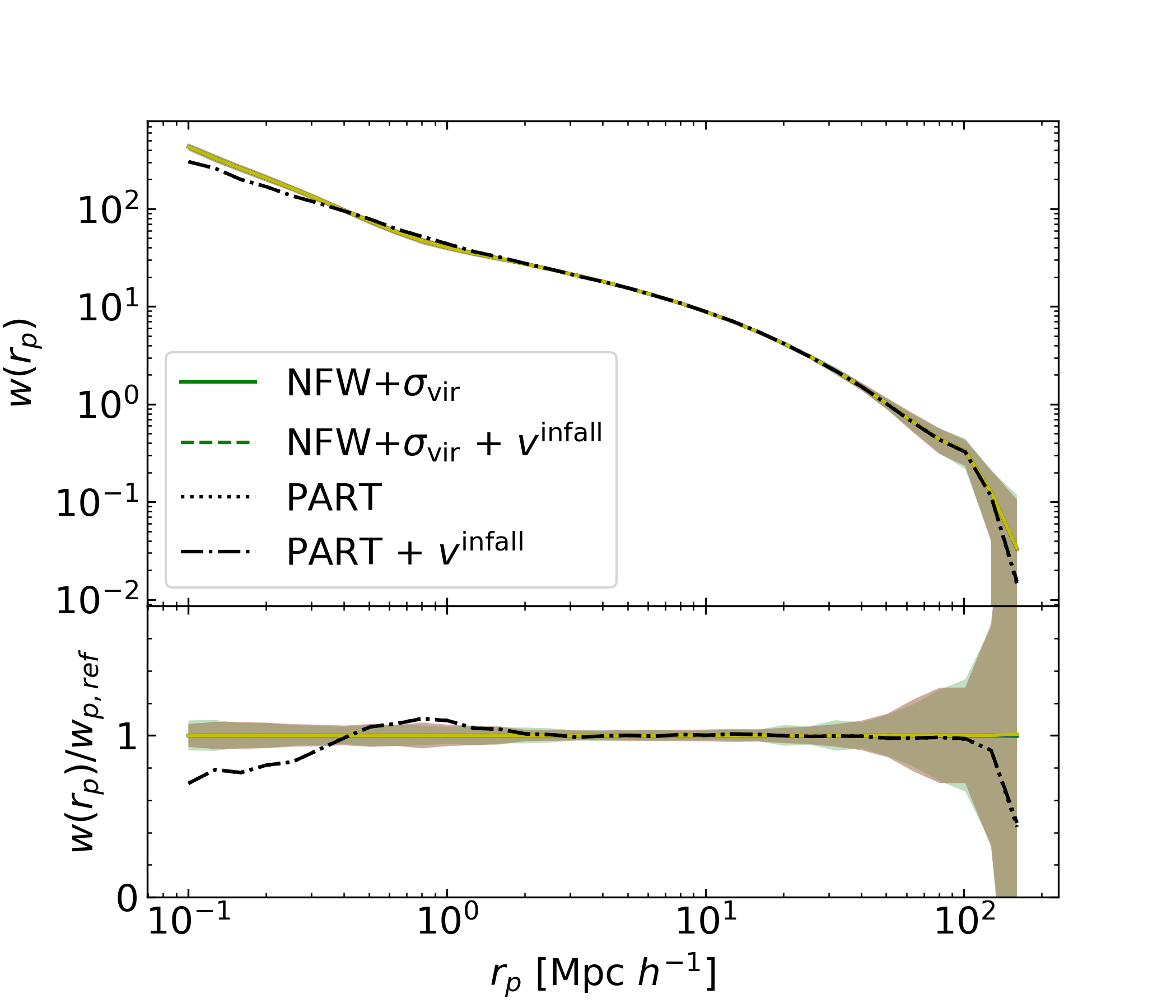}   
    \includegraphics[trim=0 5 45 40,clip,width=1\linewidth]{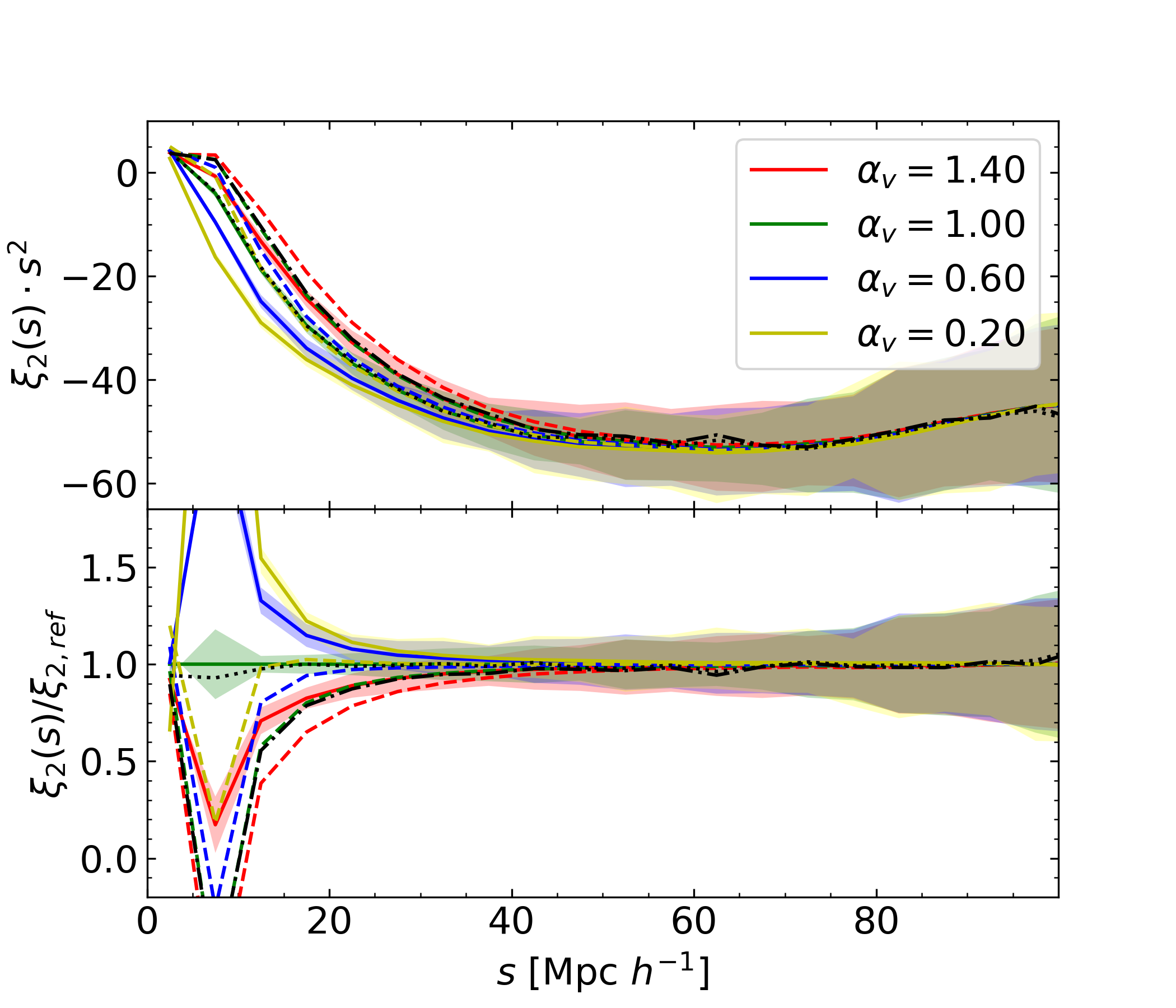}  
    \caption{  
    Similar to Fig.~\ref{fig:beta}, but for the effect of the velocity profile of the satellites. We consider a distribution given by the virial theorem (NFW+$\sigma_{\rm vir}$) and velocities as given by the particles (\textsc{part}), adding also different values of the velocity bias $\alpha_v$ (only for $\sigma_{\rm vir}$, we keep $\alpha_v=1$ for \textsc{part} in this figure) and a net infall velocity of $v_r=-500\pm200km/s$ (for both NFW and \textsc{part}). {\it Top:} Radial (from the halo centre to the satellite) velocity profiles. The area under the curves are normalised to unity. The upper vertical lines indicate the corresponding dispersion of satellite velocities around haloes along the $Z$-axis, $\sqrt{ \langle (v_{\rm gal,z}^{\rm tot} - v_z^{\rm h})^2\rangle }$. {\it Middle:} Projected correlation function and ratios with respect to the default mocks (NFW+$\sigma_{\rm vir}$,  $\alpha_v=1$). Note that most curves line up in this sub-figure, see text. {\it Bottom:} Quadrupole and ratios. Note the legends are consistent across sub-figures. 
    } 
    \label{fig:vel}
\end{figure}
\subsection{Velocity distributions}
\label{sec:velocities}

The remaining choice to make is the assignment of velocities to galaxies. For the central galaxies we simply assume the same velocity as the halo. Whereas for the satellites we consider several options: 
\begin{itemize}
   
    \item \textbf{Virial theorem}. (default)
    When using NFW for the position profiles, we follow \citet{BryanNorman} for the veolicity profiles, also used in \citet{Carretero15,Avila18}:
    
    \begin{equation}
        \sigma_{\rm vir} = 476\cdot0.9 [\Delta_{\rm vir} E^2(z)]^{1/6} \Big( \frac{M}{10^{15} M_\odot h^{-1} } \Big)^{1/3} km/s \, ,
    \end{equation}
    with $E(z)=H(z)/H_0$.
    Note that this scaling is already predicted by the virial theorem.
    And we assign:
    \begin{equation}
    \begin{split}
        v_i^{\rm gal} \curvearrowleft \mathcal{N}(v_i^{\rm h},\sigma_{\rm vir}) \quad
        {\rm for}\, i={x,y,z} \, ,
        \label{eq:vt}
    \end{split}
    \end{equation}
    with $\mathcal{N}(\mu,\sigma)$ a normal distribution with mean $\mu$ and variance $\sigma^2$ and with $\vec{v}_h$ representing the velocity of the halo. 
    
    \item  \textbf{Particle velocity}
    
    When using particles for the satellite positions (\Sec{sec:profile}), we also assign the velocity of the particles to the satellite galaxies. 
    
    \item \textbf{Velocity bias}
    
    The dispersion of velocities of dark matter particles is \textit{a priori} expected to be different than that of galaxies and subhaloes. For this reason, we include a velocity bias $\alpha_v$ in the velocity assignment: 
    \begin{equation}
    \vec{v}_{\rm sat} =  \vec{v}_{\rm h} + \alpha_v (\vec{v}_{\rm DM} - \vec{v}_{\rm h}) \, .
    \label{eq:vb}
     \end{equation}
    This equation is directly applicable when using particles. When using the virial theorem, the velocity bias may be understood as a rescaling of the velocity dispersions: 
    \begin{equation}
        v_i^{\rm gal} \curvearrowleft \mathcal{N}(v_i^{\rm h}, \alpha_v \cdot \sigma_{\rm vir}) \, .
    \end{equation}
    
    \item \textbf{Infall velocity}
    
     We can split the galaxy velocity with respect to the halo in two components, radial $v_r$ (defined along the line between the halo centre and the galaxy position) and angular $v_\phi$: 
     \begin{equation}
         \vec{v}_r = (\vec{v}^{\rm gal} - \vec{v}^{\rm h}) \cdot \vec{u}_r \, ,
     \end{equation}
     
     \begin{equation}
         \vec{v}_\phi = (\vec{v}^{\rm gal} - \vec{v}^{\rm h}) - \vec{v}_r \,
     \end{equation}
     with
    \begin{equation}
         \vec{u}_r = \frac{ \vec{r}_{\rm sat} - \vec{r}_{\rm h} }{ | \vec{r}_{\rm sat} - \vec{r}_{\rm h} |  } \, .
    \end{equation}
     
     \citet{orsi2018} studied the velocity distributions of star-forming galaxies from a semi-analytical model of galaxy formation and evolution. They found that the radial component of the velocity, $v_r$, of star-forming galaxies has two contributions: a Gaussian centred at $0$ equivalent to that described by \Eq{eq:vt} ($v_r \curvearrowleft \mathcal{N}(0,\sigma_{\rm vir})$) and a net infall velocity towards the center of the halo, that follow approximately:
     
    \begin{equation}
         v^{\rm infall} \curvearrowleft \mathcal{N}(-500 km/s,200km/s) \, .
    \end{equation} 
    
    We include this contribution in the mock generation (when specified) by adding a new component to the velocity:
    
    \begin{equation}
        \vec{v}^{\rm gal}_{\rm tot} = \vec{v}^{\rm gal} + v^{\rm infall}\cdot \vec{u}_r \, .
    \end{equation}
    
\end{itemize}

 In the top panel of \Fig{fig:vel} we show the radial velocity distribution of the model ELGs generated with the velocity models detailed above. We see on one hand how the velocity bias $\alpha_v$ affects the width of the velocity distribution and on the other hand how adding the infall velocity ($v^{\rm infall}$) shifts the centre of the distribution from $v_r \sim 0$ to $v_r \sim -500$km/s. This is a more extreme case  than that described in \citet{orsi2018}, where they found a combination of the two peaks (at $v_r \sim 0$ and $v_r \sim -500 km/s$). We expect that for a case similar to that described in \citet{orsi2018}, the results on galaxy clustering will be in between the two cases presented here (with and without $v^{\rm infall}$). 

The distribution of the particle velocities follows a distribution with a width similar that derived from the virial theorem, although slightly skewed towards negative values. This is expected for the particles, as not all of them are virialised and some are expected to currently being accreted. 

The effect of the velocity distributions on the projected correlation function (middle of Fig.~\ref{fig:vel}) is negligible. The only appreciable differences are actually due to the differences reported for the positions of satellite galaxies for NFW and particle profiles (see~\Sec{sec:profile}). 

As expected, it is in the quadrupole where the main differences appear, due to variations in the velocity profiles. We find that the quadrupole is more suppressed (closer to zero) for higher velocity dispersion (higher $\alpha_v$), as expected by the Finger-of-God effect. 
The effect of adding a net infall velocity also suppresses the quadrupole power. This is expected, as this additional peculiar velocity uncorrelated with the large scale structure also erases clustering along the line-of-sight. 

We added in the top sub-figure of \Fig{fig:vel} some vertical lines indicating the dispersion of velocities of satellite galaxies around haloes along the $Z$-axis (arbitrarily chosen as the line-of-sight) $\sqrt{ \langle (v_{\rm gal,z}^{\rm tot} - v_z^{\rm h})^2\rangle}$. This allows us to  quantify to first order the Finger-of-God effect, and the ordering if these lines can be identified with the ordering of the quadrupoles at the mildly non-linear scales. 

We remark that this dispersion is different to $\sqrt{\langle v_r \rangle }$, we checked the ordering would be quite different in that case and unrelated to the quadrupoles. When taking an arbitrary line-of-sight (e.g. the Z-axis), the dispersion due to the addition of \vinfall, appears a factor $1/3$ smaller due to projection effects. This does not occur for the viral velocities whose dispersion occur in all the 3 dimensions. 

The quadrupole from models using the particle information are within 1$\sigma$ of that from assuming NFW profiles, and their ratios are very close to 1. This is remarkable, since the velocity assignment follows different procedures, and their velocities profile (top panel) showed some differences. 

On the other hand, paying attention to the exact shape of the quadrupole, one can find subtle but statistically significant differences in the shape induced by the effect of velocity bias and the effect of infall velocities. For example, the $\alpha_v=0.2 $ and \vinfall case (yellow dashed) follows closely the line of the standard case ($\alpha_v=1$ and $v_r=0$, green solid) for $s>12Mpc/h$, where the effect of the overall velocity dispersion is already apparent. These curves, however, diverge at smaller scales. Many other subtleties could be found exploring in more detail the small scales, however, it is unclear that we could gain any further intuition from basic principles. 

We also analysed the hexadecapole $\xi_4(s)$, and found some differences at small scales due to the Finger-of-God effect. The differences among mocks were qualitatively similar to the results found in the quadrupole, but statistically less significant, which is why we omit the hexadecapole in the figures. 

Additionally, in Appendix \ref{sec:fourier} we show the clustering in Fourier space of all the mocks presented in Sections~\ref{sec:hod}, \ref{sec:pdf}, \ref{sec:profile}, \ref{sec:velocities}. When exploring the effect of velocity profile choices, we also show the hexadecapole of the power spectrum. 

\section{Fitting the eBOSS data}
\label{sec:results}

\begin{table*}
    \centering
    \begin{tabular}{l|c|c|c|c|c|c|c|r|r|r|r}
             Mock   & HOD & $f_{\rm sat}$ & $\beta$ & $K$ & $\alpha_v$ & profile & $v^{\rm infall}$ & $\chi^2_{\ wp}$ & $\chi^2_{\ 2}$ & $\chi^2_{\ 0}$ & $\chi^2_{\ \rm tot}$ (bins: 14+3+5) \\
             \hline
             0     & HOD-3 & \textbf{0.22} & 0 & 1 & 1 & {\sc NFW} & 0 & 21.7 & 1.1 & 5.0 & 31.3 \\
             \hline
             1     & HOD-3 & \textbf{{0.56}} & \textbf{N-I} & 1 & 1 & {\sc NFW} & 0 & 17 & 0.3 & 4.7 & 24 \\
             2     & HOD-3 & \textbf{{0.51}} & 0 & \textbf{{0.25}} & 1 & {\sc NFW} & 0 & 7.2 & 0.3 & 5  & 12.7 \\
             3     & HOD-3 & \textbf{{0.21}} & 0 & 1 & \textbf{1.5} & {\sc NFW} & 0 & 22 & 0.5 & 5.3 & 28.3 \\
             4     & HOD-3 & \textbf{{0.21}} & 0 & 1 & \textbf{1.0} & {\sc NFW} & -500 & 22 & 0.5 & 5.0 & 28 \\
             \hline
             5     & HOD-3 & \textbf{{0.36}} & \textbf{0.0} & 1 & 1 & PART  & 0 & 15.5 & 0.7 & 4.7 & 23 \\
             6     & HOD-3 & \textbf{{0.44}} & 0 & \textbf{0.4} & 1 & PART  & 0 & 8 & 0.3 & 4.6 & 13.5 \\
             7     & HOD-3 & \textbf{{0.26}} & 0 & 1 & \textbf{1.2} & PART  & 0 & 15 & 0.2 & 4.6 & 21.4 \\
              8     & HOD-3 & \textbf{{0.26}} & 0 & 1 & \textbf{0.8} & PART & -500 & 16 & 0.9 & 4.3 & 21.2 \\
             \hline
             9     & HOD-3 & \textbf{{0.48}} & \textbf{0.10} & \textbf{0.15} & 1 & NFW & 0 & 6 & 0.3 & 4.9 & 10.9 \\
             10     & HOD-3 & \textbf{{0.21}} & \textbf{0.0} & 1 & \textbf{{1.5}} & NFW & 0 & 22 & 0.5 & 5.3 & 28.3 \\
             11     & HOD-3 & \textbf{{0.51}} & 0 & \textbf{0.25} & \textbf{1.0} & NFW & 0 & 7.2 & 0.3 & 5  & 12.7 \\
             \hline
             12 & HOD-1 & \textbf{0.40} & \textbf{N-I} & 1 & 1 &  {\sc NFW} & 0 & 17.9 & 0.3 & 4.7 & 25 \\
             13 & HOD-1 & \textbf{0.43} & 0 & \textbf{0.25} & 1 & {\sc NFW} & 0 & 7 & 0.3 & 5.0 & 12.4 \\
             14 & HOD-1 & \textbf{{0.18}} & 0 & 1 & \textbf{1.6} & {\sc NFW} & 0 & 22 & 0.3 & 5.5 & 28.6 \\
             \hline
             15 & HOD-2 & \textbf{0.70} & \textbf{N-I} & 1 & 1 &  {\sc NFW} & 0 & 21 & 0.3 & 4.9 & 28.4 \\
             16 & HOD-2 & \textbf{0.70} & 0 & \textbf{0.25} & 1 & {\sc NFW} & 0 & 8.1 & 0.3 & 4.8 & 13.8 \\
             17 & HOD-2 & \textbf{{0.22}} & 0 & 1 & \textbf{1.5} & {\sc NFW} & 0 & 22 & 0.2 & 5.4 & 29.1 \\
             \hline

    \end{tabular}
    \caption{List of best fit mocks under different assumptions. For each best fit, we write from left to right: a number in order to label it (Mock), the mean HOD choice (HOD), the fraction of satellites (\fsat), the model parameters $\beta$, $K$, $\alpha_v$, the profile choice, the choice of infall velocity and the $\chi^2$ of the mock with respect to the data for the projected correlation function, the quadrupole, the monopole and the combined $\chi
   ^2$. Variables that are set free are in {\bf bold} . For the details on the modelling see \Sec{sec:mocks}, and for a description of the fits see \Sec{sec:results}.}
    \label{tab:fits}
\end{table*}

\begin{figure}
    \centering
    \setlength\lineskip{0pt}
    \includegraphics[trim = 10 10 20 40,clip,width=\linewidth]{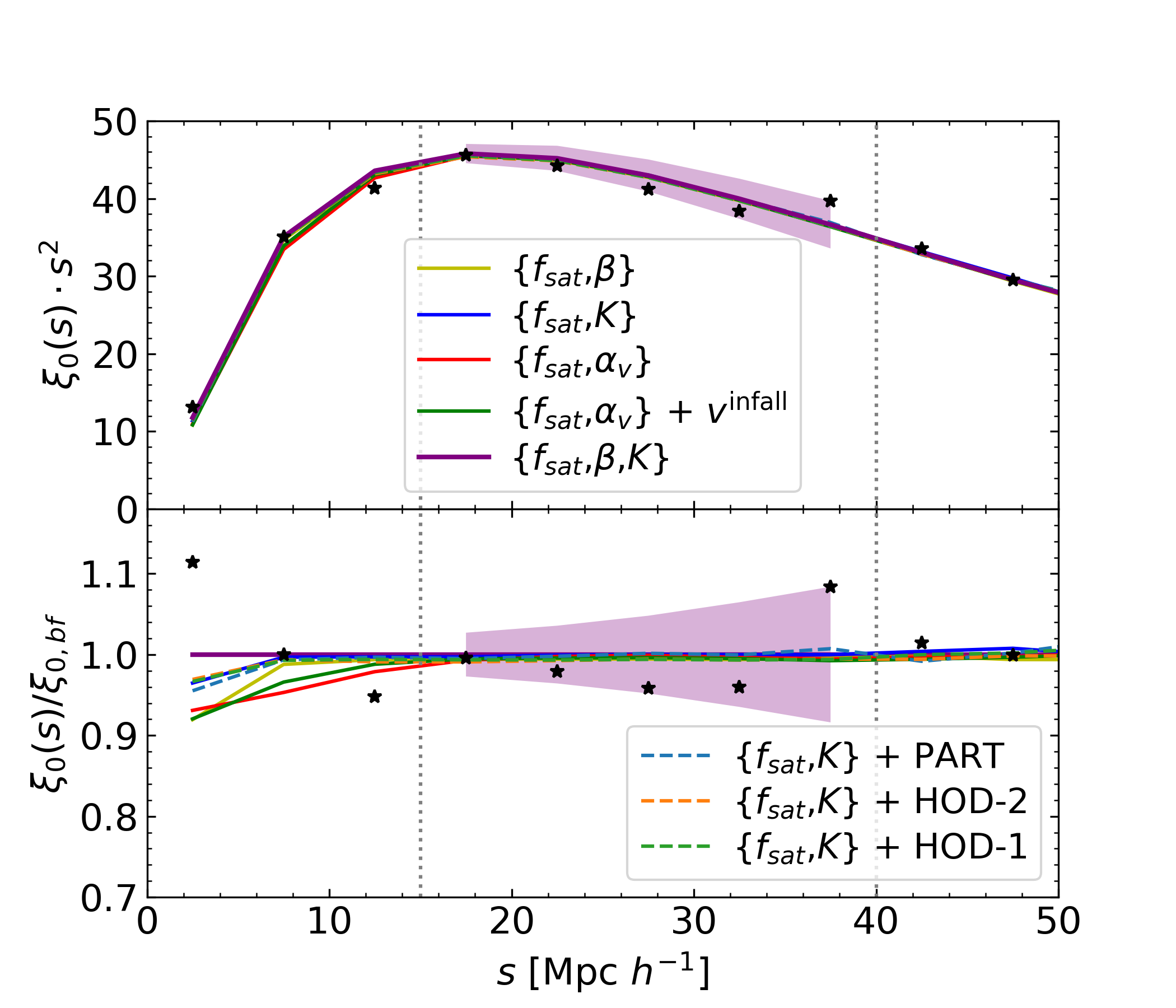}
    \includegraphics[trim = 10 10 20 40,clip,width=\linewidth]{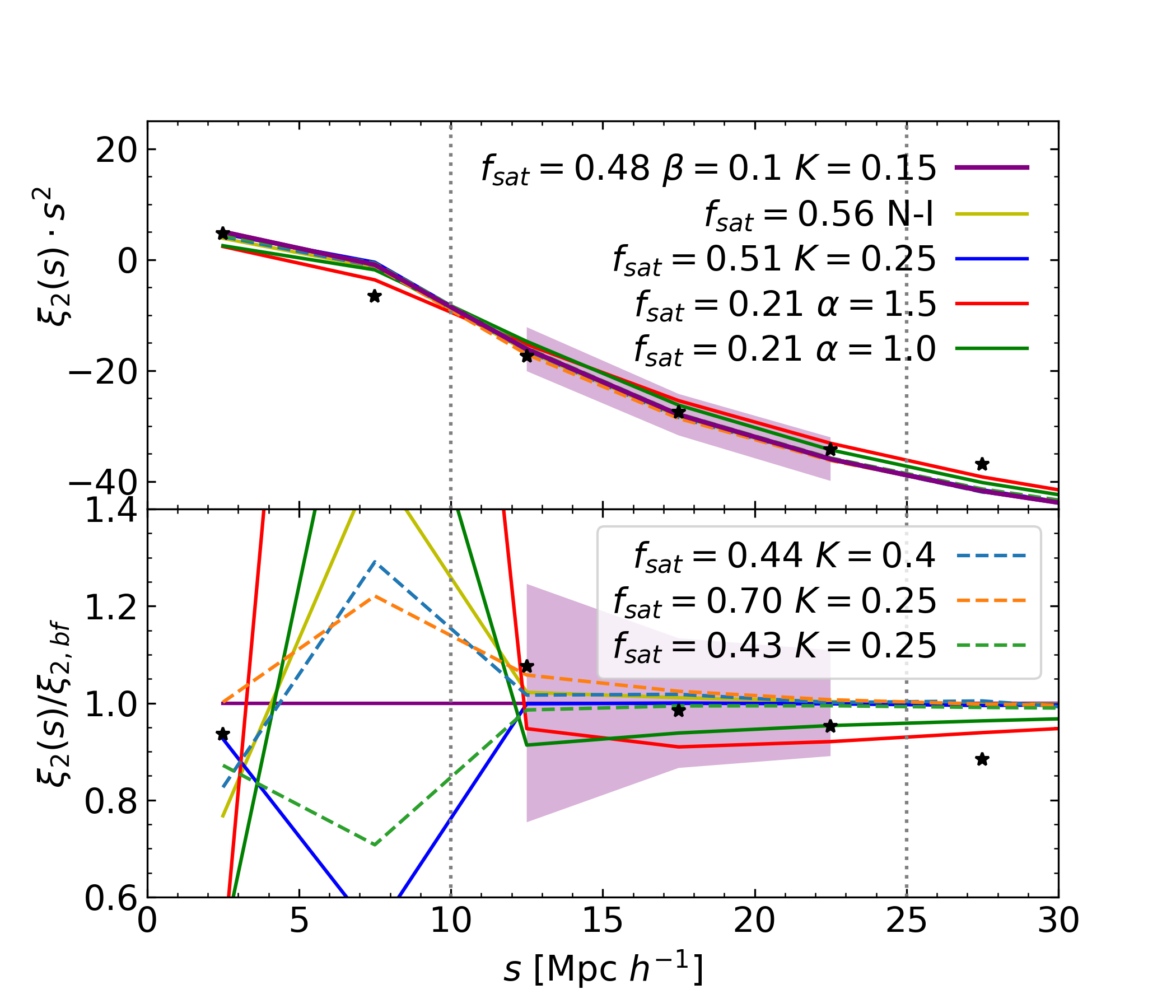}
    \includegraphics[trim = 10 5 20 40,clip,width=\linewidth]{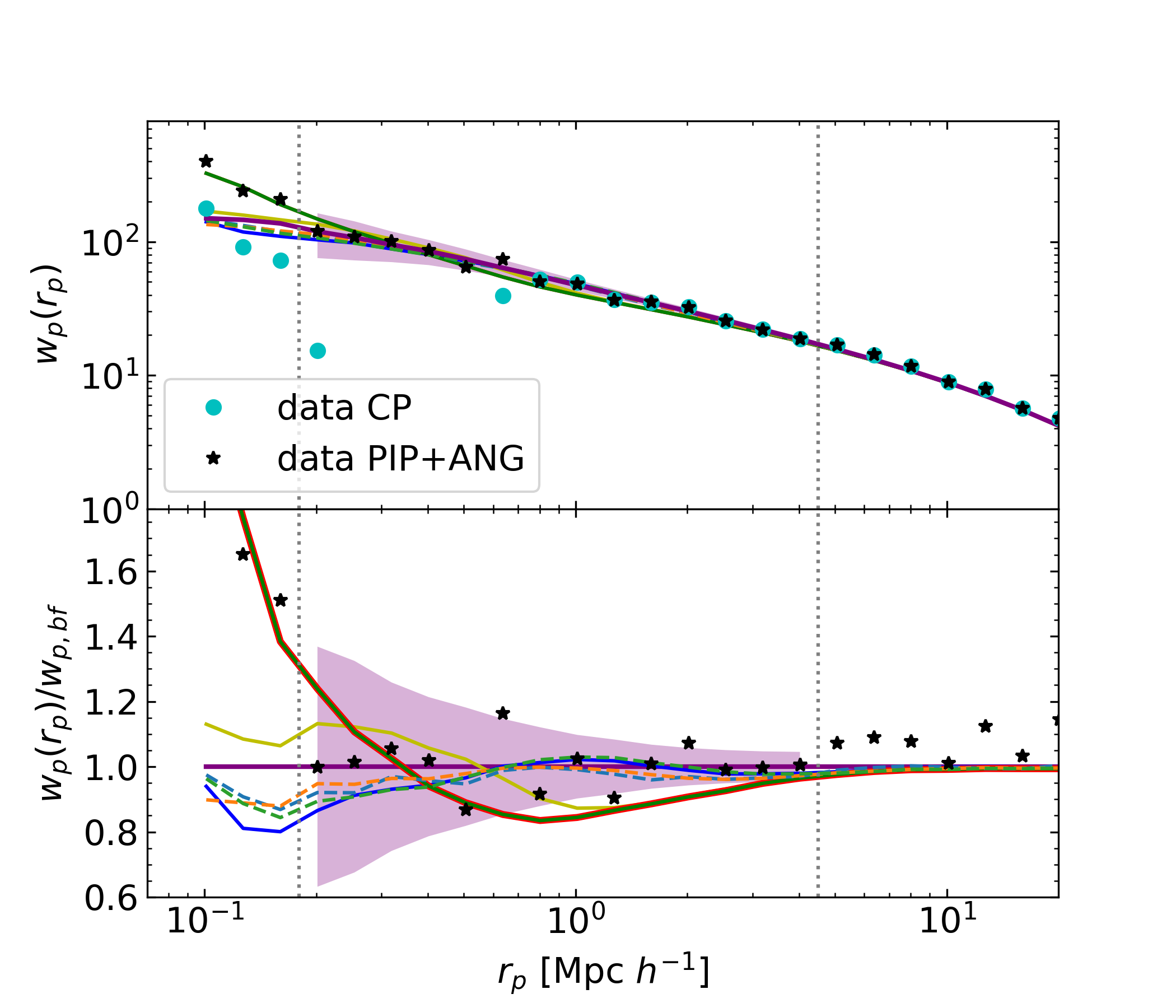}
    \vspace{-0.5cm}
    \caption{2-point correlation functions ($\xi_0$, $\xi_2$, $w_p$) of the mocks that best fit the eBOSS data, under different assumptions  mocks  1, 2, 3, 4, 6, 13, 16, 11 of \Tab{tab:fits}), as explained in Sections \ref{sec:1d},\ref{sec:2d}, \ref{sec:parts}, \ref{sec:3d}. The stars show the clustering of eBOSS data corrected with the PIP+ANG weights (Sec. \ref{sec:PIP}). Points show the eBOSS data with the CP corrections instead of the ANG+PIP for $w_p$ (the only case where differences are noticeable). We use as reference model the mock 11 for the ratios and the shaded area represents the error bars derived from the diagonal of the covariance matrix in \Eq{eq:cov}. The vertical lines represent the interval of scales considered for the fits. 
    }
    \label{fig:bestfit}
\end{figure}

In this section we describe the optimisation procedure followed to find the HOD models that produce the mock catalogues with clustering closest to the eBOSS ELG data.

\subsection{Optimisation}

In the previous section we have presented the wide variety of HOD mocks generated for this work. Here we explore the parameter space of the HOD models and constrain it with the observational data presented in \Sec{sec:data}.
We construct our data vector, $\vec \theta$, as a combination of the monopole, the quadrupole and the projected correlation function, at different scales (see also~\Sec{sec:data}):

\begin{equation}
\begin{split}
& \theta_{0,2,r_p} = \{ \xi_0(s_0),\, \xi_2(s_2), \, w_p(r_p),\}\, \\ 
& \forall\ \{ 15<s_0<40; \ 10<s_2<25; \ 0.02\leq r_p \leq 4.5 \} \ [{\rm Mpc}/h] .
\label{eq:scales}
\end{split}
\end{equation}

We choose these scales so that the information in each statistics are complementary. The range of scales considered are enclosed by two dotted vertical lines in Fig.~\ref{fig:bestfit}, which is described in detail in~\Sec{sec:bestfit}. The $w_p$ is cut at $r_{p,\rm max}=4.5$Mpc/$h$ so that it mostly contains information from the 1-halo term (see Figs. \ref{fig:2PCF_fsat}, \ref{fig:beta} and \ref{fig:K}) given that the amplitude of the 2-halo term is set by the bias, which is fixed.
As shown in Fig. \ref{fig:bestfit}, below $r_{p,\rm min}=0.02$Mpc$/h$ there is sudden change in the behaviour of $w_p(r_p)$ for the observational data. This might be due to some not fully accounted for systematic errors. As we have covered a wide range of the 1-halo term at these small scales, which are well below the fibre collision diameter scale (Section \ref{sec:catalogue}), we do not to include points below $r_p \leq r_{p,\rm min}= 0.02$Mpc$/h$. 

For the monopole, we only use quasi-linear scales ($15<s<40$), which for most cases will not be affected by the choice of HOD parameters (once $b$ is fixed), but it does help ruling out some extreme models. 
For the quadrupole, we also use quasi-linear scales. In this case those scales enter mildly in the 1-halo velocity term, as we saw in \Sec{sec:velocities} that these scales are already affected by choices of the velocity profiles. As we explained in \Sec{sec:data}, there are some systematics that have not been removed in this study, because the way they were eliminated in \citet{tamone2020} would imply a big change for the definition and interpretation of the quadrupole. For this reason we use $s_{2,\rm max}=25 Mpc/h$, that is the scale at which the effect starts to appear.

We note that the scale choices previously mentioned can affect the results that we find in the subsections below, so these have to be interpreted carefully. In Appendix \ref{sec:scales}, we look at changes in the scale cuts, finding qualitatively similar results to the fiducial choices. In the absence of systematic errors one would not need to worry about choosing certain scales, and any redundant information would be accounted for by the covariance matrix. Hence, the study in Appendix \ref{sec:scales} must be understood as a consistency check in order to search for possible systematics not accounted for. We also prefer to choose shorter data vectors because with larger and more correlated data vectors the inversion of the covariance matrix becomes numerically noisier or even unfeasible by standard methods. 

We now explore the HOD models by creating 27 $(1{\rm Gpc}/h)^3$ mock catalogues at each parameter space point, following a grid in parameter space that is refined at later iterations. We compute the different 2-point correlation functions of these mocks. Their mean will be our theory vector $\vec{\theta}_{\rm th}$ at a given point in the explored parameter space, for which we compute the $\chi^2$ against the data vector $\vec{\theta}_{\rm data}$: 

\begin{equation}
    \chi^2 = \big( \vec{\theta}_{\rm th} - \vec{\theta}_{\rm data} \big)^T C^{-1} \big( \vec{\theta}_{\rm th} - \vec{\theta}_{\rm data} \big) \, ,
    \label{eq:chi2}
\end{equation}
with $C$ the covariance matrix of $\theta$. Our best fit is then defined as the point in parameter space that minimises $\chi^2$. 

In order to compute $C$ we use the 1000 {\sc EZmocks} presented in \citet{EZmocks}.
We use a version of the mocks that include the eBOSS geometry, but no observational systematics (in particular no fibre assignment). The effect of the observational systematics is expected to be minor compared to the need of rescaling explained below.
We reanalyse these mocks with the {\sc Outer Rim} fiducial cosmology in order to follow the procedure used with the observational data. This changes the amplitude of the clustering statistics with respect to using the true cosmology of the {\sc EZmocks}. There is also a miss-match in the amplitude of $w_p$ at scales below the fibre collision scale regardless of the choice of cosmology. This is expected as the {\sc EZmocks} were not tuned to match the clustering at highly non-linear scales. 
For this reason, we rescale the covariance matrix by 

\begin{equation}
    C_{i j} = C_{i j}^{\rm EZ} \frac{\theta_i^{\rm OR} \cdot \theta_j^{\rm OR} }{\theta_i^{\rm EZ} \cdot \theta_j^{\rm EZ} } \, ,
    \label{eq:cov}
\end{equation}
where $C^{\rm EZ}$ is the raw  covariance matrix from the {\sc EZmocks}, $\theta^{\rm EZ}$ the mean of the correlation function of the {\sc EZmocks} and $\theta^{\rm OR}$ the correlation function of one of our best fit {\sc Outer Rim} mocks (mock 9 in \Tab{tab:fits}, with the lowest $\chi^2$). This rescaling is derived from the assumption that the correlation matrix of the {\sc EZmocks}, their relative uncertainty and the amplitude of the clustering of the {\sc Outer Rim} mock are all correct. We do not expect this choice of covariance matrix to affect strongly the best fit values. In fact, we also did the same analysis with a diagonal covariance inferred directly from the standard deviation of the $1$Gpc$/h$-subboxes of  mock 9 in \Tab{tab:fits}, rescaled to the eBOSS volume (as done in \Sec{sec:stat} for the errorbars). For this case we found results for the best fits similar to the ones shown here, but with artificially small $\chi^2$ contours, given that no correlation between points was taken into account.

\begin{figure*}
    \includegraphics[trim=18 3 40 22,clip,width=0.5\linewidth]{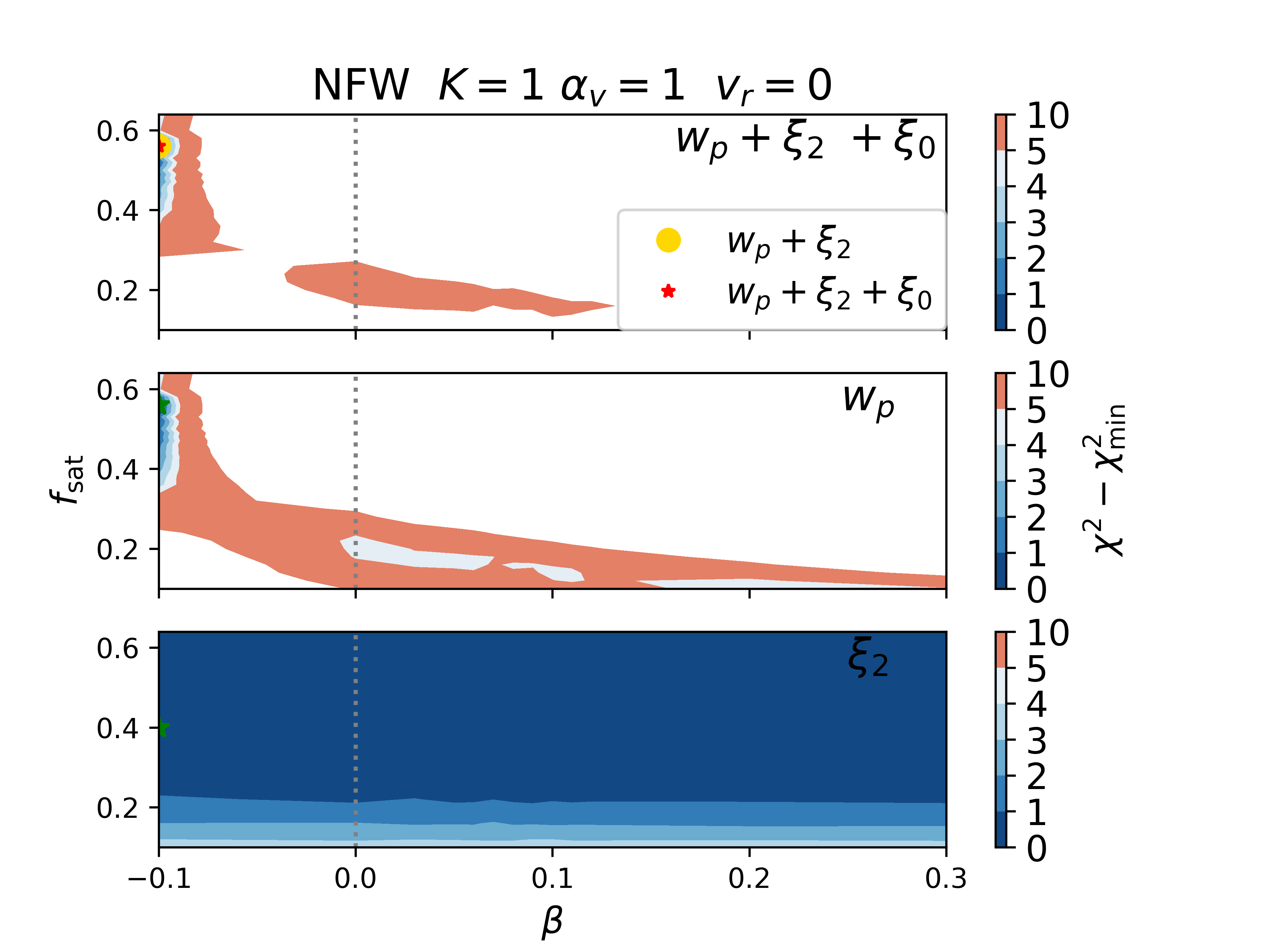}\includegraphics[trim=15 3 40 22,clip,width=0.5\linewidth]{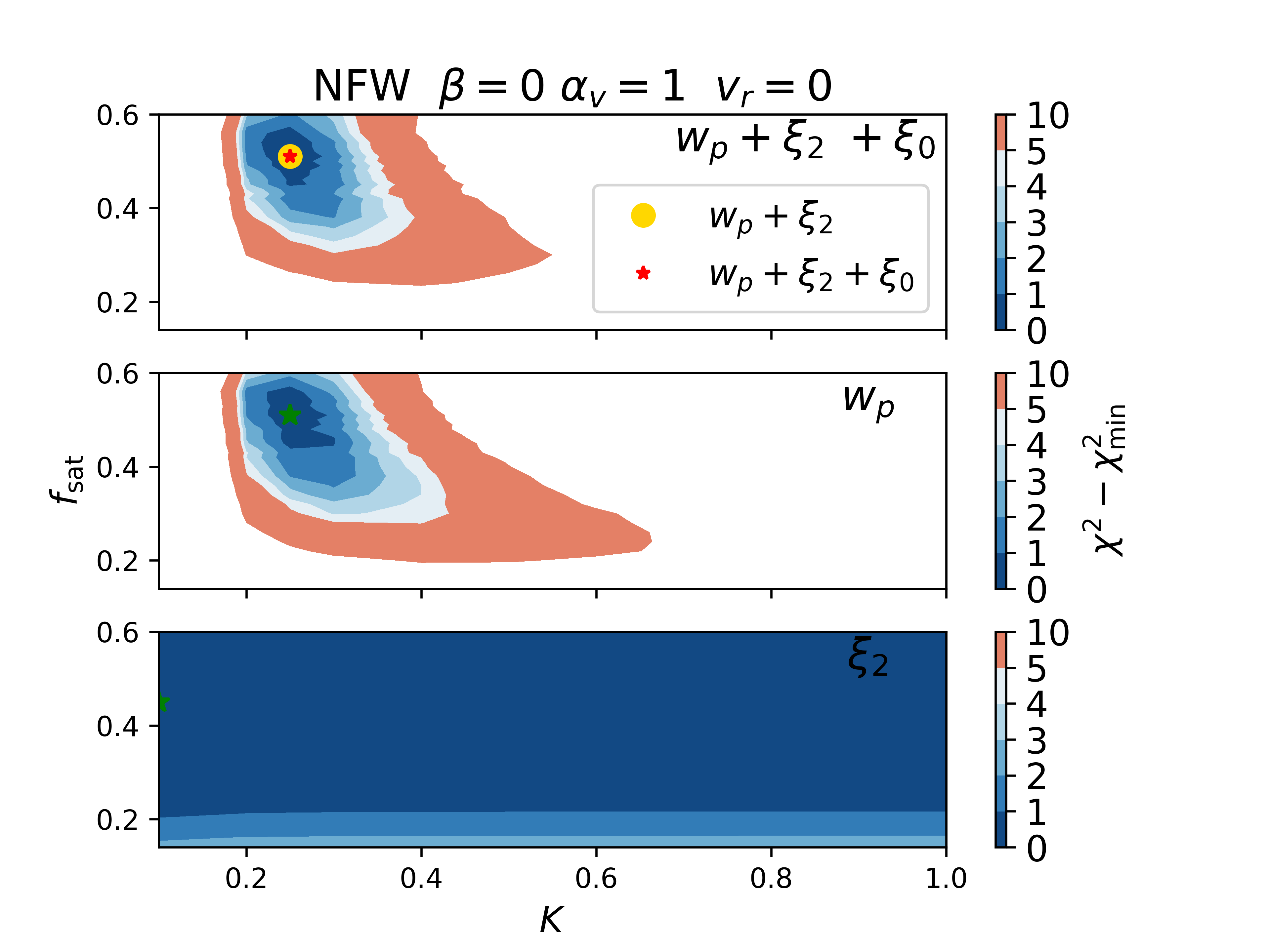}
    \includegraphics[trim=18 3 40 22,clip,width=0.5\linewidth]{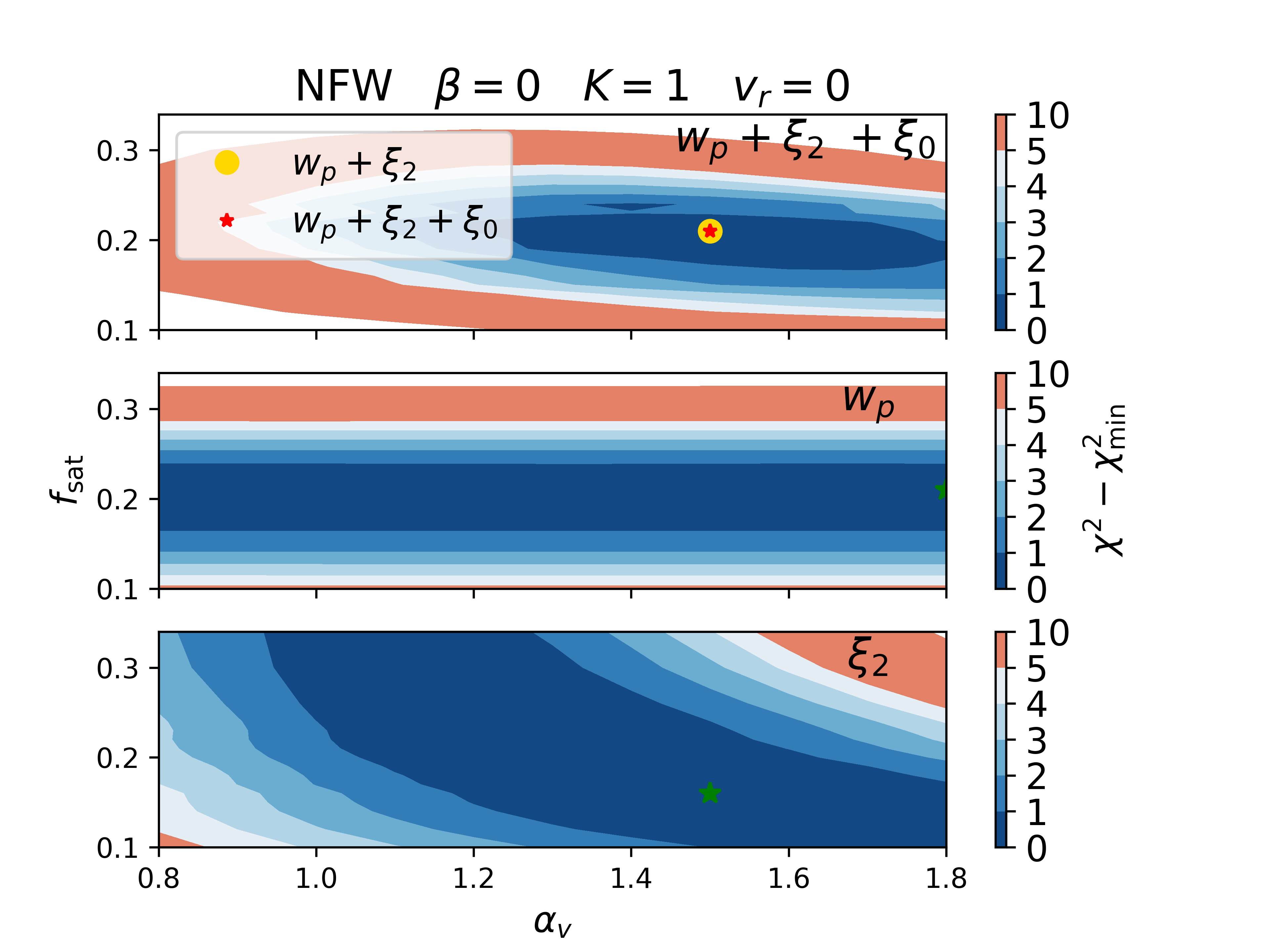}\includegraphics[trim=15 3 40 22,clip,width=0.5\linewidth]{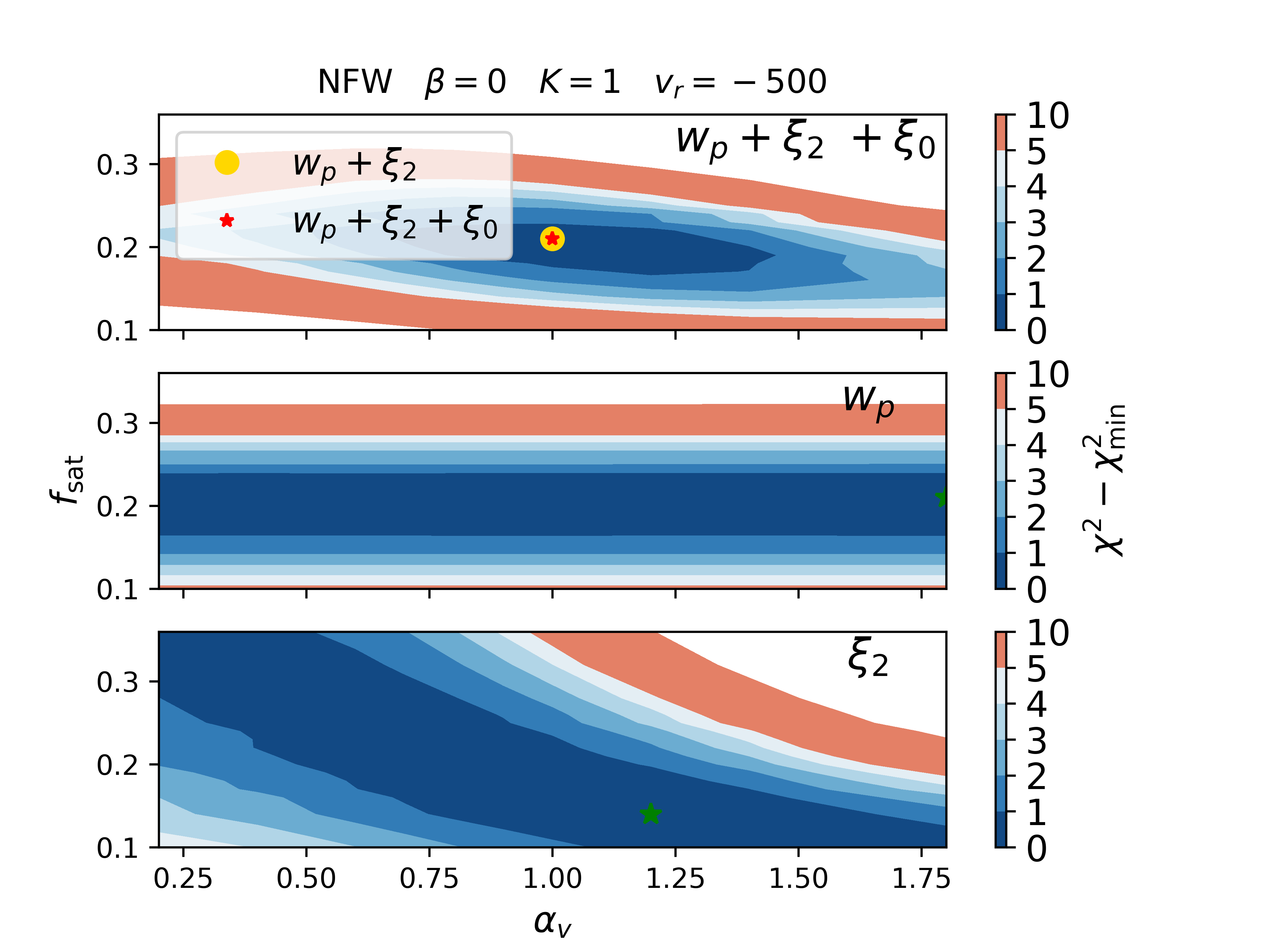}
    \caption{ $\chi^2$ contours for the baseline + one free parameter (different in each sub-figure) model. {\it Top-Left:} $\beta$ is set free, with $\beta=0$ representing the Poisson distribution, $\beta>0$ a negative binomial distribution and $\beta=-0.1$ (arbitrary choice for the representation)  the nearest-integer distribution (see~\S~\ref{sec:pdf}). {\it Top-Right:} The concentration of the satellite profiles are rescaled by a free parameter $K$, $c\to K \cdot c$ (see~\S~\ref{sec:profile}). {\it Bottom-Left:} The velocity dispersion is allowed to vary, $\sigma_{\rm vir} \to \alpha_v \sigma_{\rm vir}$ (see~\S~\ref{sec:velocities}).  {\it Bottom-Right:} An infall velocity component is added, \vinfall$=-500\pm200 km/s$, while still letting $\alpha_v$ free. Within each sub-figure we show the $\chi^2$ component of the quadrupole (bottom panel), projected correlation function (middle panel) and the combination full fit to the monopole, quadrupole and $w_p$ (top panel, scales are defined in Eq.\ref{eq:scales}). Except for the variations specified in each sub-figure, we assume the baseline model described in \S~\ref{sec:1d}: HOD-3, NFW profile, $\beta=0$, $K=1$, $\alpha_v=1$, \vinfall=0. The models shown here assume, $n = 7\times n_{\rm eBOSS}$. The filled stars  show the minimum $\chi^2$ for the statistics used in each sub-figure. In the top panels we also include a filled circle for the minimum $\chi^2$ if we exclude the  monopole of the fit.  
    }
    \label{fig:chi2_2D}
\end{figure*}

\begin{figure*}
    \includegraphics[trim=18 3 40 22,clip,width=0.5\linewidth]{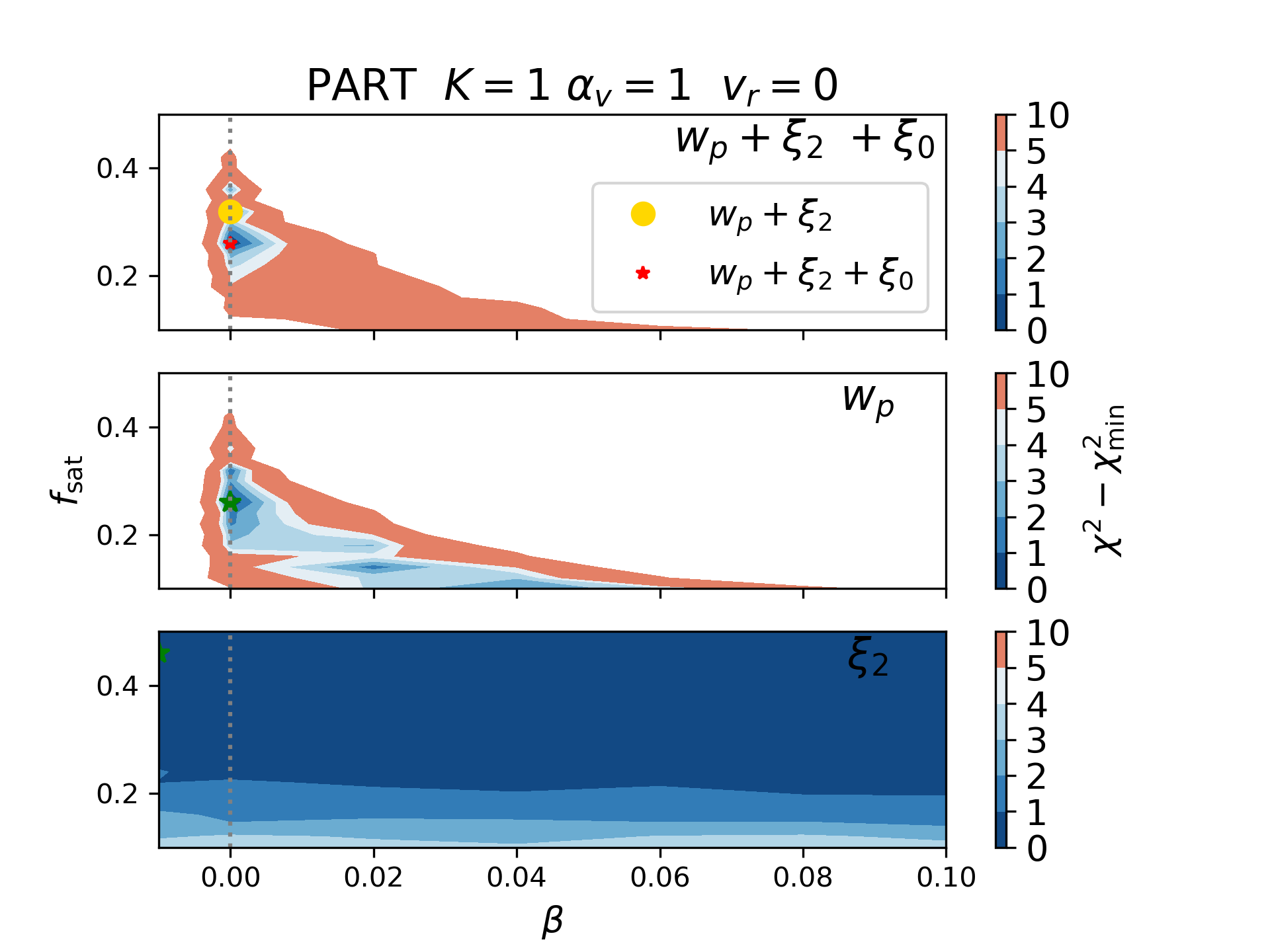}\includegraphics[trim=15 3 40 22,clip,width=0.5\linewidth]{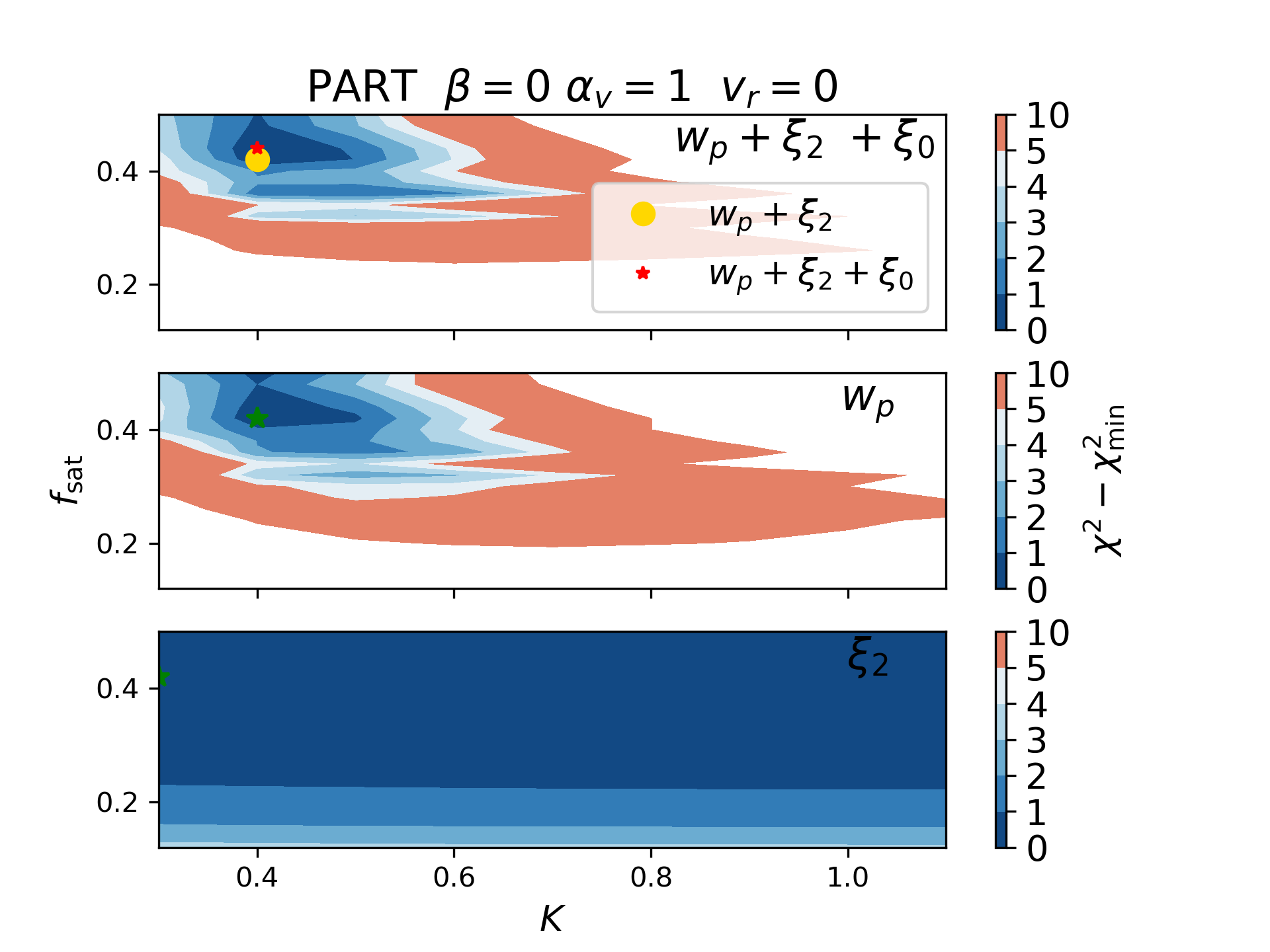}
    \includegraphics[trim=18 3 40 22,clip,width=0.5\linewidth]{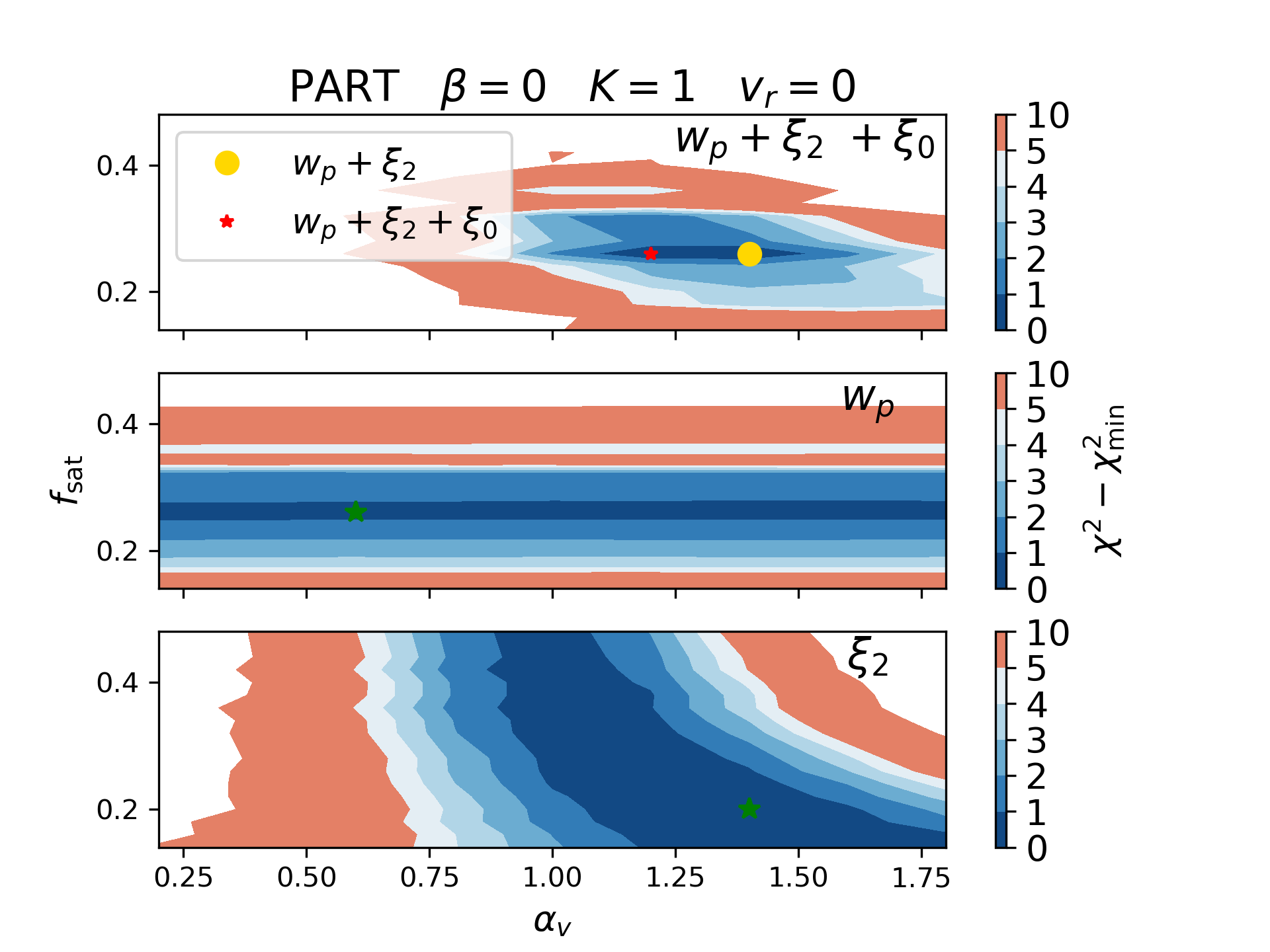}\includegraphics[trim=15 3 40 22,clip,width=0.5\linewidth]{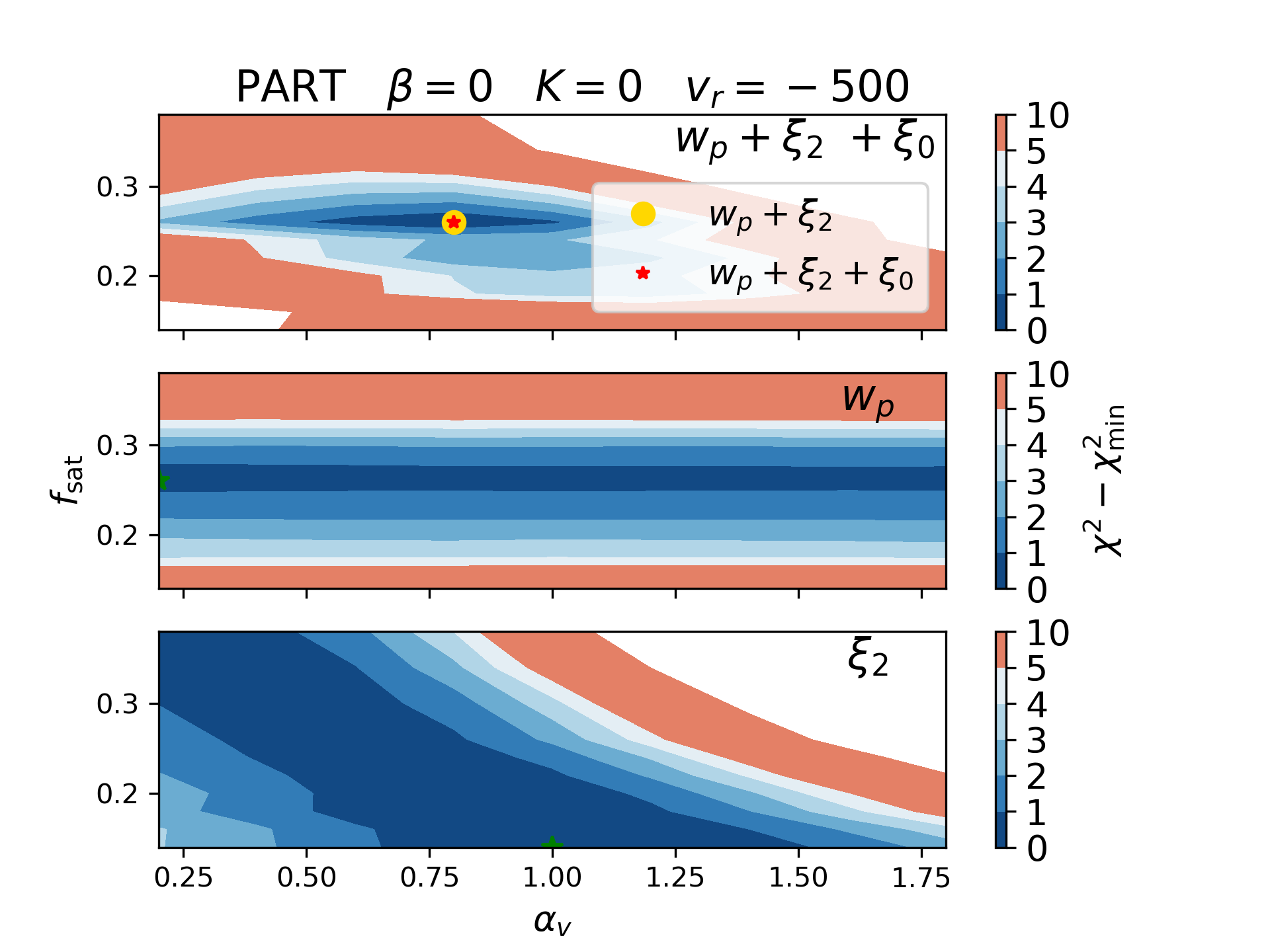}
    \caption{$\chi^2$ contours for the particle  + $f_{\rm sat}$ + one free parameter model. We follow the same structure as in \Fig{fig:chi2_2D}, but using the particles profiles and velocities as a starting point for the models. See \Sec{sec:profile} \& \Sec{sec:velocities} for the details of the differences in the modelling. Note that this figure is made from mocks with lower number density, $n = 1\times n_{\rm eBOSS}$, yielding noisier contours.
   }
    \label{fig:chi2_2D_PART}
\end{figure*}

\subsection{Baseline model}
\label{sec:1d}

Our baseline model consists of mock catalogues with the HOD-3 shape (\Sec{sec:hod}), satellite galaxies drawn from a Poisson distribution ($\beta=0$, \Sec{sec:pdf}) following a NFW profile (with  $K=1.0$, \Sec{sec:profile}) and with virial velocities  ($\alpha_v=1$ and $v^{\rm infall}=0$, \Sec{sec:velocities}). These are the default choices taken in \Sec{sec:mocks}, except that now we have not fixed \fsat, which is set to vary following \Eq{eq:fsat}.

For all the models we keep constant the linear bias, $b=b_{\rm eBOSS}$, and set the number density to $n = 7\times n_{\rm eBOSS}$. This number density is the maximum before reaching $\langle N_{\rm cen} \rangle = 1.0$ for \fsat$=0$, where the model breaks down (i.e. when the number of central galaxies becomes larger than 1). Having a larger number density reduces the noise in our theoretical model, $\vec{\theta}$, hence reducing the noise in the inferred best fit and $\chi^2$ contours. 

The best fit from this baseline model is the first entry in \Tab{tab:fits}, mock 0. It has \fsat$=0.22^{+0.02}_{-0.03}$, with error bars representing the $\Delta \chi^2 =1$ interval. The best value found for \fsat\ is close to that found by~\citet{guo19}, 13-17\%, where an earlier version of the eBOSS ELG sample was analysed with the incomplete conditional stellar mass function model presented in~\citet{guo2018}.

The baseline model gives a poor fit to the observational data, so we explore other alternatives below.

\subsection{Baseline + 1 parameter  model}
\label{sec:2d}

In this subsection, we relax the baseline model by allowing an extra degree of freedom.  The change is introduced, one at a time, in one of the three following aspects: (i) a non-Poissonian PDFs for satellite galaxies, (ii) rescaling the satellite density profile by a factor $K$, (iii) modifying the velocity dispersion by a factor $\alpha_v$. Additionally, for the $\alpha_v$ model we also consider a separate case with a net infall velocity, \vinfall. The results are summarised in the second tier of \Tab{tab:fits}, mocks 1 to 4, and in \Fig{fig:chi2_2D}. This figure shows the $\chi^2$ as a function of $f_{\rm sat}$ and another variable ($\beta$, $K$ or $\alpha_v$) for $w_p$, $\xi_2$ and the combination of \{$w_p$, $\xi_2$, $\xi_0$\}.

In the case of modifying the PDF of satellite galaxies, the nearest integer is represented with a negative $\beta$ in \Fig{fig:chi2_2D} (with an arbitrary value of $\beta=-0.1$), the Poisson distribution with $\beta=0$ and negative binomial distributions with $\beta>0$.
Remarkably, the best fit shows a preference for low scatter, with the nearest-integer PDF, and a large satellite fraction ($f_{\rm sat}=0.56$). 
We highlight that there is not a smooth transition between a Poisson and a nearest-integer PDF in terms of scatter, resulting in the \textit{break} that appears at $\beta=0$ in the top-left sub-figure of \Fig{fig:chi2_2D}\footnote{We use the \textsc{contourf} function from the \textsc{python} library \textsc{matplotlib} for these figures. This function interpolates the colour between discrete values. This is why the \textit{break} appears below $\beta=0$.}. 
In line with the effects found in \Sec{sec:pdf}, $\xi_2$ does not constrain $\beta$, but does set a lower limit for \fsat\ (with a very slight degeneracy with $\beta$). $w_p$ constrains $\beta$ to be small, with a notable degeneracy with $f_{\rm sat}$. A preference for a sub-Poissonian distribution, $\beta<0$, is at odds with the results from SAMs~\citep{jimenez2019}. However, as we will show in the following subsections, our best fit values for $\beta$ depend on other model assumptions. 
 
When we vary the profile concentration, setting $K$ free (see~\S~\ref{sec:profile}), we find a similar effect, with $\xi_2$ only constraining $f_{\rm sat}$ and $w_p$ driving the main constraints. We find the best fit for $K=0.25$ (with a step of 0.05 in the parameter space grid) and \fsat=0.52, clearly favouring profiles less concentrated than NFW, in line with previous studies (see references in \Sec{sec:profile}).
 
In the bottom-left of \Fig{fig:chi2_2D} we show what happens when allowing for a velocity bias $\alpha_v$. In this case, $w_p$ is the quantity that is insensitive to the choice of $\alpha_v$ (in lines with \Fig{fig:vel}), constraining only \fsat. $\xi_2$ shows a strong degeneracy between \fsat\ and $\alpha_v$. When combining both, we find the best fit at \fsat$=0.21$ and $\alpha_v=1.5$. For the NFW profiles, $\alpha_v$ represents a deviation from the galaxy velocity dispersion found in \citet{BryanNorman}. Hence, the observational data prefers an enhanced velocity dispersion within this model.  

Building upon the preference for a larger velocity dispersion, we also include a net infall velocity of \vinfall$\sim-500 km/s$ (see \Sec{sec:velocities}), letting $\alpha_v$ to also vary. The bottom-right panel of \Fig{fig:chi2_2D} shows similar results to the previous case, but with a preference for lower $\alpha_v$ values. The constraints from $w_p$ remain the same and those from $\xi_2$ shift and get distorted. In this case, we get a best fit of $\alpha_v=1.0$ and the same fraction of satellites \fsat$=0.21$. This suggest that including \vinfall$\sim-500 km/s$ enhances the Finger-of-God effect, in a similar way as with $\sim 0.5 \sigma_{\rm vir}$. We note again that the modelling for \vinfall adopted here is more extreme than that of \citet{orsi2018}.

Out of the 4 extensions to the baseline model considered here, the modification of the satellite density profile (with the concentration controlled by K) yields the best fit to the data, with a reduced $\chi^2$ below unity (we do not show explicitly the reduced $\chi^2$ as it can be derived from the data already provided in Table \ref{tab:fits}). We consider combinations of these extensions in \Sec{sec:3d}.

\subsection{Particles + 1 parameter  model}
\label{sec:parts}

In this section we repeat the variations described in \Sec{sec:2d}, but using the particle position and velocity profiles, \parts, instead of the NFW profile and virial theorem velocities. 
A lower density $n = 1\times n_{\rm eBOSS}$ is used, as the computation is much more demanding in this case. With this choice, we also minimise the cases for which we run out of particles. However, this choice increases the noise in the modelling of $\theta$, giving, in turn, noisier contours. 

The results are summarised in \Fig{fig:chi2_2D_PART} and the third tier of \Tab{tab:fits}, mocks 5 to 8. The best fits for these models, follow roughly the results described in \Sec{sec:2d}, with some differences: 

\begin{itemize}
    \item In the \{\fsat, $\beta$\} plane, this time the data prefer the Poisson PDF ($\beta=0$), together with a lower fraction of satellites, following a similar degeneracy as that seen in \Sec{sec:2d}.
    
    \item Results in the other planes (\{\fsat, $K$\}, \{\fsat, $\alpha_v$\}, \{\fsat, $\alpha_v$+\vinfall\}) show qualitatively similar results to the NFW case. 
    
    \item The best \parts\ fit give $K=0.4$, which, although closer to the baseline model ($K=1$) than the \nfw\ case, is clearly preferring less concentrated profiles, $K<1$. The differences in the profiles, seen in the top panel of \Fig{fig:K}, change the preferred value of $K$. The satellite fraction is also somewhat lower than for the \nfw\ case.
    
    \item The best fit $\alpha_v$ gets shifted by $\Delta \alpha_v \sim -0.3$ (-0.2 for the \vinfall\ case) when using \parts\ profiles.

    \item For the \parts\ models, $\alpha_v$ corresponds to the definition of galaxy velocity bias: the ratio of the velocity dispersion of galaxies to the one of dark matter particles. The value found here, $\alpha_v=1.2$, is compatible with that found in sub-haloes in simulations \citep{vbias}.
    
    \item For those models with $\alpha_v$ set as a free parameter (with or without \vinfall), the \parts\ models give better fits than for the \nfw\ ones. For the other cases, the $\chi^2$ has similar values.
    
\end{itemize}

As for both NFW, with \parts\ profiles we find the effect of \vinfall\ to be mostly equivalent to a shift in $\alpha_v$. Thus, we will not explore further the \vinfall$=-500$ case in the following subsections or in Appendix \ref{sec:scales}.


\subsection{Baseline + 2 parameter  mode}
\label{sec:3d}

\begin{figure}
    \includegraphics[trim = 2 0 45 20,clip,width=\linewidth]{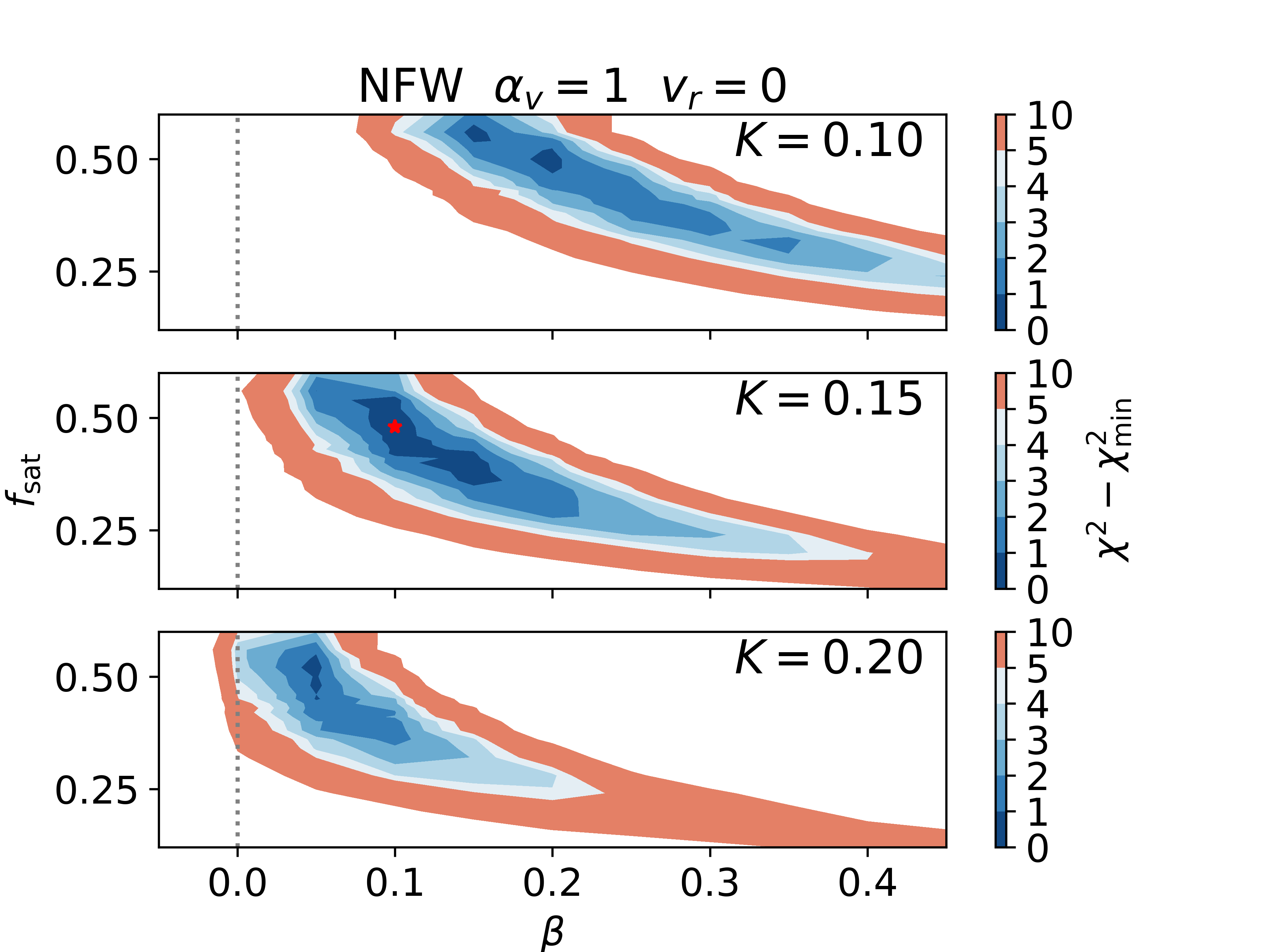}
    \includegraphics[trim = 10 0 45 20,clip,width=\linewidth]{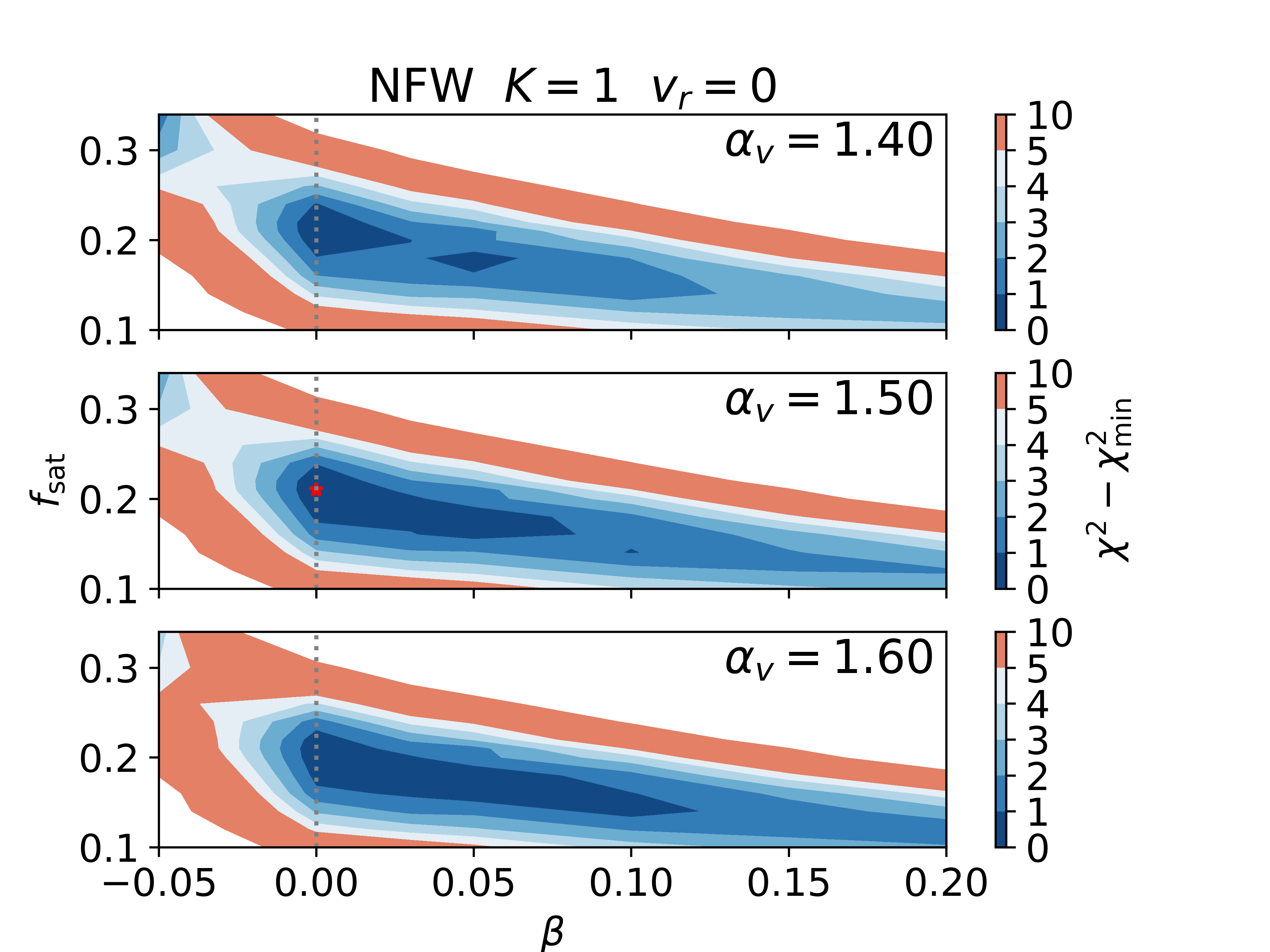}
    \includegraphics[trim = 10 0 45 20,clip,width=\linewidth]{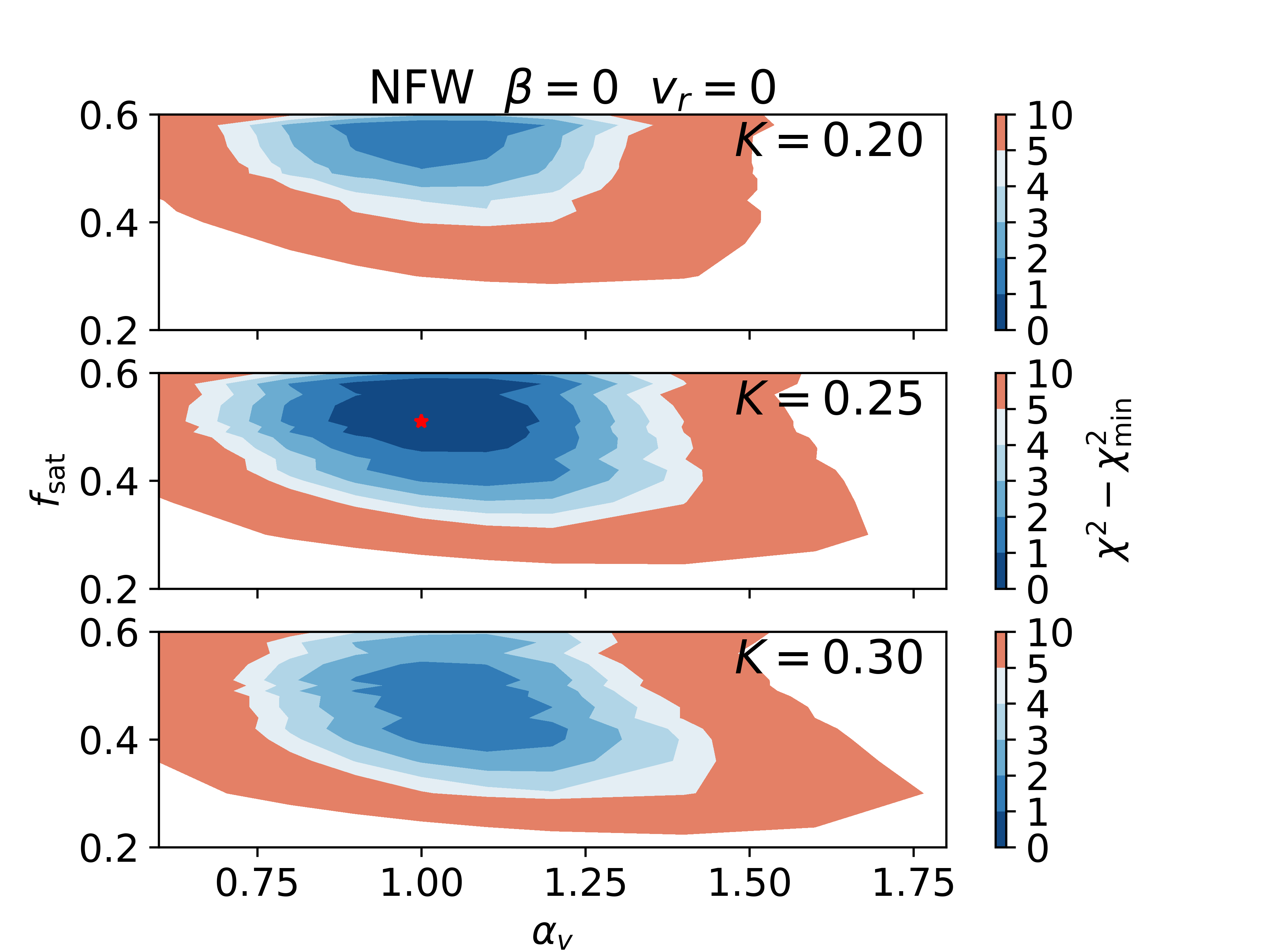}
    \caption{$\chi^2$ contours for the baseline + 2 parameters model. These two parameters are \{$\beta$ \& $K$\}, \{$\beta$ \& $\alpha_v$\},  \{$K$ \& $\alpha_v$\} for the top, middle and bottom sub-figures, respectively. In this figure, we only show the combined (\{$\xi_0,\xi_2,w_p$\}) $\chi^2$. Within sub-figures, panels are used to represent an extra dimension, which is set constant within the panel, and varied across panels. The star represents the minimum of the $\chi^2$. 
    }
    \label{fig:chi2_3D}
\end{figure}

We keep increasing the level of complexity of the model by setting free $f_{\rm sat}$ and two other parameters, while fixing the rest to the default choices, including a NFW profile. 
We set free at once the following groups of parameters: \{\fsat, $\beta$, $K$\}, \{\fsat, $\beta$, $\alpha_v$\} and \{\fsat, $\alpha_v$, $K$\}. The results from these fits are summarised in \Fig{fig:chi2_3D} and the fourth tier of \Tab{tab:fits}, mocks 9 to 11. We highlight some of the main results:

\begin{itemize}
    \item When both $\beta$ and $K$ are free, the data prefer positive $\beta$ and $K$ gets even smaller ($K=0.15$, see Table \ref{tab:fits}).
    
    \item \{$f_{\rm sat}=0.48$, $K=0.15$, $\beta=0.1$\} gives the best fit of all models considered.
    
    \item When both $\beta$ and $\alpha_v$ are set free, the data prefers a Poisson distribution (unlike when only $\beta$ was free, \Sec{sec:1d}) and we recover the $\alpha_v=1.5$ found in \Sec{sec:1d} (for $\alpha_V$ free).
    
    \item When both $K$ and $\alpha_v$ are varied, we recover the $K=0.25$ case (like in \Sec{sec:1d}) and the data prefer a model with the fiducial virial theorem velocity, $\alpha_v=1$.
    
    \item A general result is that a low $K$ (a less concentrated profile than NFW) is necessary to obtain a good fit to the data. 

\end{itemize}

\subsection{HOD variations + 1 parameter}

\begin{figure*}
    \centering
    \includegraphics[trim = 10 20 45 10,clip,width=0.5\linewidth]{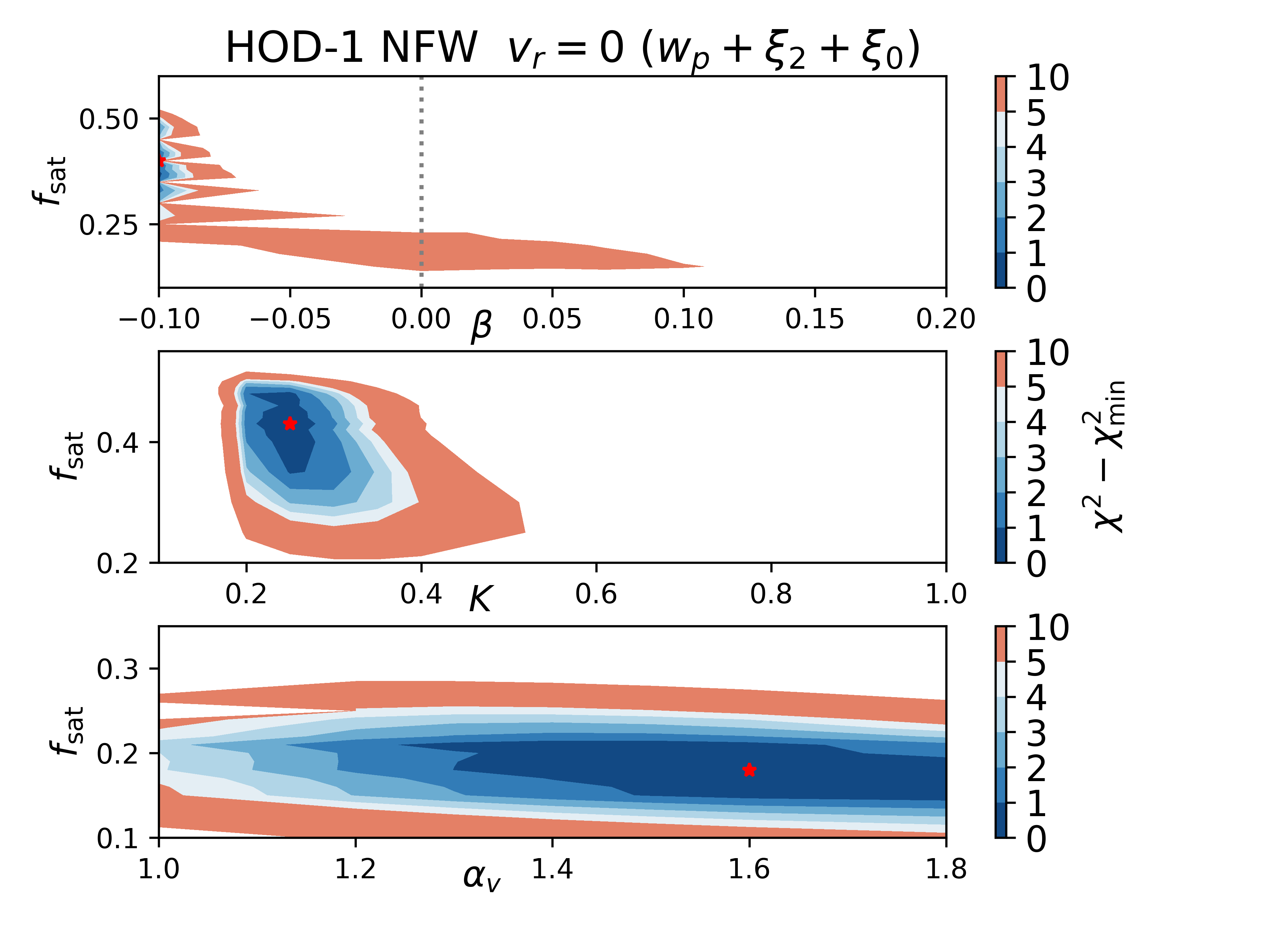}\includegraphics[trim = 10 20 45 10,clip,width=0.5\linewidth]{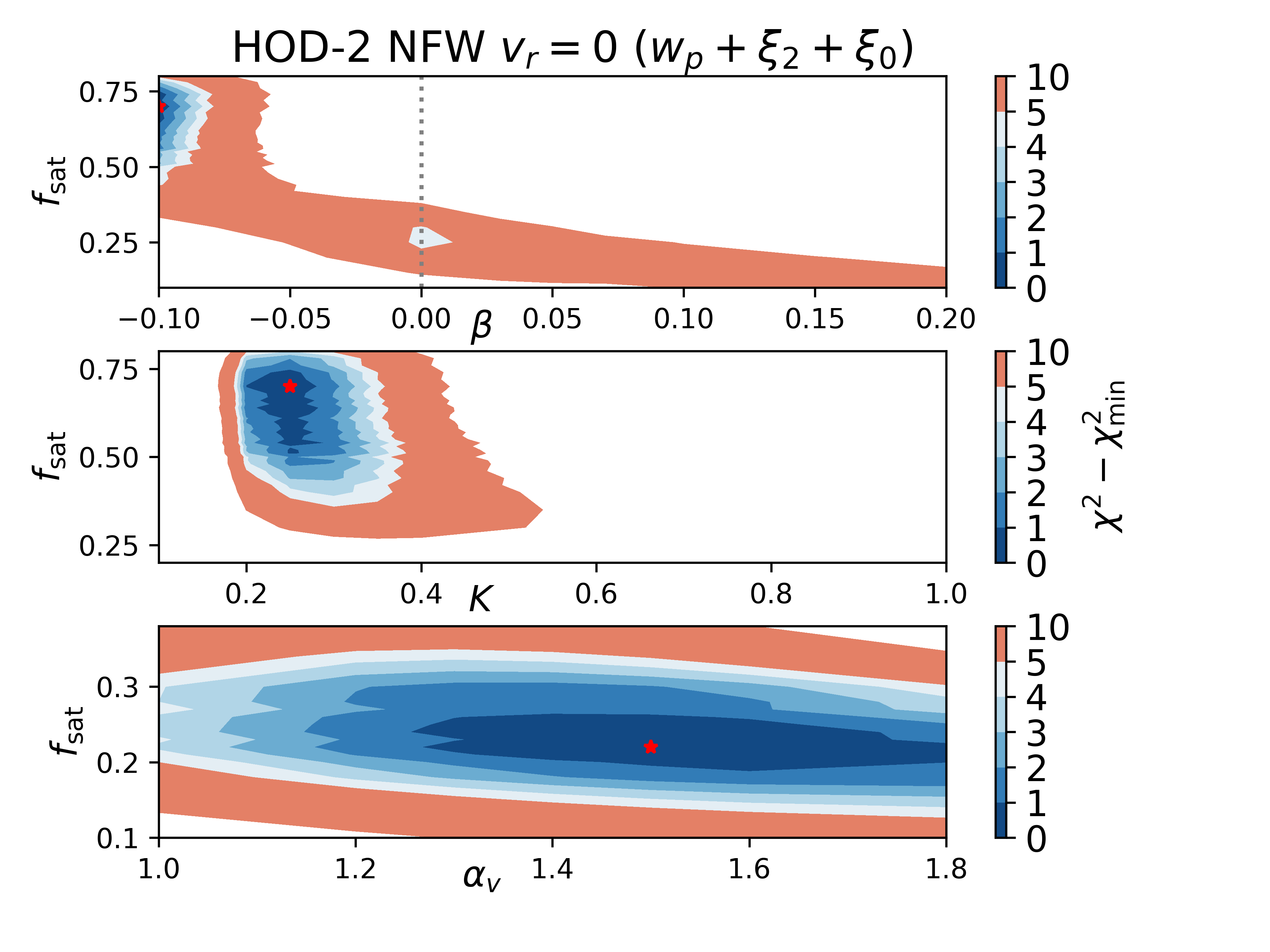}
    \caption{$\chi^2$ contours of the HOD-1 (left sub-figure) and HOD-2 (right sub-figure) mocks, constructed varying \fsat\ and one other parameter ($\beta$, $K$ or $\alpha_v$ as indicated in each panel). The rest of default choices are the same as in Fig. \ref{fig:chi2_2D}. All the panels in this Figure show results for the full data vector $w_p$+$\xi_2$+$\xi_0$.
    }
    \label{fig:chi2_HOD}
\end{figure*}

So far in this section, we have used the HOD-3 model in all cases. In this subsection, we use the  HOD-1 and HOD-2 functions (Eqs. \ref{eq:hod1}, \ref{eq:hod2}, \& Table \ref{tab:HOD}) for the mean halo occupation distributions, and explore the parameter space by leaving \fsat\ and one parameter ($\beta$, $K$ or $\alpha_v$) free.
The results are summarised in the last 2 tiers of Table \ref{tab:fits}, mocks 12 to 17, and Figure \ref{fig:chi2_HOD}.

We find that the choice of mean HOD changes the preferred fraction of satellites. 
In the case of the HOD-1, \fsat\ is always found lower than for models assuming the HOD-3. This can easily be explained by the behaviour of the $w_p$ presented in Fig. \ref{fig:2PCF_fsat}. For the same value of \fsat=0.30, the $w_p$ 1-halo term is larger for HOD-1 and lower for HOD-2. Hence, with respect to HOD-3, HOD-1 models will need a lower fraction of satellite, and HOD-2 a higher one, to match the $w_p$ from the data.

Remarkably, the best fit value of the rest of parameters ($\beta$, $K$ or $\alpha_v$) and the overall shape of the $\chi^2$-contours remain unchanged. This implies that the HOD shape is not degenerated with any parameter other than \fsat. Additionally, the $\chi^2$ shown in Table \ref{tab:fits} mildly disfavours HOD-2.

\subsection{Best Fits}\label{sec:bestfit}

Finally, we compare the correlation functions of the data to the best fit mocks in several of the most representative scenarios assumed in past subsections. Results are shown in \Fig{fig:bestfit} with the lowest $\chi^2$ mock (9 in \Tab{tab:fits}) as a reference. 

All the best fits shown in \Fig{fig:bestfit} reproduce similarly well the monopole of the data over the range of selected scales. We also find that the quadrupoles from all the mocks shown agree with the data, within the error bars. Nevertheless, we find a differentiated shape for the quadrupole for the two fits (with and without \vinfall) with $f_{\rm sat}=0.21$ and fitted $\alpha_v$. 

When studying the projected correlation function, the differences between mocks are clearer, partially because for this statistics we explore a larger range of scales. Here, we see clearly that only models with $K<1$, i.e. with profiles less concentrated than the NFW/PART default one, can fit well small scale data. We can see the differentiated shape induced in the $w_p$ for the different ingredients for the HOD (e.g. $\beta$ vs $K$ vs $\alpha_v$). We note that the HOD shape does not change much the $w_p$ once the \fsat\ is free to compensate for the excess or deficit of small scale clustering. We also see the importance of using the PIP+ANG weights at scales $r_p<1$Mpc$/h$, with the traditional CP weights resulting in a lower clustering \citep[more details in][]{pipeBOSS}.

\section{Conclusions}\label{sec:conclusions}

In this paper we study a series of Halo Occupation Distribution (HOD) models for star-forming Emission Line Galaxies (ELGs) motivated by theoretical and observational studies in the literature. 
We create mock ELG catalogues using \ors haloes at $z=0.865$. This is one of the largest ($L=3 $Gpc$/h$) existing dark matter only $N$-Body simulations within its mass resolution range, $m_p\sim 2\cdot 10^9 M_{\odot}/h$.
Throughout this study we fix the galaxy bias of the mock ELGs to that observed in the eBOSS data catalogue and we take its number density, $n=n_{\rm eBOSS}$, as a reference. We use $n=n_{\rm eBOSS}$ for error estimations or reporting HOD parameters, and $\times7$ to $10$ higher density when computing clustering signal at a given point of the parameter space.
We make the mock catalogues available so they may be used for model testing in preparation for future surveys\footnote{\url{http://popia.ft.uam.es/eBOSS_ELG_OR_mocks}}.

We revisit the HOD model for the case of ELGs, reconsidering most of the assumptions that go into it. We consider different shapes of the mean HOD for central galaxies, $\langle N_{\rm cen} (M) \rangle$: from the classical smoothed step function (HOD-1), through a model with decaying occupation for higher masses (HOD-3) up to a model with no occupation at large masses (HOD-2). We set our default choice to HOD-3 (\S~\ref{sec:hod}) with a piece-wise Gaussian plus a power law that best fit the results from the semi-analytical model of galaxy formation and evolution presented in \citet{vgp2018}. For the mean HOD of satellite galaxies, $\langle N_{\rm sat} (M) \rangle$, we always use a power-law, as is typically done in the literature. We allow three HOD parameters to vary in order to control the number density $n$, the bias $b$ and the fraction of satellites $f_{\rm sat}$, while fixing the scaling relations for the other parameters (see Table~\ref{tab:HOD}).
We highlight that one basic need to match the number density and the large scale bias of a ELG sample is to account for the incompleteness in stellar mass of the ELG central galaxy sample~\citep[see also][]{favole16}. This is common to other samples clearly incomplete in stellar mass such as QSOs~\citep{qso}.

We do not focus exclusively on the functions and parameters that control the shape of the mean HOD, but also on the other choices that need to be made when populating haloes with satellite galaxies. The first of these choices is the probability distribution function $P(N_{\rm sat}|\langle N_{\rm sat} \rangle)$. We consider a Poisson distribution as our default ($\beta=0$), but we also consider a Negative Binomial with greater scatter ($\beta > 0$) and a nearest-integer function ($\beta<0$). 

We place satellite galaxies either following a NFW profile or the particle distribution within haloes. We allow for a rescaling of the profiles as we expect ELGs to follow different profiles than dark matter. We assign velocities either using the virial theorem for the NFW profiles or simply the particle velocities. On top of that we allow for a velocity bias and for the inclusion of a net infall velocity. 

With different combinations of the above choices we construct a range of HOD models.  We study how each of these models affect the clustering of mock galaxies, mainly via the projected correlation function $w_p$ and the quadrupole $\xi_2$. We find that these statistics help separating different effects. Whereas the fraction of satellites affects both statistics, the PDF and the position assignment affect mostly the projected correlation. The velocity assignment mostly affects the quadrupole. We also studied the monopole, but it shows nearly no variations on the linear scales because we fixed the bias to that of the data.

In \Sec{sec:results} we fit to the eBOSS ELG data, mocks produced with different HOD models. Some general findings are: 

\begin{itemize}
    \item In all the analysed scenarios, the observational data prefers dispersed profiles for ELGs ($K=0.15-0.4$).
    \item We also find a mild preference (lower $\chi^2$) for particle profiles as opposed to NFW ones. However, this preference goes away if we let $K$ vary.
    \item We find some preference for a positive velocity bias ($\alpha_v>1$), i.e. a larger velocity dispersion of satellite galaxies around the halo velocity, although once $K$ is set free we recover the $\alpha_v=1.0$ scenario (no velocity bias).
    \item The PDF preference depends a lot on the rest of assumptions. We find that negative, positive and vanishing $\beta$ are preferred in different scenarios.
    \item The shape of the mean HOD does have some effect on the clustering but can be mostly compensated by increasing or decreasing the fraction of satellites. After that change in \fsat, the effect of the HOD shape is found subdominant. The data slightly disfavours HOD-2. 
    \item For HOD-3, the fraction of satellites found to match the clustering vary from $f_{\rm sat}\sim 0.21$, for the cases where both $K$ and $\beta$ are fixed to their default values ($1$ and $0$, respectively), to $f_{\rm sat}\sim 0.50$ when either of those parameters are let vary. In the latter case, $f_{\rm sat}$ rises to $0.70$ when assuming HOD-2 and it goes down to $\sim 0.40$ for HOD-1. The change of profile (to PART) also affects \fsat.    
    \item The key ingredient to match the data seems to be the profile rescaling with a factor $K$, $c\to K \cdot c$ (see~\S~\ref{sec:profile}). 
\end{itemize}

We find that small scale clustering strongly depends on the details of how we place satellite galaxies within haloes. These details may be more relevant than the shape of the mean HOD, which is the quantity many studies in the literature put the focus on. 

The general results obtained here, such as ELGs distributing in more disperse profiles than NFW, are expected to also be applicable to star-forming galaxies at the studied redshifts, independently of their particular selection. Thus, this work is relevant for DESI, that will select ELGs also based on their $\left[\ion{O}{\,\sc II}\right]$ flux, but also Euclid and other surveys targeting star-forming galaxies at $z\sim 1$ in different ways from eBOSS. 

This study shows what scenarios of the ELG - dark matter relation are preferred or ruled out by the observational data. 
These findings have implications for the modelling of physical processes that shape the formation and evolution of ELGs. Studies like this one, can give us a unique insight of the physics of galaxy formation and evolution of ELGs. Such study could also provide information on other samples that will be available with current and future cosmological surveys. 

In this study we did not include galaxy assembly bias, i.e. the dependence of the galaxy clustering on properties other than the halo mass. This is an effect widely seen in model galaxy~\citep[e.g.][]{zehavi2018}. Although several observational studies have concluded that galaxy assembly bias is not a strong source of systematic uncertainty~\citep{tinker2011,kilian2019}, others have found evidence of galaxy assembly bias~\citep{obuljen2020}. Exploring such effect is beyond the scope of this work, partly because we only had access to limited information of the FoF halo catalogue. Additionally, other approaches, such as sub-halo abundance matching, might be more adequate for such purpose~\citep[e.g.][]{contreras2020}. 
Another effect that has not been considered in this study 
is the conditional probability of $N_{\rm sat}$ on whether or not the central galaxy is an ELG. This is known as galactic conformity and has been studied elsewhere \citep[e.g.][]{kauffmann2013,hearin2015,lacerna2018,alam2019}. 

In this study we have not explored the change of the selection function and galaxy evolution within the redshift range of this ELGs sample. The results from \citet{guo19} indicate that the variation in number density at different redshifts has the largest effect on the derived eBOSS ELG mean HODs. This suggests that the shape and, likely, distribution of satellite galaxies, could be maintained, while adjusting the target number density. Similar results are obtained for model ELGs~\citep{vgp2018}. We defer to the future a more in depth study of the evolution of ELG samples.

Our results could be sensitive to the choice of fiducial cosmology. In order to assess that, we would need other {\sc Outer Rim}-size simulations at different cosmologies. This is something beyond the scope of this project, but stage-IV surveys already in preparation might need to consider this.

The eBOSS ELG program, with the largest ELG sample to date serves as a bridge from stage-III to stage-IV experiments, where ELGs will be crucial. ELGs probe, on average, lower halo masses compared to LRGs and have a more complicated selection function. This posed the question whether ELGs would have in turn a complicated relation to dark matter that could have implications when interpreting their anisotropic clustering to understand cosmology. This study probes a very wide variety of plausible scenarios within our current knowledge of ELG formation and evolution. The mocks presented here have been analysed following the same procedure used to derived the eBOSS ELG BAO+RSD measurements \citep{demattia2020,tamone2020,elg} (see Appendix \ref{sec:cosmo} and \citealt{alam2020} for more details), finding no evidence of any bias on the derived parameters within the statistical errors provided by the setup of \ors, which is much lower than the eBOSS uncertainties.

If we want to extract the full cosmological potential of future surveys, we will need to consider smaller and smaller scales in the analysis. Studies like this one will need to be carried out in order to validate the correct extraction of cosmological information and to test ways to disentangle cosmology from baryon physics when interpreting galaxy clustering.

\section*{Data Availability}

A selection of mocks, including those used for Figs. \ref{fig:2PCF_fsat}, \ref{fig:beta}, \ref{fig:K} \& \ref{fig:vel}, is currently available here: \url{http://popia.ft.uam.es/eBOSS_ELG_OR_mocks}. Other mocks may be provided upon request.

\section*{Acknowledgements}

The authors thank Sergio Contreras for comments and Sesh Nadathur for sharing his modified version of \textsc{cute}. Santiago Avila is supported by the MICUES project, funded by the European Union's Horizon 2020 research programme under the Marie Sklodowska-Curie Grant Agreement No. 713366 (InterTalentum UAM). VGP acknowledges support from the European Research Council (ERC) under the European Union's Horizon 2020 research and innovation programme (grant agreement No 769130).
SA is supported by the European Research Council through the COSFORM Research Grant (\#670193). EMM was funded by the European Research Council (ERC) under the European Union's Horizon 2020 research and innovation programme (grant agreement No 693024).

Funding for the Sloan Digital Sky Survey IV has been provided by the Alfred P. Sloan Foundation, the U.S. Department of Energy Office of Science, and the Participating Institutions. SDSS-IV acknowledges support and resources from the Center for High-Performance Computing at the University of Utah. The SDSS web site is www.sdss.org. SDSS-IV is managed by the Astrophysical Research Consortium for the 
Participating Institutions of the SDSS Collaboration including the 
Brazilian Participation Group, the Carnegie Institution for Science, 
Carnegie Mellon University, the Chilean Participation Group, the French Participation Group, Harvard-Smithsonian Center for Astrophysics, Instituto de Astrof\'isica de Canarias, The Johns Hopkins University, Kavli Institute for the Physics and Mathematics of the Universe (IPMU), University of Tokyo, the Korean Participation Group, Lawrence Berkeley National Laboratory, Leibniz Institut f\"ur Astrophysik Potsdam (AIP),  
Max-Planck-Institut f\"ur Astronomie (MPIA Heidelberg), Max-Planck-Institut f\"ur Astrophysik (MPA Garching), Max-Planck-Institut f\"ur Extraterrestrische Physik (MPE), 
National Astronomical Observatories of China, New Mexico State University, 
New York University, University of Notre Dame, Observat\'ario Nacional / MCTI, The Ohio State University, Pennsylvania State University, Shanghai Astronomical Observatory, 
United Kingdom Participation Group, Universidad Nacional Aut\'onoma de M\'exico, University of Arizona, University of Colorado Boulder, University of Oxford, University of Portsmouth, University of Utah, University of Virginia, University of Washington, University of Wisconsin, Vanderbilt University, and Yale University.

This work used the Sciama High Performance Computing cluster which is supported by the Institute of Cosmology and Gravitation and the University of Portsmouth. This research used resources of the National Energy Research Scientific Computing Center, a DOE Office of Science User Facility supported by the Office of Science of the U.S. Department of Energy under Contract No. DE-AC02-05CH11231. This work has benefited from the public available programming language {\sc python}.



\bibliographystyle{mnras}
\bibliography{biblio}



\appendix
\section{Clustering in Fourier Space}
\label{sec:fourier}

\begin{figure}
    \centering
    \includegraphics[trim=5 25 35 48,clip, width=\linewidth]{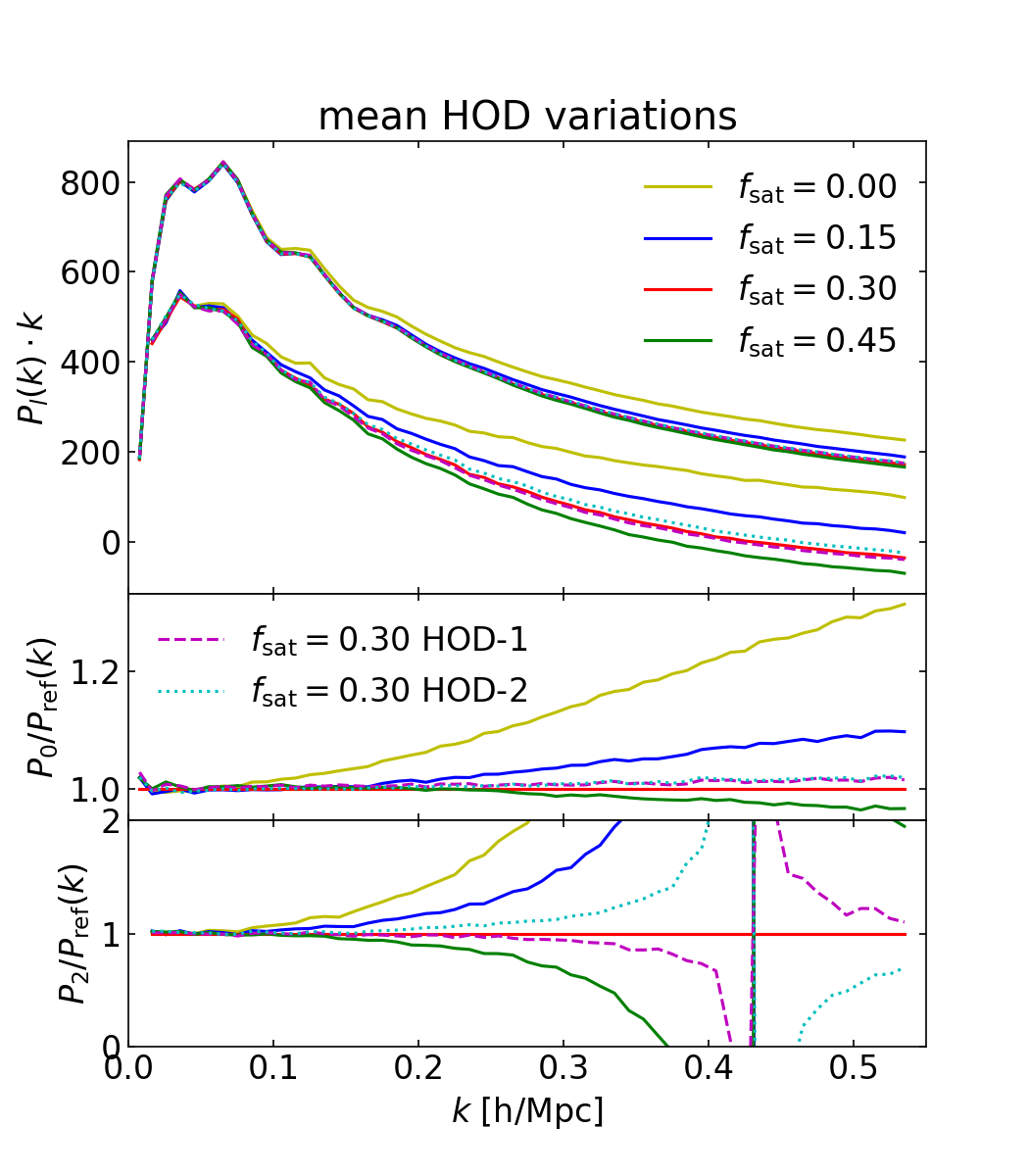}
    \caption{Power Spectrum multipoles of mean occupation halo variations: HOD shapes (HOD-1, HOD-2, HOD-3), and fraction of satellites (\fsat). These correspond to the same mocks as shown in \Fig{fig:2PCF_fsat}. In the top panel we present both the monopole (upper set of lines) and quadrupole (lower set of lines). In the middle/lower panel we show the ratio of the monopole/quadrupole to the reference model: \{HOD-3, \fsat$=0.30$, NFW, $K=1$, $\beta=0$, $\alpha_v=1$\}.
    }  
    \label{fig:Pk_HOD}
\end{figure}

\begin{figure}
    \centering
    \includegraphics[trim=5 25 35 48,clip, clip,width=\linewidth]{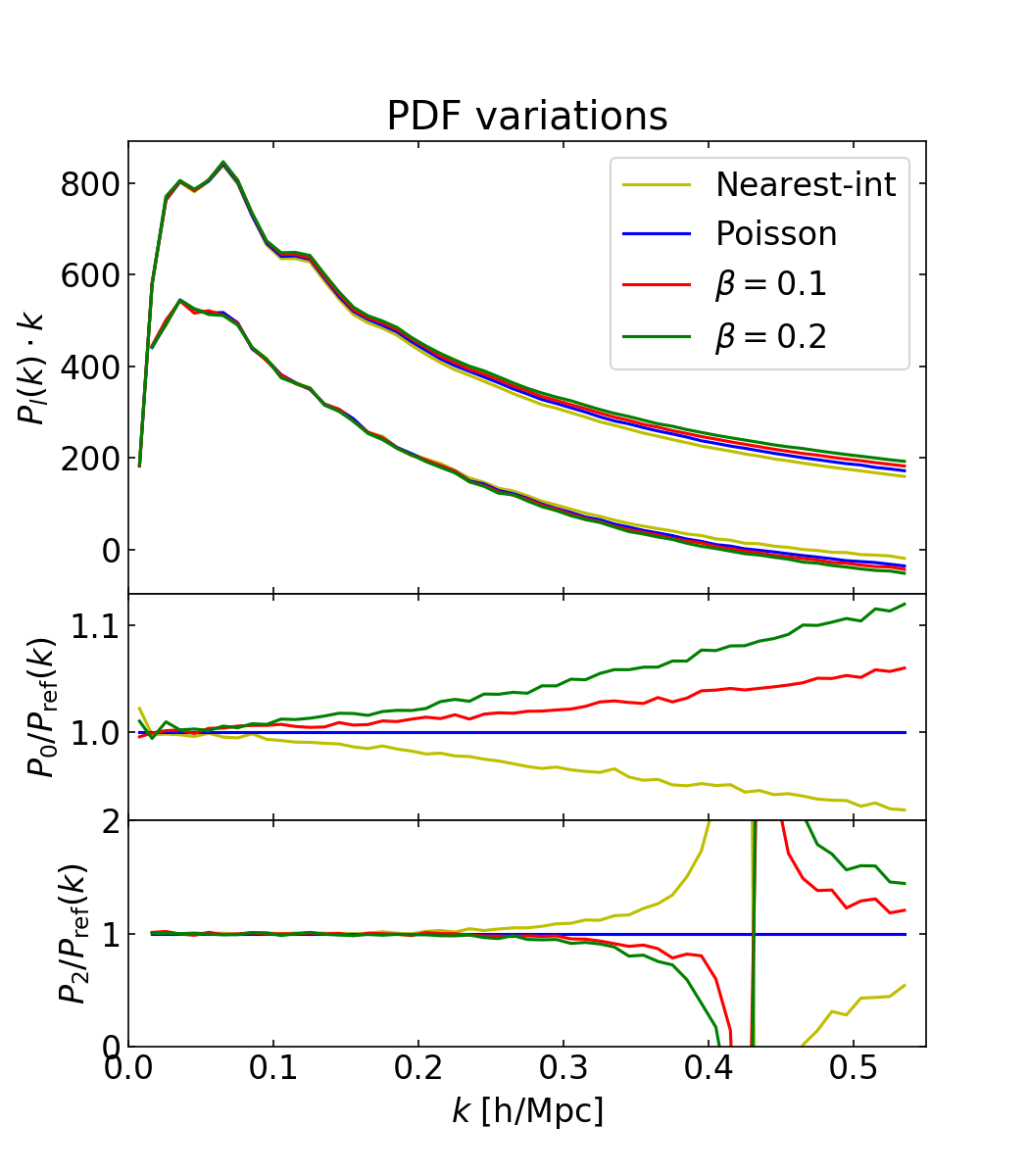}
    \caption{Power Spectrum multipoles of mocks with different Point Distribution Function Variations: nearest-integer, Poisson ($\beta=0$) or Negative Binomial ($0<\beta<1$)). These correspond to the mocks shown in \Fig{fig:beta}. In the top panel we present both the monopole (upper set of lines) and quadrupole (lower set of lines). In the middle/lower panel we show the ratio of the monopole/quadrupole to the reference model: \{HOD-3, \fsat$=0.30$, NFW, $K=1$, $\beta=0$, $\alpha_v=1$\}. 
    }     
    \label{fig:Pk_PDF}
\end{figure}

\begin{figure}
    \centering
    \includegraphics[trim=0 25 35 48,clip, width=\linewidth]{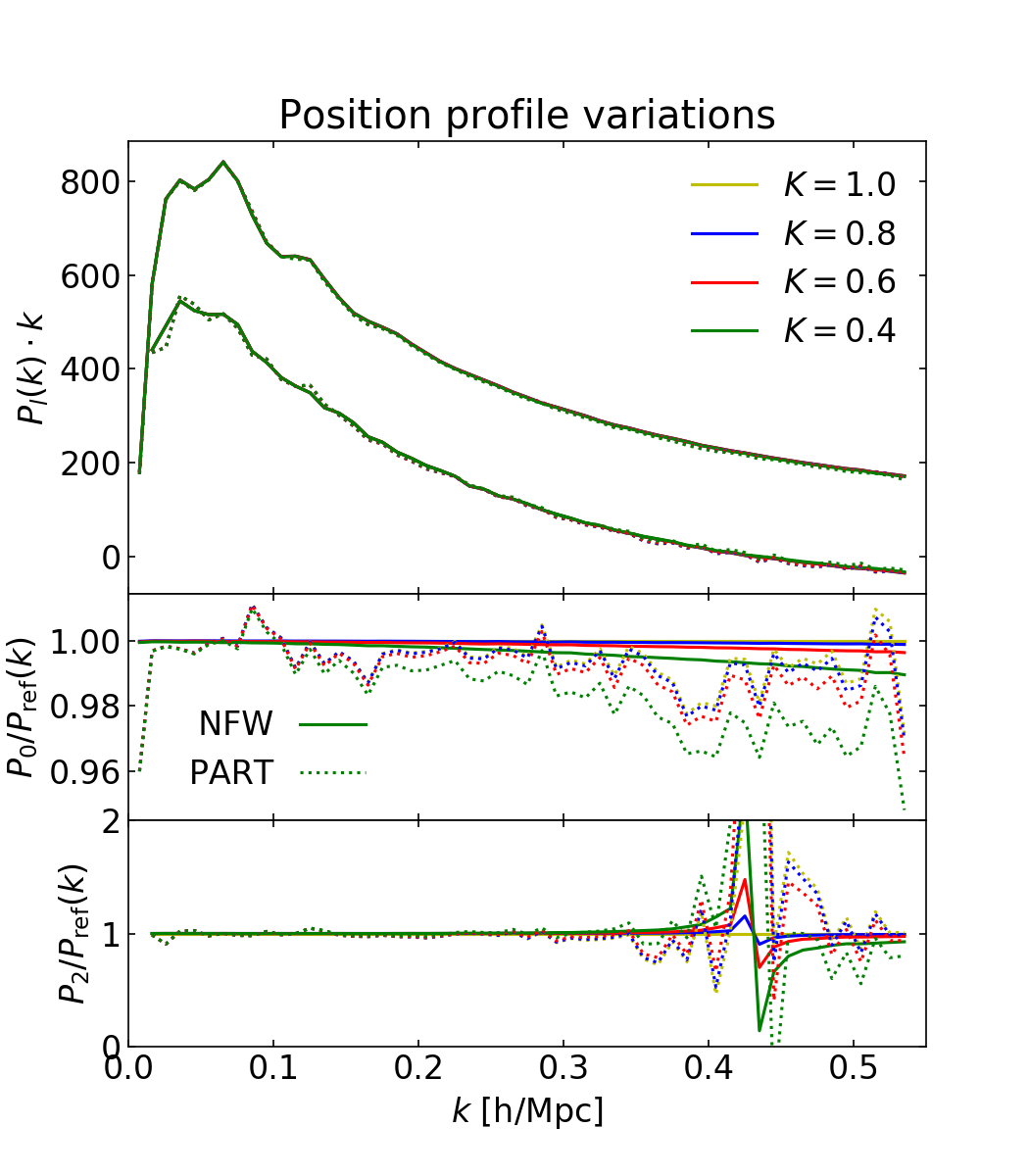}
    \caption{Power Spectrum multipoles of mocks with different position assignment (NFW vs \parts\ and different values of $K$). These correspond to the same mocks as shown in \Fig{fig:K}. In the top panel we present both the monopole (upper set of lines) and quadrupole (lower set of lines). In the middle/lower panel we show the ratio of the monopole/quadrupole to the reference model: \{HOD-3, \fsat$=0.30$, NFW, $K=1$, $\beta=0$, $\alpha_v=1$\}. 
    }
    \label{fig:PK_pos}
\end{figure}

\begin{figure}
    \centering
    \includegraphics[trim=5 35 35 60,clip, width=\linewidth]{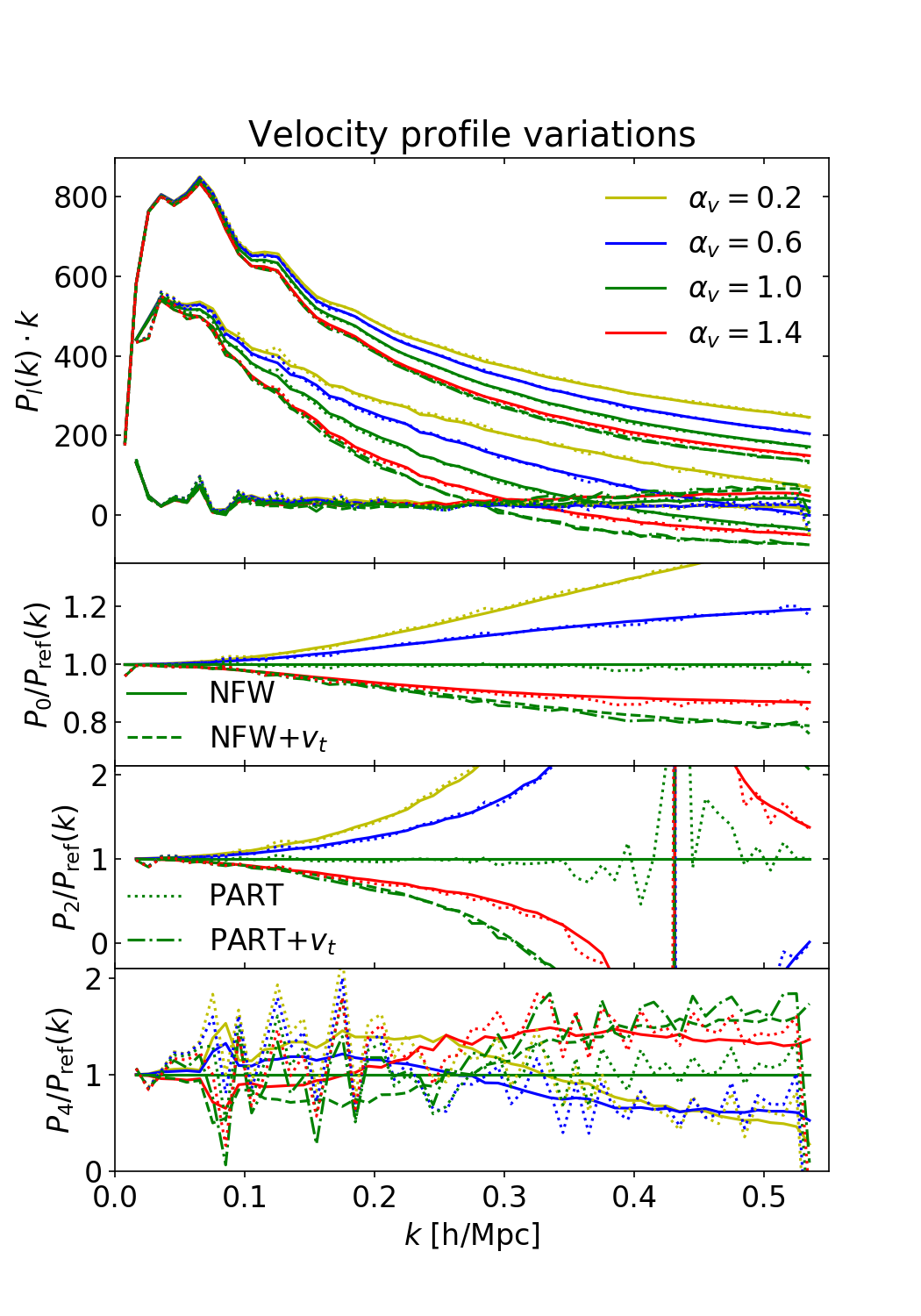}
    \caption{ Power Spectrum multipoles of mocks with different velocity assignment (NFW vs \parts, different values of $\alpha_v$ and optionally $v_t=v^{\rm infall}$) as indicated in the legend. Due to visualisation purposes we show slightly different combinations of choices with respect to \Fig{fig:vel}. In the top panel we represent  the monopole (upper set of lines), quadrupole (middle set of lines) and hexadecapole (lower set of lines). In the lower panels we show the ratio of the monopole ($P_0$), quadrupole ($P_2$) and hexadecapole ($P_4$) to the reference model: \{HOD-3, \fsat$=0.30$, NFW, $K=1$, $\beta=0$, $\alpha_v=1$\}.
    }
    \label{fig:PK_vel}
\end{figure}

For completeness, in this Appendix, we analyse the clustering of the mock catalogues presented in \Sec{sec:mocks} (Figs. \ref{fig:2PCF_fsat}, \ref{fig:beta}, \ref{fig:K}, \ref{fig:vel}) in Fourier space (Figs. \ref{fig:Pk_HOD}, \ref{fig:Pk_PDF}, \ref{fig:PK_pos}, \ref{fig:PK_vel}). The power spectrum multipoles are computed in the $L=3$Gpc$/h$ box using periodic conditions and redshift space distortions along the $Z$-axis. We use the {\sc nbodykit} code \citep{nbodykit} to compute the power spectrum multipoles $P_\ell(k)$ using a grid size of $512^3$ and Triangular-Shape-Cloud mass assignment together with interlacing \citep{interlacing}. This configuration gives a Nyquist frequency of $k_{\rm Ny}=0.54 h/{\rm Mpc}$.  We refer to \citet{demattia2020} for further details. 

Qualitatively, the results are similar to those found in \Sec{sec:mocks}, but in some cases the information is spread differently in $k$ space. We find that the effects that did not change the quadrupole in configuration space, $\xi_2$, have a very small effect in $P_0$ and $P_2$ at the scales shown here. This is very clear in \Fig{fig:PK_pos}, where profile variations barely change the power spectrum multipoles. In those lines, in \Fig{fig:Pk_PDF}, we find a mild effect of the PDF onto the multipoles. On the other hand, the effects that did change the $\xi_2$ have a strong effect on $P_2$ but are also relevant for $P_0$. This is clearly seen when varying the fraction of satellites (\Fig{fig:Pk_HOD}) or the velocity profiles (\Fig{fig:PK_vel}). 
The above findings can be summarised by saying that power spectrum multipoles, within the explored scales, are affected only by the Finger-of-God effect from the 1-halo term.

In \Fig{fig:PK_vel}, for completeness, we also show the hexadecapole $P_4(k)$. We find the the dependence on the satellite velocity assignment scheme is relatively low. We note that \citet{demattia2020} finds that, for the eBOSS uncertainty, the hexadecapole is compatible with zero. 

A remarkable result is that, whereas for the analysis in configuration space we could clearly split the 1-halo contributions from the large scales signal used for cosmology, in $P_\ell(k)$ the 1-halo does affect modes with $k\sim 0.1 h/Mpc$, which are scales that are also used BAO and RSD analysis. 
Despite this, we show in Appendix \ref{sec:cosmo}, that this has a negligible effect on  the derived cosmological constraints, as the effects are absorbed by the nuisance parameters.

\section{Cosmological Constrains}
\label{sec:cosmo}

In this Appendix, we present the results of fitting a Redshift-Space-Distortion and anisotropic Alcock-Paczynski model (with $f\sigma_8$, $\alpha_\parallel$, $\alpha_\perp$ as free parameters) to the mock catalogues presented in \Sec{sec:mocks} based on their power spectrum multipoles ($P_l(k)$, Appendix \ref{sec:fourier}). These mocks have been analysed following the methodology used for the eBOSS data in \citet{demattia2020}. The model considered here combines Regularised Perturbation Theory with the Taruya, Nishimichi and Saito (TNS) RSD model \citep{TNS,TNSRPT}.

A representative subset of the mocks presented in this work is also shown in \citet{alam2020}, together with other complementary $N$-Body mocks. 
In that paper we not only show the results from fitting those mocks in Fourier space, but also in configuration space $\xi_\ell$, following the methodology used for the data in \citet{tamone2020}. We refer to \citet{alam2020} for details on the way the fits presented here were performed, with the only difference being that here we only report results using the $Z$-axis as the line of sight. 

\citet{alam2020} also reports a systematic error budget due to possible theory uncertainties of (\{ 3.3\%, 1.8\%, 1.5\%\} for $\{f\sigma_8,\alpha_\parallel, \alpha_\perp\}$). 
These are conservative systematic error budgets, corresponding to 2$\times \sigma_{\rm stat}$, as no significant deviation was found at the $2$-$\sigma_{\rm stat}$ level, with $\sigma_{\rm stat}$ being the statistical uncertainty on the \ors.

For completeness, here we present the fits in Fourier space of a wider range of mocks sweeping the parameter space of the HOD considered as done in the main body of this paper. We consider this analysis more necessary for the Fourier space case, as the cosmological and HOD scales mix more than in configuration space. 

\begin{figure*}  \includegraphics[trim=60 0 100 20,clip,width=1\textwidth]{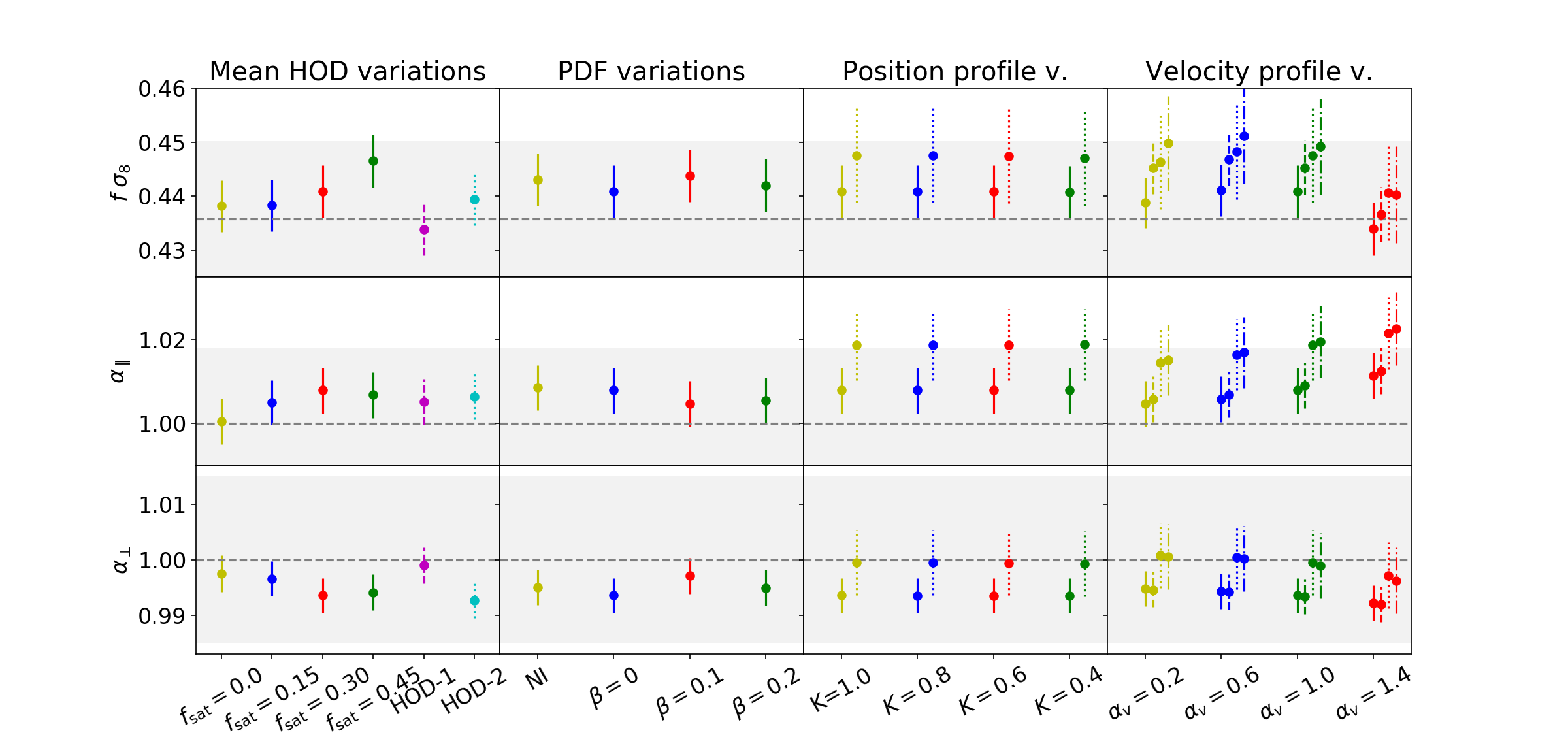}
  \caption{Cosmological parameters fits for different mocks presented in this paper. On the top row we show the product of the growth rate and the normalisation of the power spectrum $f\sigma_8$, on the middle row of panels we show the Alcock-Paczynski distortion along the line of sight $\alpha_\parallel$ and on the bottom row we show the Alcock-Paczynski parameter on the angular direction $\alpha_\perp$. From left to right we show the same effects as in Figs. \ref{fig:Pk_HOD}, \ref{fig:Pk_PDF}, \ref{fig:PK_pos} \& \ref{fig:PK_vel}, respectively, following the same line-styles. The error bars show the statistical uncertainty of the best fit. The horizontal grey dashed line represents the true value for \ors, and the grey band the systematic error reported in \citet{alam2020}. 
  }
\label{fig:cosmo}
\end{figure*}

\Fig{fig:cosmo} shows the results all the fits, following the same notation and line-styles as in Figs. \ref{fig:Pk_HOD}, \ref{fig:Pk_PDF}, \ref{fig:PK_pos} \& \ref{fig:PK_vel}. We find that the fits are consistent with the {\sc Outer Rim} cosmology (dashed horizontal line) within the systematic error budget, marked by the grey bands in \Fig{fig:cosmo}.

From \Fig{fig:cosmo} it is clear that the cosmological constraints are robust against the different details of the HOD modelling, and the same result is found in \citet{alam2020}. We also note that the theoretical systematic error reported is an order of magnitude lower than the statistical error of eBOSS. 

\section{Scale cuts}
 \label{sec:scales}

\begin{figure*}
    \centering
    \includegraphics[trim = 15 0 50 20,clip,width=0.33\linewidth]{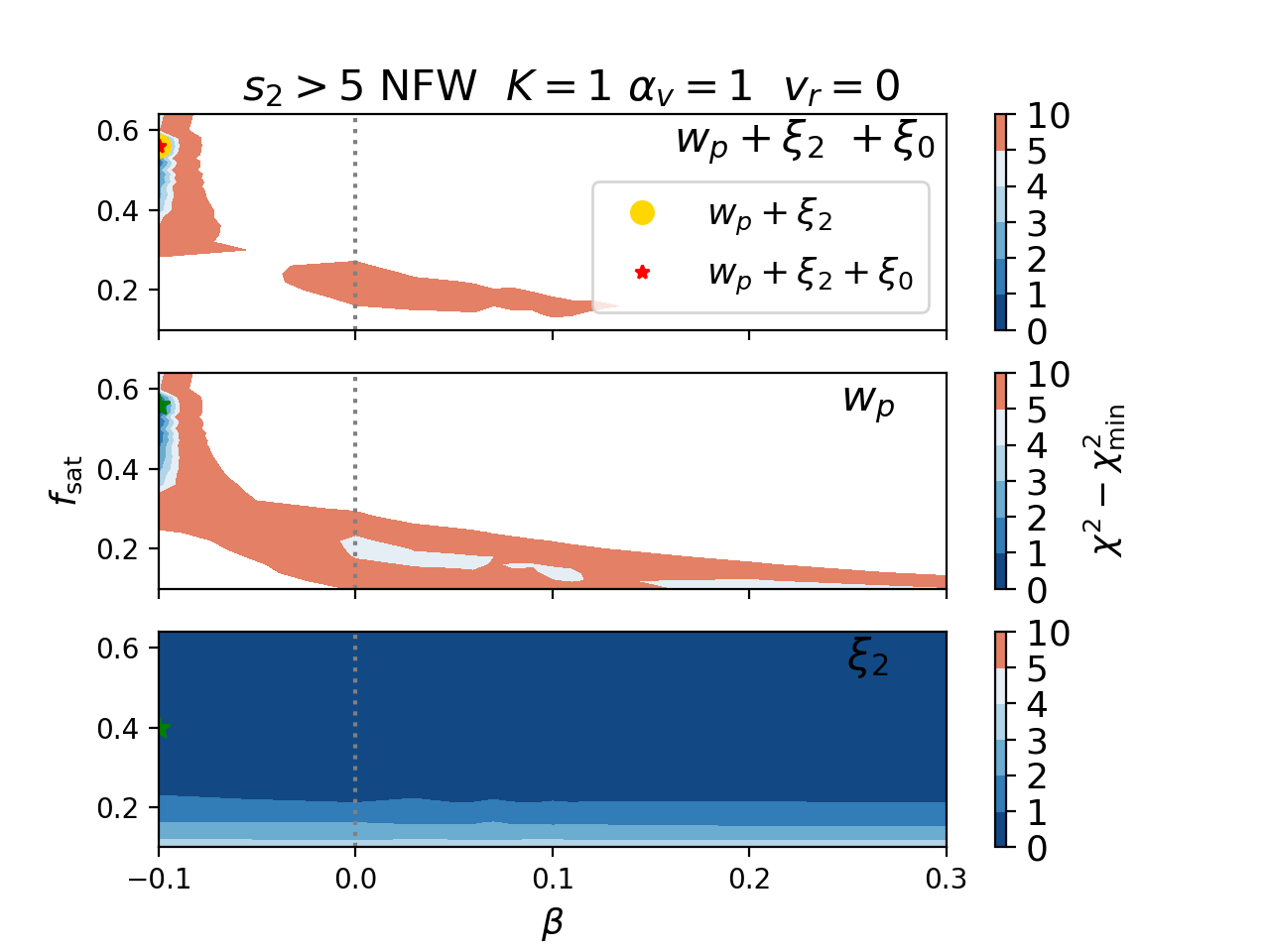} \includegraphics[trim = 15 0 50 20,clip,width=0.33\linewidth]{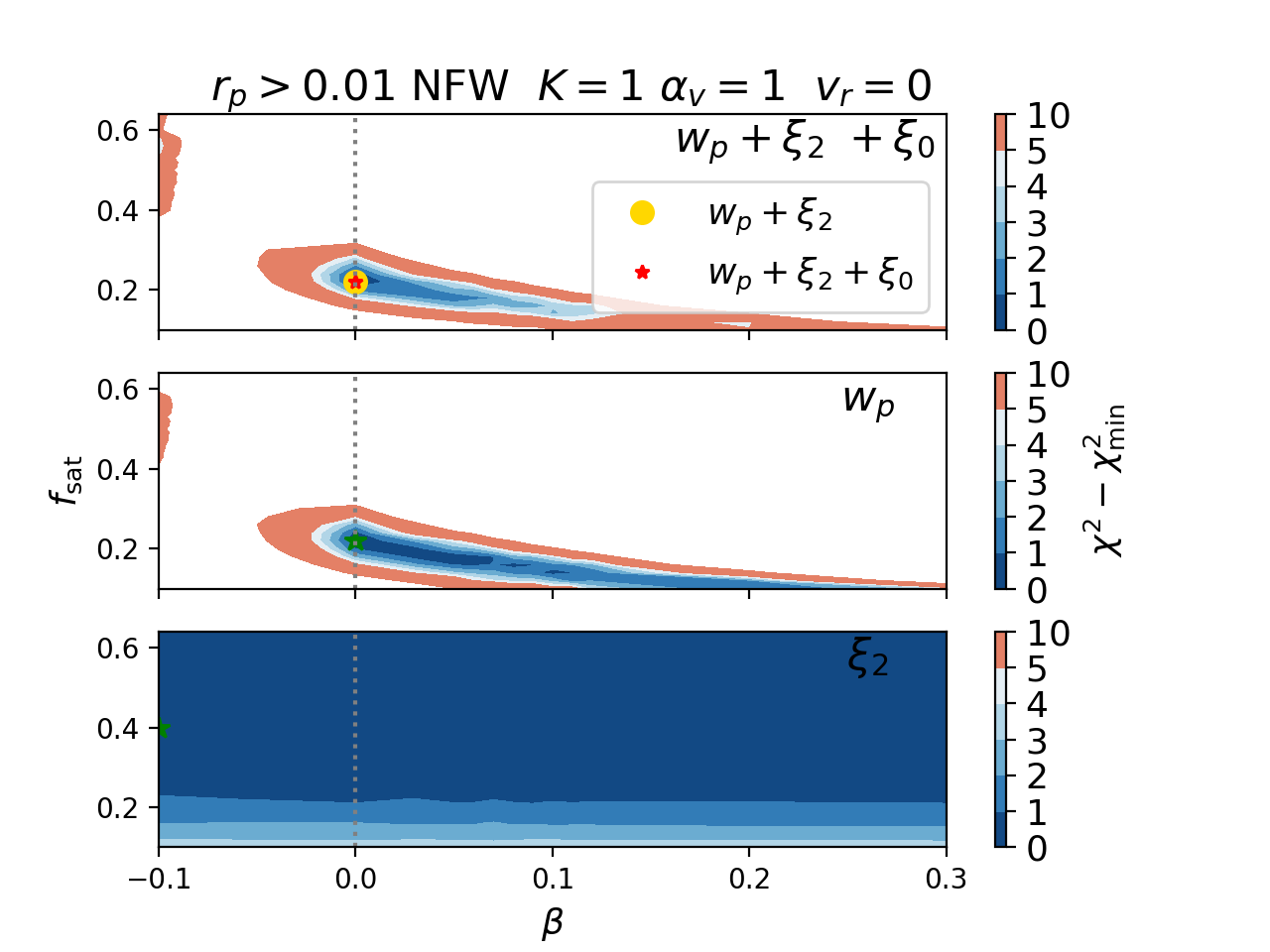} \includegraphics[trim = 15 0 50 20,clip,width=0.33\linewidth]{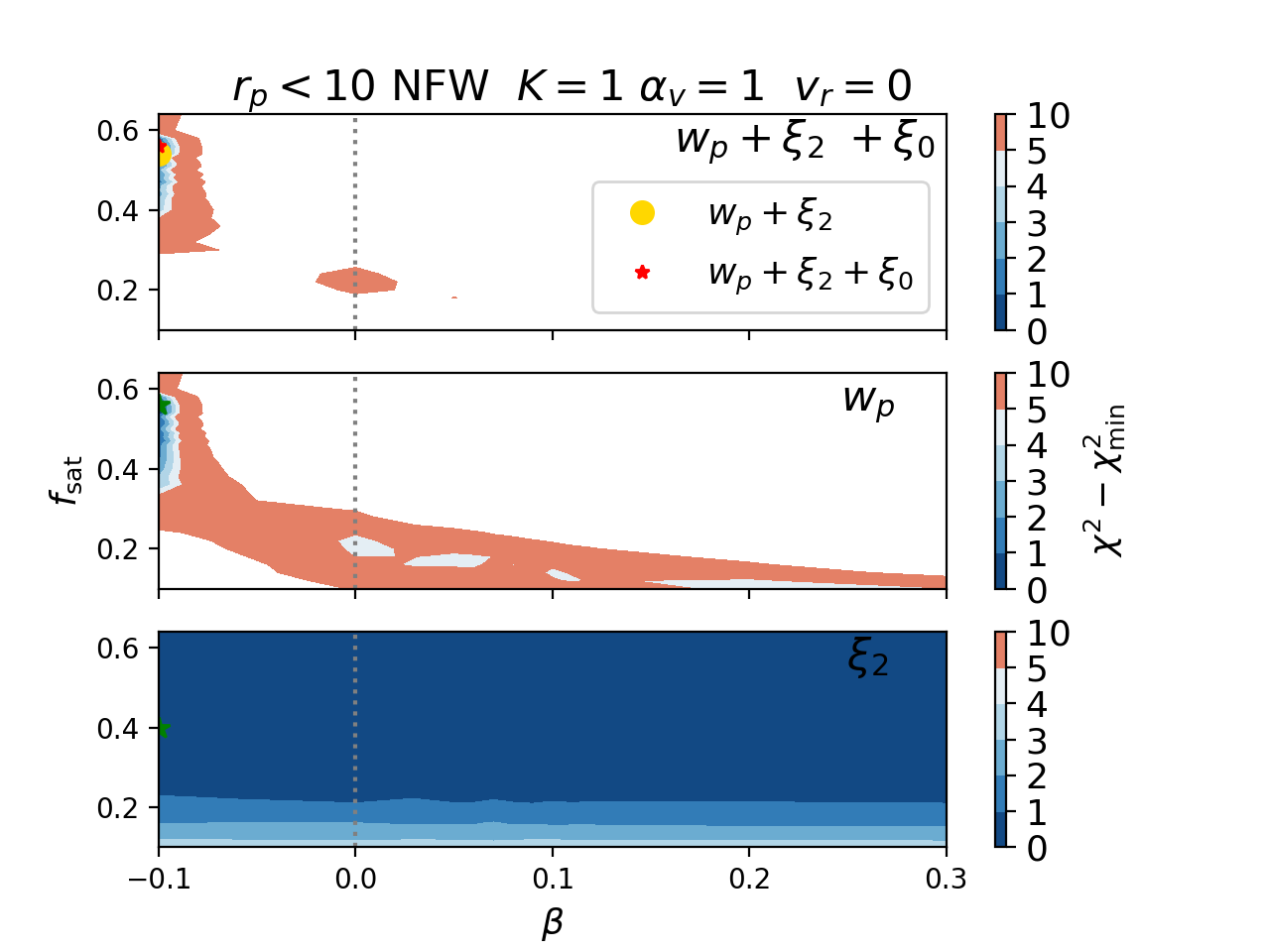}
    
    \includegraphics[trim = 15 0 50 20,clip,width=0.33\linewidth]{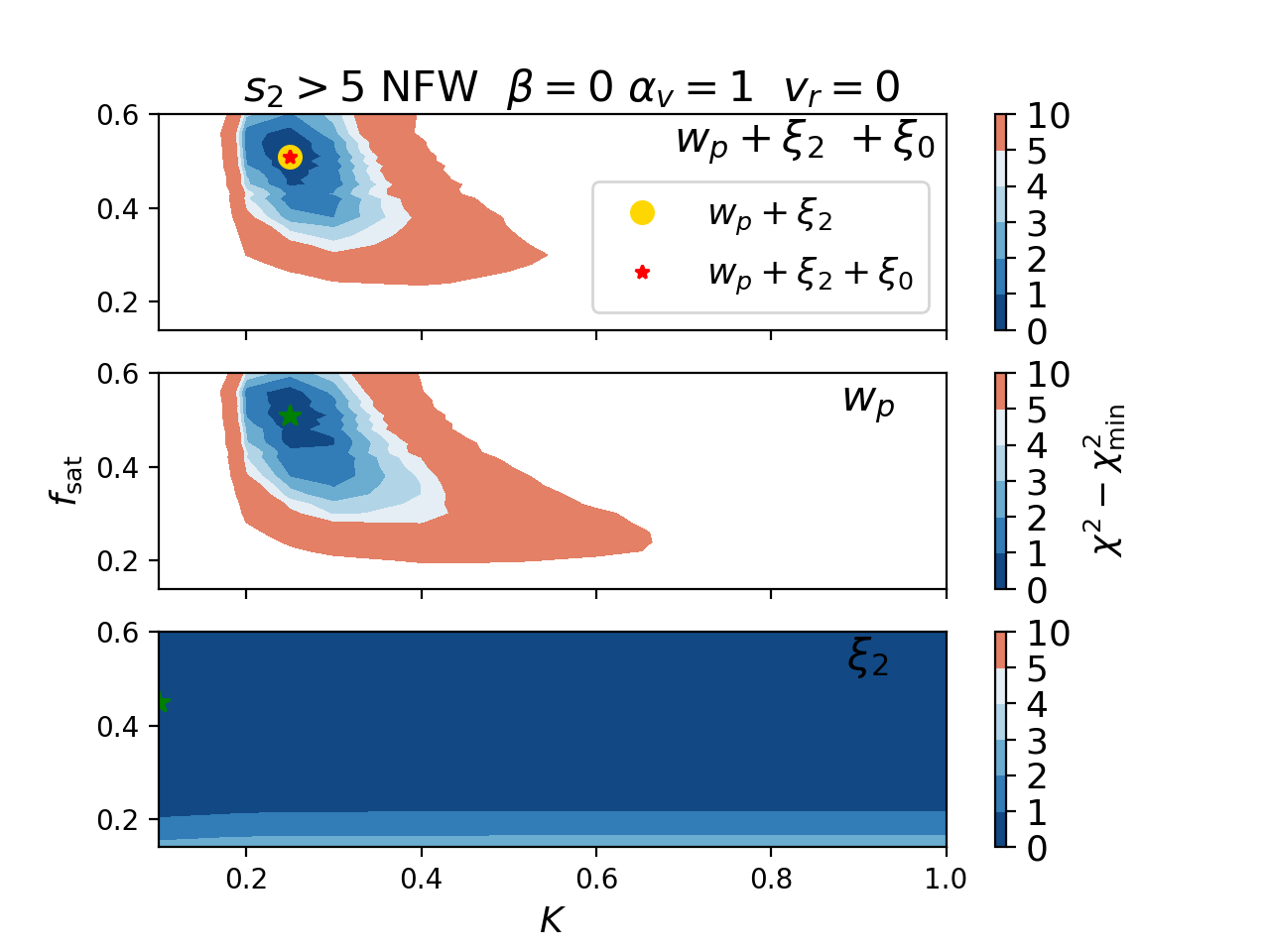} \includegraphics[trim = 15 0 50 20,clip,width=0.33\linewidth]{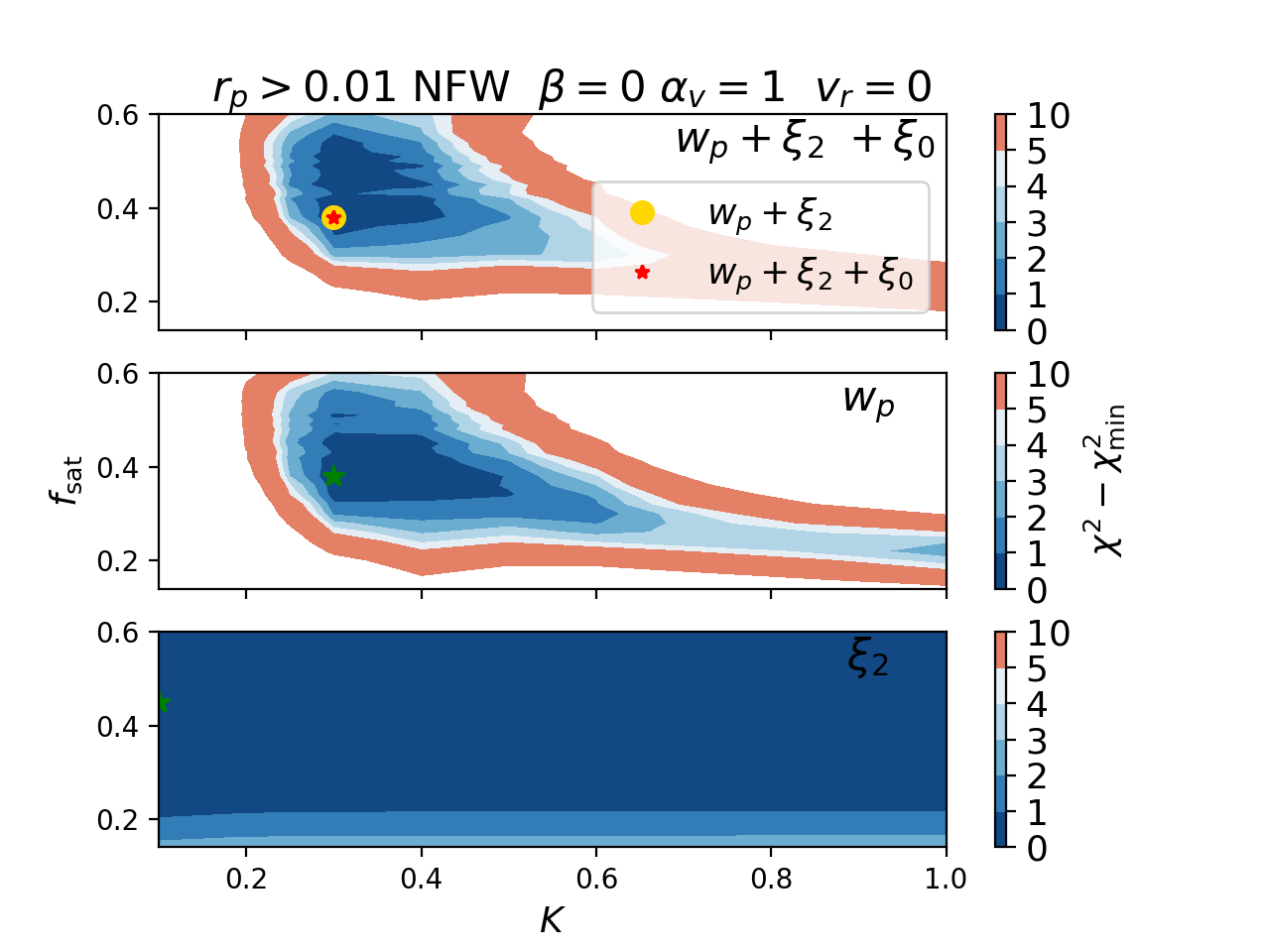} \includegraphics[trim = 15 0 50 20,clip,width=0.33\linewidth]{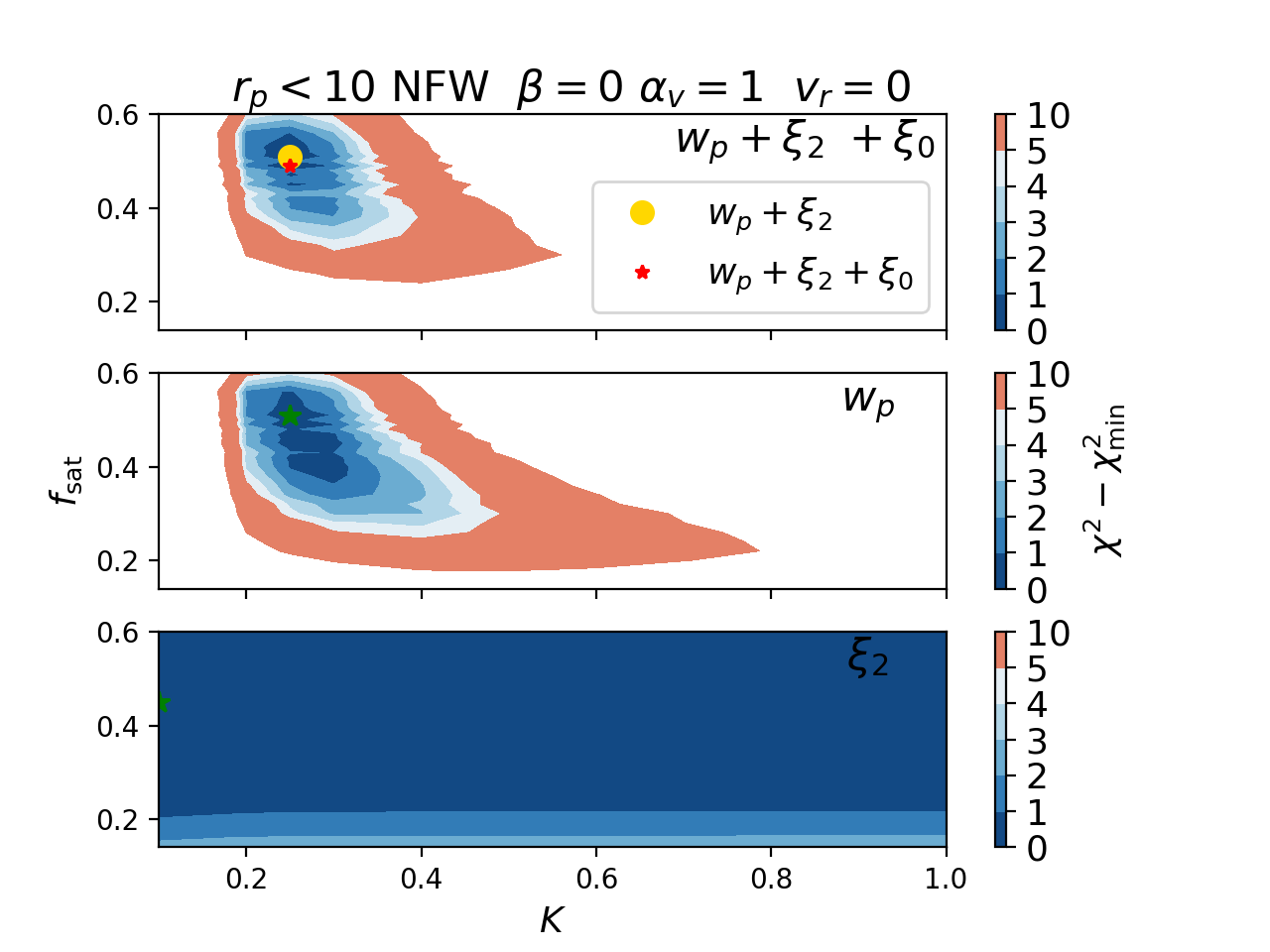}
    
\includegraphics[trim = 15 0 50 20,clip,width=0.33\linewidth]{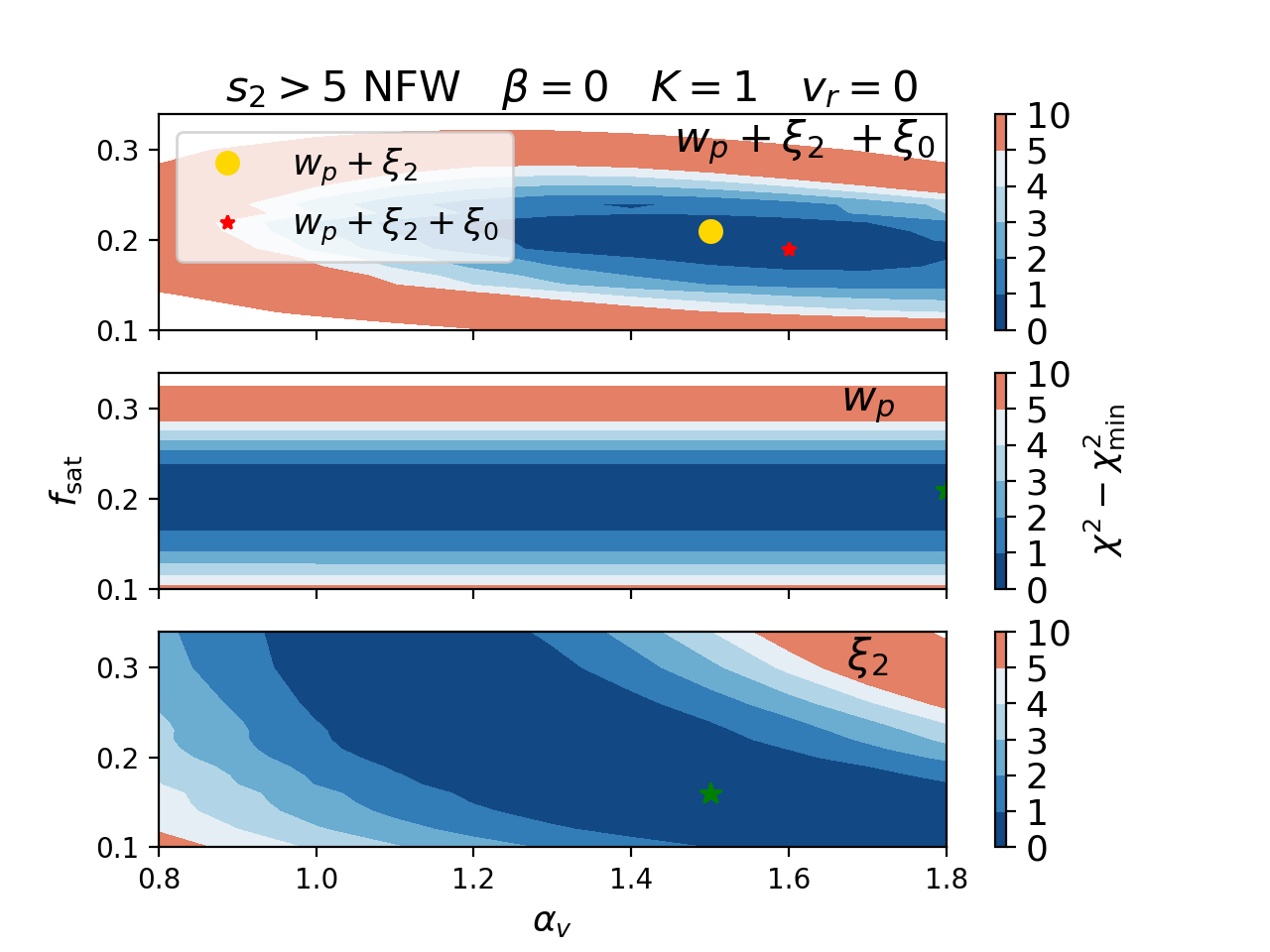}\includegraphics[trim = 15 0 50 20,clip,width=0.33\linewidth]{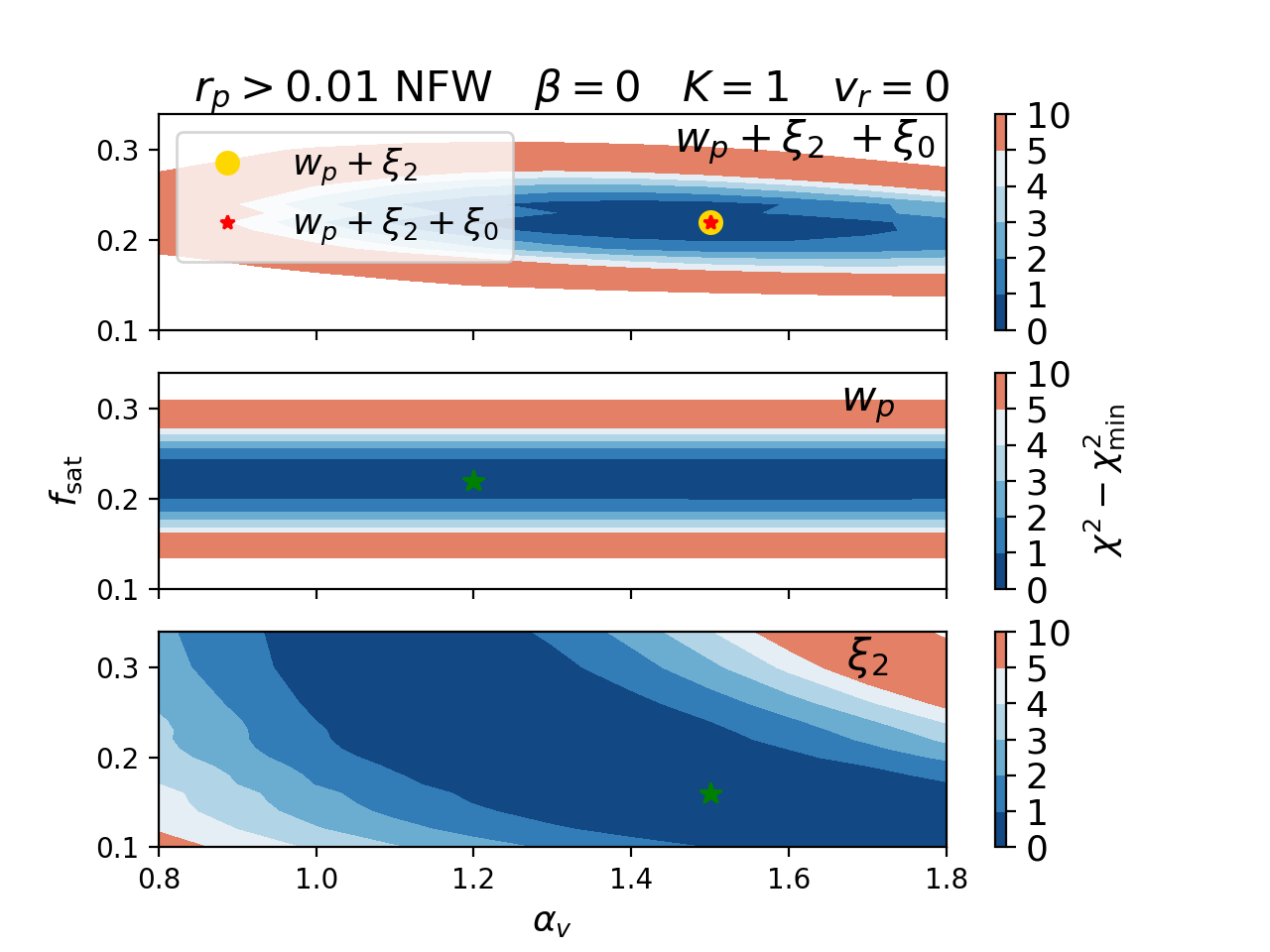}\includegraphics[trim = 15 0 50 20,clip,width=0.33\linewidth]{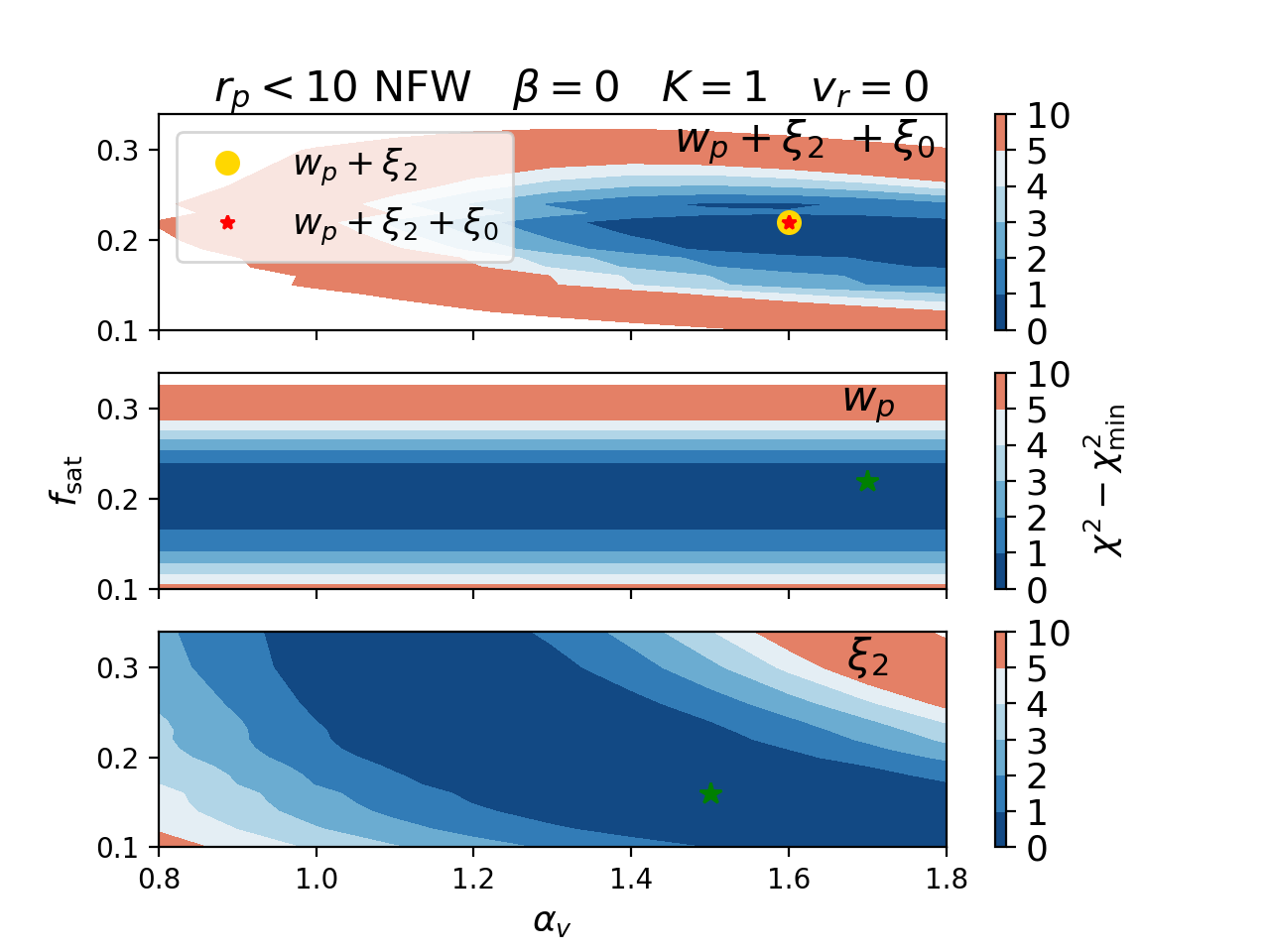}
    \caption{$\chi^2$ contours for varying scale cuts. The fiducial choices of scale cuts are $r_{p,\rm min}=0.02$, $r_{p, \rm max}=4.5$, $s_{2,\rm max}=10$, $s_{2, \rm max}=25$ in Mpc$/h$ (as in Sec. \ref{sec:results}), whereas we set $s_{2,\rm max}=5$ in the left column of sub-figures, $r_{p,min}=0.01$ in the central column and $r_{p,\rm max}=10$ in the right column. We redo the same plots as in Fig. \ref{fig:chi2_2D} with varying \fsat\ and $\beta$ (top row of sub-figures),  with varying \fsat\ and $K$ (middle row), and varying \fsat\ and $\alpha_v$ free (bottom row).  
    }
    \label{fig:scales}
\end{figure*}

 In \Sec{sec:results}, when fitting the HOD mocks to the data, we chose the scale cuts defined by Eq. \ref{eq:scales}. Although we justified these choices, we explore here what would have happened if we chose different scale cuts. Since we focused in the constraining power of the quadrupole and projected correlation function and their complementarity we also focus here in those statistics. 
 
For the quadrupole $\xi_2(s_2)$, we are limited on the upper side by the systematics at $s_{2,\rm max}=25$ Mpc$/h$ (See Sec. \ref{sec:data}), whereas on the lower side the choice was relatively more arbitrary. Hence, we also test here imposing $s_{2,\rm min}=5$Mpc/$h$. Results are shown in the left column of \Fig{fig:scales}. The differences of these plots with respect to the original ones in \ref{fig:chi2_2D} are marginal, hence, finding consistency. 

For the projected correlation function $w_p (r_p)$, we change both the upper and lower scale limits. We show the results of changing $r_{p,\rm min}$ to 0.01 Mpc/$h$ in the middle column of \Fig{fig:scales}. The differences for the varying \{\fsat + $\alpha_v$\} case are marginal, whereas it has a larger effect on both the varying \{\fsat + $\beta$\} and varying \{\fsat + $K$\} cases, where the influence of $w_p$ is larger. By reducing the $r_{p,\rm min}$ to 0.01 we are including in the fit  the first three $r_p$ points of Fig. \ref{fig:bestfit}, that show a sudden change of behaviour with respect to the rest of points, reason why they were removed in the first place. Hence, this larger effect on the scale cut is expected. Despite this larger change, the differences are  smaller than those found across Sec. \ref{sec:results} for different choices of modelling. 

Finally, if we consider moving the upper limit of $w_p(r_p)$ to $r_{p,\rm max}=10$Mpc/$h$ (right column of sub-figures in \Fig{fig:scales}), we find marginal differences. 

In summary, our conclusions are robust against changes in the considered scale range for the observational data.


\bsp	
\label{lastpage}

\end{document}